\documentclass[useAMS,usenatbib,english,a4paper,fleqn]{mnras}
\usepackage{newtxtext}
\usepackage{mathptmx}
\usepackage{txfonts}
\usepackage[T1]{fontenc}
\usepackage{ae,aecompl}

\usepackage{graphicx}	% Including figure files
\usepackage{amsmath}	% Advanced maths commands
\usepackage{amssymb}	% Extra maths symbols
\usepackage{subfig}      %
\usepackage{caption}      
\usepackage{natbib}
\usepackage{booktabs}
\usepackage[normalem]{ulem}
\usepackage{cleveref}
\crefname{figure}{Fig.}{Figs.}
\crefname{table}{Table}{Tables}
\crefname{chapter}{Chapter}{Chapters}
\crefname{section}{Section}{Sections}
\newcommand{\units}[1]{\, \mathrm{#1}}
\newcommand{\unitstx}[1]{\mathrm{#1}}
\newcommand{\diff}{\mathop{}\!\mathrm{d}}

\newcommand{\equnp}[1]{eq.~\ref{eq:#1}}

\newcommand{\subrfig}[1]{\protect\subref{fig:#1}}

\newcommand{\orcid}[1]{\href{https://orcid.org/#1}{\includegraphics[scale=0.08]{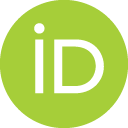}}}
\newcommand{\rgal}{r_\mathrm{gal}}
\newcommand{\rhGas}{r_{1/2,\mathrm{gas}}}

\newcommand{\tcool}{t_{\mathrm{cool}}}
\newcommand{\tff}{t_{\mathrm{ff}}}
\newcommand{\ts}{t_\mathrm{s}}
\newcommand{\Mbh}{M_{\mathrm{BH}}}

\newcommand{\mproton}{m_{\mathrm{p}}}
\newcommand{\kboltz}{k_{\mathrm{B}}}
\newcommand{\zeq}[1]{\mbox{$z=#1$}}

\newcommand{\msun}{\units{M_\odot}}
\newcommand{\Rv}{R_{\mathrm{200,c}}}   %{R_{\mathrm{vir}}}
\newcommand{\Mv}{M_{\mathrm{200,c}}}   %{M_{\mathrm{vir}}}
\newcommand{\Vv}{V_{\mathrm{vir}}}
\newcommand{\Tv}{T_{\mathrm{vir}}}

\graphicspath{ {figs/} }
\DeclareGraphicsExtensions{.pdf,.png,.svg}
\DeclareCaptionFormat{cont}{#1 (cont.)#2#3\par}

\title[AGN Feedback in IllustrisTNG ]{Ejective and preventative: the IllustrisTNG black hole feedback and its effects on the thermodynamics of the gas within and around galaxies}

\author[Zinger et al.]{Elad Zinger$^1$\thanks{E-mail:\href{mailto:elad.zinger@mail.huji.ac.il}{elad.zinger@mail.huji.ac.il}}\orcid{0000-0002-6316-3996},Annalisa Pillepich$^1$, Dylan Nelson$^2$\orcid{0000-0001-8421-5890}, Rainer Weinberger$^3$\orcid{0000-0001-6260-9709},\newauthor 
R\"udiger Pakmor$^2$\orcid{0000-0003-3308-2420}, Volker Springel$^2$\orcid{0000-0001-5976-4599}, Lars Hernquist$^3$, Federico Marinacci$^4$\orcid{0000-0003-3816-7028} \newauthor and Mark Vogelsberger$^5$\orcid{0000-0001-8593-7692}\\
\\
$^{1}$Max-Planck-Institut f\"ur Astronomie, K\"onigstuhl 17, D-69117 Heidelberg, Germany\\
$^{2}$Max-Planck-Institut f\"ur Astrophysik, Karl-Schwarzschild-Str.\@ 1, 85748 Garching, Germany\\
$^{3}$Institute for Theory and Computation, Harvard-Smithsonian Center for Astrophysics, 60 Garden Street, Cambridge, MA 02138, USA\\
$^{4}$Department of Physics \& Astronomy, University of Bologna, via Gobetti 93/2, 40129 Bologna, Italy\\ 
$^{5}$Kavli Institute for Astrophysics and Space Research, Massachusetts Institute of Technology, Cambridge, MA 02139, USA}

\date{Accepted XXX. Received YYY; in original form ZZZ}

\pubyear{2020}

\begin{document}
\pagerange{\pageref{firstpage}--\pageref{lastpage}}
\maketitle
\label{firstpage}

\begin{abstract}
Supermassive black holes (SMBHs) which reside at the centres of galaxies can inject vast amounts of energy into the surrounding gas and are thought to be a viable mechanism to quench star-formation in massive galaxies. Here we study the $10^{9\--12.5}\msun$ stellar mass central galaxy population of the IllustrisTNG simulation, specifically the TNG100 and TNG300 volumes at \zeq{0}, and show how the three components -- SMBH, galaxy, and circumgalactic medium (CGM) -- are interconnected in their evolution. We find that gas entropy is a sensitive diagnostic of feedback injection. In particular, we demonstrate how the onset of the low-accretion BH feedback mode, realised in the IllustrisTNG model as a kinetic, BH-driven wind, leads not only to star-formation quenching at stellar masses  $\gtrsim10^{10.5}\msun$ but also to a change in thermodynamic properties of the (\emph{non}-star-forming) gas, both within the galaxy and beyond. The IllustrisTNG kinetic feedback from SMBHs increases the average gas entropy, within the galaxy and in the CGM, lengthening typical gas cooling times from $10\--100\units{Myr}$ to $1\--10\units{Gyr}$, effectively ceasing ongoing star-formation and inhibiting radiative cooling and future gas accretion. In practice, the same AGN feedback channel is simultaneously `ejective' and `preventative' and leaves an imprint on the temperature, density, entropy, and cooling times also in the outer reaches of the gas halo, up to distances of several hundred kiloparsecs. In the IllustrisTNG model, a long-lasting quenching state can occur for a heterogeneous CGM, whereby the hot and dilute CGM gas of quiescent galaxies contains regions of low-entropy gas with short cooling times.  
\end{abstract}

% Select between one and six entries from the list of approved keywords.
% Don't make up new ones.

\section{Introduction}\label{sec:intro}
When characterized within the parameter space of colours/SFR vs.\@
magnitude/stellar mass, galaxies in the observed Universe are
generally found in one of two regions: the star-forming `blue cloud'
comprised of galaxies with an abundance of young stars and the
quiescent `red sequence' in which the star formation has been halted
\citep[e.g.\@][]{Kauffmann2003,Baldry2004,Balogh2004,Hogg2004}. Observations
show that these populations are already well-established around $z\sim
1$ \citep{Bell2004,Cooper2007} and that the `blue cloud' is mostly
populated by low-mass galaxies, while massive galaxies lie in the
`red sequence' with the transition occurring at a stellar mass
of $\sim 3\times 10^{10}\msun$ \citep{Kauffmann2003}.

When comparing the populations of `central' galaxies, i.e.\@ the most
massive galaxies within a dark matter host, to satellite galaxies, one
finds a different quenched fraction at fixed stellar mass in each
category \citep[e.g.\@][]{Bosch2008,Wetzel2012,Geha2012,Wetzel2013}. This indicates that different physical processes govern the transition from star-forming to quenched in each class, with the quenching of satellites being driven by their environment \citep{Wang2009,Peng2010,Woo2013}.

While for the quenching of satellite galaxies a host of physical ``environmental'' mechanisms have been proposed,  e.g.\@  gas removal by ram-pressure \citep{Gunn1972,McCarthy2008,Zinger2018} or tidal forces
\citep{Gnedin2003,Villalobos2014}, the processes which terminate the
formation of new stars in the central galaxy population must originate from within the galaxy itself \citep{Peng2010,Liu2019}. In this study we focus on central galaxies and thus exclude the effects of ``environmental'' processes in the transition of galaxies from star-forming to quenched. 

For quenching of central galaxies, energy injection (feedback) is often invoked as the main driver, either from supernov\ae (SNe) or from the supermassive black holes (SMBHs) that reside in the centres of galaxies. SMBHs are found in galaxies at very high redshifts \citep{Banados2018} as well as in our local Universe \citep[e.g.][]{Reines2015}. When accreting mass, SMBHs can inject energy into their host galaxies and are known as Active Galactic Nuclei (AGN). 

While the energy feedback from stars and SNe can be an important factor
in shaping the evolution of low-mass galaxies and galaxies in general
\citep{Dekel1986,Hopkins2011}, from a theoretical perspective it is
widely accepted that, in galaxies at and above the \mbox{$\sim3\times10^{10}\msun$} stellar mass transition scale, AGN
feedback plays an important role in regulating star formation and
possibly in quenching galaxies
\citep{DiMatteo2005,Silk2013,Somerville2015,Penny2018}. More
specifically, theoretical studies with numerical cosmological
simulations and semi-analytic models that follow galaxy evolution
with and without AGN feedback 
\citep[e.g.\@][]{McCarthy2011,Dubois2016,Kaviraj2017} show
that the quenching of \emph{entire populations} of galaxies in a manner consistent with observations is only possible when AGN feedback is included. This is especially
true at the high-mass end, where feedback from supernov\ae 
and massive stars is ineffective in shutting off star formation
\citep{Bower2006,Weinberger2017,Choi2018}. As such, AGN feedback is a critical component of most galaxy formation models in both semi-analytic
models and large-volume simulations \citep{ Vogelsberger2014,
  Schaye2015,McCarthy2017,Bower2017,Weinberger2017}.
\begin{table*}
\centering
  	 \begin{tabular}{@{}lcccccc@{}}
	 \toprule
	 Name & $L_\mathrm{box}\,[\unitstx{cMpc}]$ & $m_{\mathrm{baryon}}\,[10^{6}\msun]$ &$\epsilon_\mathrm{gas,min}\,[\unitstx{ckpc}]$& \multicolumn{3}{c}{Mean cell size (std.\@ dev.) of non-SF gas (\zeq{0}) $[\unitstx{kpc}]$ } \\ 
	 %&  &   &  & \multicolumn{3}{c}{ } \\
        & & & & Galactic Gas & Inner CGM & Outer CGM  \\ 
         \midrule
        TNG100 & 110.7  & 1.4 & 0.19 & 1.8 (0.98) & 5.7 (2.7)   & 10.4 (3.1) \\
        TNG300 & 302.6  & 11  & 0.37 & 4.1 (2)    & 12.1 (5.5) &  21.3 (6.3) \\
 	\bottomrule 
	\end{tabular}
	 \caption{The physical parameters of the TNG100 and TNG300
           simulation runs: size of simulation box, the target mass
           gas cells which is roughly equal to the average initial
           mass of stellar particles, and the minimum comoving value of the adaptive gas gravitational softening. The final three column show the mean size of the non-star-forming gas cells at \zeq{0} in the gas components we study (see \cref{sec:gasComponents}).}% The size of a given cell is defined to be $(m/\rho)^{1/3}$.}
	 \label{tab:TNGproperties}
 \end{table*}
 
From an observational perspective, the association of AGN feedback to the
quenching of star formation is more nuanced. On the one hand, AGN
feedback may not account for quenching in all massive galaxies and
other secular processes have been invoked to explain the quenched
population of central galaxies at the high-mass end
\citep{Smethurst2016}. On the other hand, while there are strong
observational indications that AGN feedback can lead to quenching
\citep{Hickox2009,Cheung2016}, strong AGN outflows have also been
observed in massive star-forming galaxies, at intermediate redshifts
\citep{Genzel2014,FoersterSchreiber2014}. Yet, the existence of
observed scaling relations between the large-scale properties of the
host galaxies -- e.g.\@ luminosity, stellar mass, and stellar velocity
dispersion -- and the mass of the SMBH suggest that the evolution of
the latter and its host galaxy are intertwined \citep[][and references
  therein]{Kormendy2013,Heckman2014}.

Though SMBHs comprise a small fraction of the mass and volume of a
galaxy, their effects can be seen and measured in the galaxy as well
as out to hundreds of kiloparsecs in the surrounding gas
\citep{Birzan2004,Fabian2012}. In galaxy clusters, for example, the
energy from BH feedback is known to be a key player in compensating
the high cooling rates inferred within the Intra-Cluster Medium (ICM)
and in setting the thermodynamic conditions in the central regions
of the cluster \citep{McNamara2007,Boehringer2010,Prasad2015}.
Various observations indicate that the thermodynamics of the gaseous
atmospheres surrounding massive galaxies (i.e.\@ not just those at the
centres of the most massive clusters) are also regulated by the energy
injected by black holes \citep{Nulsen2009,Werner2019}. In
atmospheres heated in such a manner the cooling and accretion of gas
onto the galaxy can be suppressed and thus future epochs of star formation
in the central galaxy can also be suppressed. Through this `preventative'
feedback \citep{Somerville2015}, the AGN not only quenches the galaxy
but can maintain its quenched state for long periods of time.

Observations of AGN suggest two main modes of activity: AGN in which
the mass accretion rates is high, exceeding a few percent of the
Eddington rate, are dominated by radiation in IR, optical, UV all the
way to X-ray wavelengths, originating from the luminous gas in
the black hole accretion disk
\citep{Crenshaw2010,VillarMartin2011,Woo2016,Rupke2017}. This energy
deposition mode, mediated by radiation-driven winds, is often called the
`quasar' mode. In the other case, linked to low accretion rates,
mechanical energy is deposited in the surrounding gas, often resulting
in radio jets or large synchrotron emitting plasma lobes. This
feedback mode is often referred to as the `radio' or `mechanical' mode
\citep{Fabian2012,McNamara2007,Shin2016}. From an observational
perspective, a general picture has emerged whereby the first mode of
feedback, the high-accretion channel, is responsible for triggering
the quenching of the galaxies, while the second mode, at low accretion
rates, is often referred to as a `maintenance' mode 
\citep{Mathews2003,Best2005,Rangel2014}. Still, this separation
remains uncertain and so does the nature of the feedback for
quenching star formation; i.e.\@ whether the energy injection from the
SMBHs effectively heats up or vacates the gas within the galaxies and
their surroundings and how these phenomena complement one another at
different mass scales and cosmic epochs.

While there has been extensive study of the physical processes that
connect the accretion of gas onto the SMBHs to their energy output \citep[e.g.\@][]{Yuan2014} the subject is still not fully understood, especially when it comes to the coupling of the feedback energy to the surrounding gas. Complicating matters further, some galaxies have been found with signatures of both
extended radio emission and a bright X-ray source
\citep{Yuan2008,Coziol2017}. The murkiness surrounding the basic
physical process which take place on the BH scale \citep{King2015},
and the challenges of sharing the output energy with the gas over large
scales \citep{McNamara2007}, makes modelling AGN feedback in galaxy
simulations particularly challenging.

Simply injecting the right amount of energy, which in itself is uncertain, is not enough. The details of modelling AGN feedback
are also quite important, as the comparison between the
Illustris \citep{Vogelsberger2014,Vogelsberger2014a} and the
IllustrisTNG \citep[\emph{The Next
    Generation,}][]{Marinacci2018,Naiman2018,Nelson2018a,Pillepich2018a,Springel2018}
simulation results clearly shows. Though very similar in the overall
energy injected by the AGN, the differences in implementation of the
feedback in the two models led to several key
differences. The `bubble model' \citep{Sijacki2007} employed in the
Illustris suite and parameterised as in \citealt{Vogelsberger2013}
\citep[but see][]{Henden2018} resulted in too-low gas fractions in
galaxy groups and low-mass galaxy clusters due to an over-efficient
expulsion of the hight density gas \citep{Genel2014}, while at the same time
generating too-high stellar masses in the central galaxies. These
issues have been resolved and the overall agreement with observations
has improved with the new AGN Model \citep{Weinberger2017} of the
IllustrisTNG simulations. Another example of this is found in
\citet{Davies2019} who compared the halo gas-fractions in the
IllustrisTNG and EAGLE \citep{Schaye2015} simulations: differences in
the implementations of the AGN feedback resulted in a very different
halo gas content for halo masses of \mbox{$\sim 10^{12.5}\msun$} and below. Future observational measures of this property, in this mass range, will be a powerful tool to discriminate between various models.

In this study, we put forward a comprehensive description of the role
AGN feedback plays in setting the physical and thermodynamic
conditions of the gaseous component in the galaxy and in the CGM
within the framework of the IllustrisTNG simulations (hereafter TNG)
and the AGN feedback model therein implemented. We study the entire
(resolved) galaxy population afforded by the TNG100 and TNG300
simulations (galaxy stellar masses in the $10^{9-12.5}\msun$ range)
and focus exclusively on central galaxies at \zeq{0}.

Within the TNG framework, earlier works by
\citet{Weinberger2017,Weinberger2018,Nelson2018a,Terrazas2020} have
demonstrated a close connection between the suppression of
star formation in massive galaxies and the BH feedback in its
kinetic mode, indicating the crucial role played by the latter in
establishing the quenched population in the TNG simulations. The
phenomena emerging from the underlying effective theory for galaxy
formation include the appearance of a population of quenched galaxies
also at intermediate/high redshifts \citep{Weinberger2018,Donnari2019}
and the good match of the shapes and widths of the red sequence and
blue cloud in comparison to SDSS galaxies \citep{Nelson2018a}.  This
base level of realism in the properties of the simulated galaxy
population enables us to gain insight into the workings of AGN
feedback in the real Universe. 

We take these notions further by examining the importance of the AGN
energy injections in TNG, particularly the kinetic BH-driven winds, in
setting the thermodynamic state of the gas in the galaxy itself and
beyond in the CGM. As a case in point, \citet{Nelson2018} have shown how this feedback model affects the abundance of highly ionized Oxygen atoms, OVI, which are an observational tracer of the thermodynamic state of the CGM.

In this way we gain a better understanding of the gas conditions which bring a galaxy to be quenched -- 'ejective feedback' -- and the conditions in the CGM which are conducive for long-term quenching, i.e.\@ 'preventative feedback'. In our analysis we take care to address the multi-phase nature of the CGM \citep{Tumlinson2017}, and to explore how it affects our findings.

The paper is organised as follows: In \cref{sec:sims} we present the 
aspects of the TNG simulation suite most important to our
study, including the details of the AGN model, as well as specifying
the galaxy sample used and the particular physical properties we
consider. In \cref{sec:results} we explore how the changes in
thermodynamic conditions of the galactic and halo gas, particularly
the gas entropy, mirror the transition of the galaxy from
star-forming to quenched, and how these relate to the AGN
feedback. The multi-phase nature of the CGM is also addressed in this
context and we explore the implications of our findings on the future
evolution of the galaxies in \cref{sec:discuss}. In \cref{sec:summary} we
summarise the results of our analyses.

\section{Methods}\label{sec:sims}
\subsection{Simulations}\label{sec:tng}
The IllustrisTNG project\footnotemark is a suite of
magneto-hydrodynamic cosmological simulations carried out in three
simulation volumes of varying size and resolution that, when combined,
provide an extremely large statistical sample of well-resolved
galaxies, groups of galaxies and galaxy
clusters.\footnotetext{\href{http://www.tng-project.org/}{http://www.tng-project.org/}}

The simulations were carried out with the AREPO code
\citep{Springel2010,Pakmor2013}, coupled with the TNG physics
model \citep{Weinberger2017,Pillepich2018} which has been shown to
reproduce, with a reasonable level of accuracy, myriad observational
properties of galaxies
\citep{Marinacci2018,Naiman2018,Nelson2018a,Pillepich2018a,Springel2018}.

The simulation suite consists of three volumes roughly of side-length 50,
100 and 300 Mpc, dubbed TNG50, TNG100, and
TNG300. The simulations are run within the framework of the
$\Lambda$CDM cosmological model with the cosmological parameters based
on the \citet{Planck2016} data: cosmological constant
\mbox{$\Omega_\mathrm{\Lambda}= 0.6911$}, matter density
\mbox{$\Omega_\mathrm{m}=\Omega_\mathrm{dm}+\Omega_\mathrm{b}=0.3089$},
with a baryonic density of \mbox{$\Omega_\mathrm{b}= 0.0486$}, Hubble
parameter $h = 0.6774$, normalisation \mbox{$\sigma_\mathrm{8} =
  0.8159$}, and spectral index \mbox{$n_\mathrm{s} = 0.9667$}.

In \cref{tab:TNGproperties} we detail, for both the TNG100 and TNG300
volumes, the target mass of the gas cells (which is roughly equal to the average initial mass of stellar particles), and the minimum comoving value of the gravitational softening for the gas elements. In addition, we give the average size of the gas cells\footnotemark within each of the three gas components we study in this paper (see \cref{sec:gasComponents}), averaged over all objects in our sample (\cref{sec:galSample}). The full simulation data of TNG100 and TNG300 is publicly available \citep[see][for details]{Nelson2019a}.
\footnotetext{Since the AREPO code employs an unstructured mesh scheme, we use a representative cell-size scale of $(m_\mathrm{cell}/\rho_\mathrm{cell})^{1/3}$.}

The computational and physical models for unresolved astrophysical processes
incorporated in the TNG simulations have been presented in
\citet{Weinberger2017,Pillepich2018}, and are updates to the earlier
Illustris model \citep{Vogelsberger2013,Torrey2014}. As
such, we forgo an in-depth description of these here, and
present in the following sub-sections a short description of those aspects of the
models that pertain to the analysis at hand; namely the AGN
feedback model and the star formation recipe, as well as a short
description of the galaxy sample herein analysed.

\subsection{AGN feedback in the TNG model}\label{sec:AGNmodel} 
BHs are formed in the simulation based on the host-halo relationship
\citep{DiMatteo2008,Sijacki2009}: a BH is seeded in any friends of friends (FoF) halo whose
mass exceeds $7.3\times10^{10}\msun$ (if the halo does not already
possess a BH), with an initial seed mass of $6.2\times
10^6\msun$. Once seeded, BHs can grow by accreting gas from their close vicinity, the `feedback region', which also defines the gas cells into which feedback energy is deposited. 
The size of the feedback region in the \zeq{0} snapshots is of order a few kiloparsecs: \mbox{$\sim 2.2\units{kpc}$} in TNG100 and \mbox{$\sim 3.6\units{kpc}$} in TNG300 on average. In addition, BH-BH mergers occur whenever one BH is within the `feedback region' of the other \citep{Sijacki2007}. The BHs are `pinned' to the local minimum of the potential field, which in most cases coincides with the global minimum of its host halo.

The theoretical treatment of AGN feedback accounts for two modes of energy injection, set by the accretion rate of matter onto the BH \citep{Begelman2014}. The high accretion mode or thermal mode is associated with efficient accretion through a disk and feedback in this state is implemented as an injection of \emph{thermal} energy into the surrounding gas. The low accretion mode or kinetic mode occurs at lower accretion rates in which the accretion disk is in a radiatively inefficient state \citep{Blandford1999,Yuan2014}, and results in a \emph{kinetic} injection of energy. Two feedback modes acting at different BH accretion rates have been called, traditionally, the `quasar' and `radio' modes, respectively. However, within the TNG model the kinetic mode is distinct from the observed jets that have been typically associated with radio galaxies at the centres of clusters. Hence, in what follows we label the two modes as the thermal and kinetic modes.

The feedback energy injected in each of the modes is 
\begin{equation}\label{eq:HAM_fb}
    \dot{E}_\mathrm{high/low}=\epsilon_\mathrm{high/low}\dot{M}_\mathrm{BH}c^2,
\end{equation}
where $\dot{M}_\mathrm{BH}$ is the accretion rate onto the BH. For the thermal mode, \mbox{$\epsilon_\mathrm{high}=0.02$} is the efficiency with which
energy couples to the surrounding gas. The energy is injected continuously as
thermal energy in a kernel-weighted manner within the gas cells in the `feedback region'.

For the kinetic mode, the energy coupling factor has a maximal value\footnotemark of
\mbox{$\epsilon_\mathrm{low}=0.2$}. This energy is imparted as a momentum boost to the gas cells in the feedback region (with no mediation by `wind-particles') in a series of
discrete `injection events', which occur once the accumulated energy output from the BH
has exceeded a certain threshold. In each injection event the
direction of the momentum kick is chosen randomly, thus the energy
injection is isotropic when averaged over many injection events, and,
by extension, momentum conservation also holds in a time-averaged
sense. \footnotetext{To prevent runaway kinetic
  feedback due to ever-decreasing gas densities, a `floor' is set for
  the energy coupling parameter. }
% set to \mbox{$\epsilon_\mathrm{low}=\mathrm{min}\left(0.2,20\rho/\rho_\mathrm{SFthresh}\right)$}.}%,  \citep{Vogelsberger2013}.}
% % \rho_thresh = 0.13 cm^-3

The feedback mode of the AGN is based on the ratio between the accretion
rate and the Eddington accretion rate. The AGN feedback is in the thermal mode
when the ratio is above a threshold value $\chi$:
\begin{equation}\label{eq:accRatio_thresh}
f_\mathrm{Edd}\equiv\frac{\dot{M}_\mathrm{BH}}{\dot{M}_\mathrm{Edd}} \ge \chi,\quad \chi=\mathrm{min\left[0.002\left(\frac{\Mbh}{10^8\mathrm{M_\odot}}\right)^2,0.1\right]},
\end{equation} 
with $\chi$ scaling with the mass of the BH, thus favouring a
transition to the kinetic mode for higher masses. The accretion rate
of gas onto the BH is set as
\mbox{$\dot{M}_\mathrm{BH}=\mathrm{min}\left(\dot{M}_\mathrm{Bondi},\dot{M}_\mathrm{Edd}\right)$}.

In both feedback modes, the thermal and kinetic, the energy is imparted to the cells in the small `feedback region' of the BH. There is no de-coupling of these cells from the hydrodynamical forces, or from the radiative cooling, i.e.\@ the progression in time of the physical state of these cells is governed by the same magneto-hydrodynamical model as before the energy was injected. 

\begin{figure*}
  \centering
  \subfloat[Energy injected vs.\@ BH mass] {\label{fig:bhModel_bhMass}
     \includegraphics[height=6.8cm,keepaspectratio]{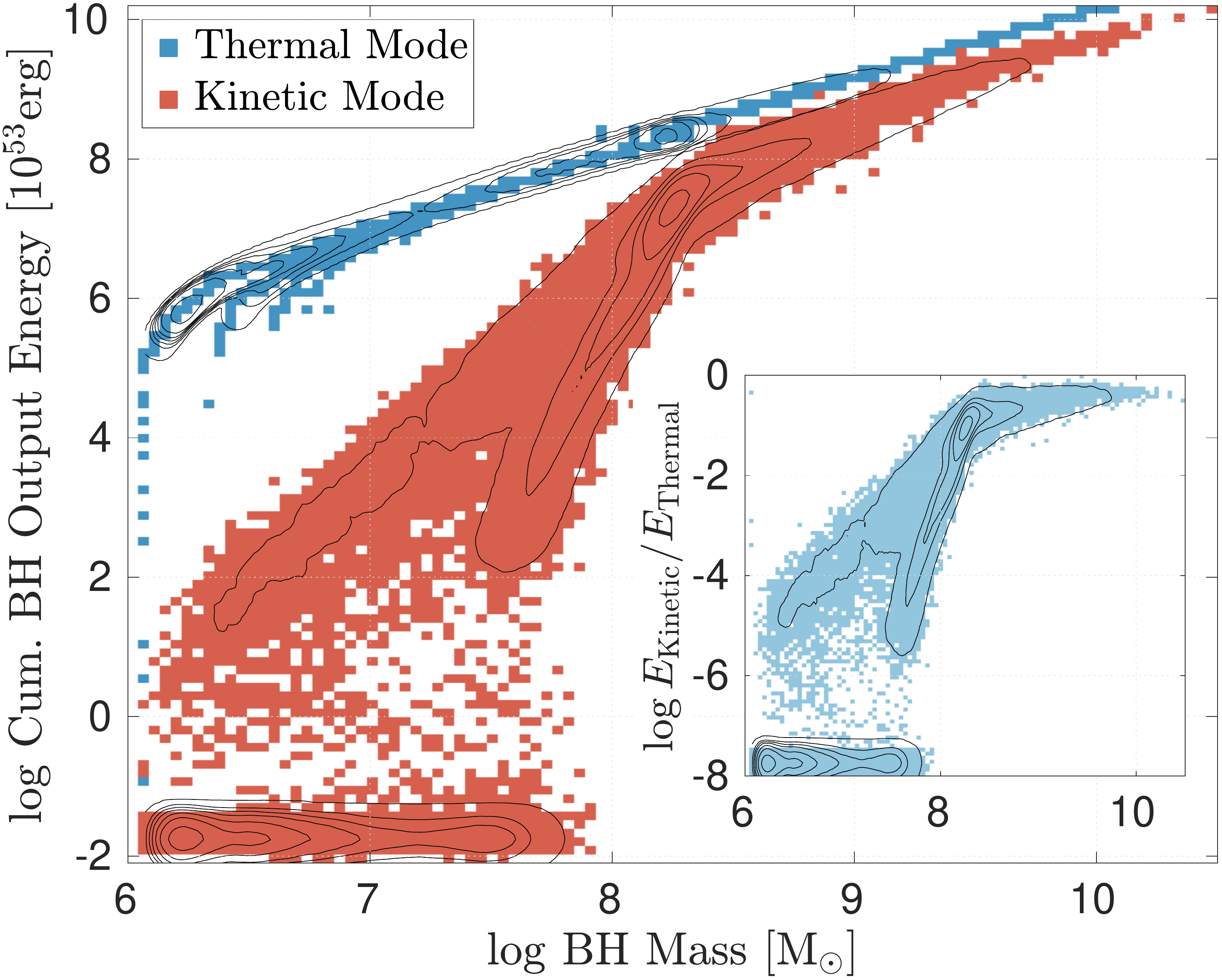}}
     \subfloat[Energy injected vs.\@ stellar mass] {\label{fig:bhModel_stellarMass}
       \includegraphics[height=6.8cm,keepaspectratio]{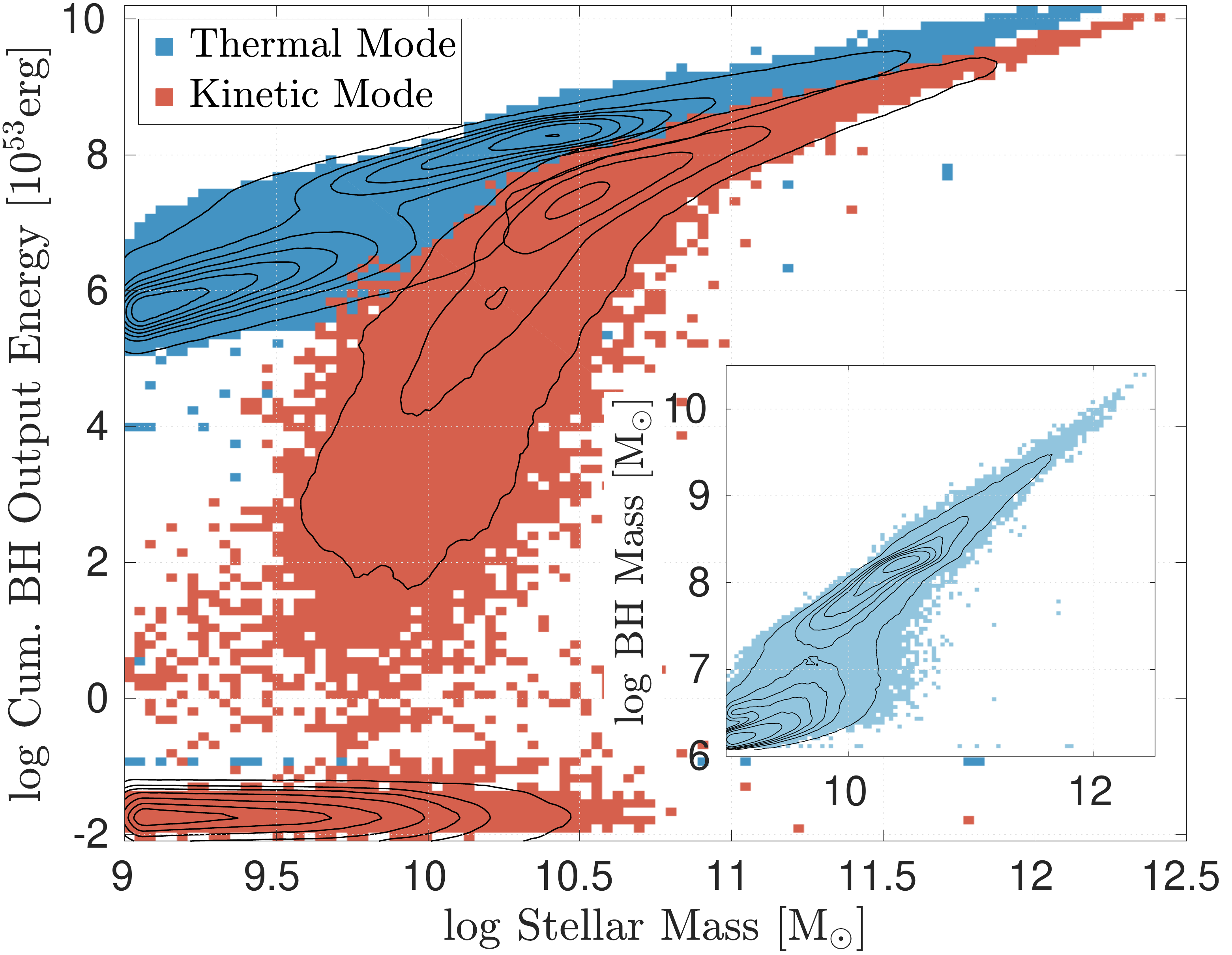}}
     \caption{BH properties and cumulative injected energy in
       different BH accretion states in the TNG simulations. The
       energy injected in the two feedback modes integrated over the
       lifetime of the central BH in a galaxy vs.\@ the BH mass, at
       \zeq{0}, is shown in panel \subrfig{bhModel_bhMass}. The inset
       shows the ratio of the energies from the two modes vs.\@ the BH
       mass. The energy vs.\@ the galaxy stellar mass is shown in
       panel \subrfig{bhModel_stellarMass}, with the BH mass to
       stellar mass relation shown in the inset. Here and throughout
       the paper we show the central galaxies of the TNG300 simulation
       (excluding back-splash galaxies at \zeq{0} -- see
       \cref{sec:galSample}). The black contours show the distribution of
       the galaxy population. Roughly 60 per cent of all galaxies in the depicted mass range have
       never had a kinetic feedback event and these have been
       placed by hand on a flat relation in the bottom of the plots.}
  \label{fig:bh_stuff}
\end{figure*}
\subsubsection{Emerging BH populations in TNG at \zeq{0}}\label{sec:BHpops}
The BH particles in the simulation record the cumulative amount of
energy released to the gas in each of the two modes (\equnp{HAM_fb})
over the entire history of the BH: \mbox{$\int
\dot{E} \diff t$}. In \cref{fig:bh_stuff}, we show the cumulative
energy of the two feedback modes of the central BH in the TNG300
central galaxies at \zeq{0} as a function of both the BH mass
(\cref{fig:bhModel_bhMass}) and the stellar mass of the host galaxy
(\cref{fig:bhModel_stellarMass}). The ratio of the two
energies vs.\@ the BH mass is shown in the
\cref{fig:bhModel_bhMass} inset. The central BH is chosen as the most
massive BH particle in the galaxy. In addition, for any given BH at
the current epoch, the cumulative energy injection includes \emph{all}
its progenitors across its history (and not only the feedback
associated with the ``main progenitor branch'').

These figures emphasise some important features of the
AGN model. For all galaxies, more energy is injected in the thermal
mode channel. In low-mass galaxies the dominant mass growth channel
for the SMBH is through the high-accretion thermal mode. Since at any time $\dot{E}_\mathrm{high}\propto \dot{M}_\mathrm{BH}$, the cumulative
energy will naturally be proportional to the total BH mass. Once a BH
is seeded at a mass of $10^{6.07}\msun$, and starts giving off energy,
we see a tight one-to-one relation between $E_\mathrm{high}$ and
$\Mbh$ form\footnotemark (\cref{fig:bhModel_bhMass}). The convolution
of this tight relation with the $\Mbh$--stellar mass relation (see
inset in \cref{fig:bhModel_stellarMass}) leads to a similar relation
with stellar mass, albeit with more scatter.  \footnotetext{In the
  low-mass end of the $E_\mathrm{high}\--\Mbh$ relation, several
  `branches' are seen to rise and join the main relation. These
  branches are cases in which BH mergers occurred before enough energy
  was released to reach the power-law relation.}

Examining the distribution of BH mass vs.\@ stellar mass (in the
inset of \cref{fig:bhModel_stellarMass}) we see a concentration of BH
masses at $\sim 10^6\msun$, close to the initial BH seed mass, found
within low-mass galaxies and another concentration at $\Mbh\gtrsim
10^{7.5}\msun$, with a dearth of BHs in between. This
distribution most likely arises due to the regulation of BH
mass growth by the SN feedback process
\citep[see also][though we note that these studies do not address the TNG model]{Dubois2015,Habouzit2017,AnglesAlcazar2017,Bower2017}. \citet{Pillepich2018} show with simulation run with variations of the TNG model, that in the low-mass galaxies,
SN feedback limits the amount of available gas for BH
accretion, keeping the BH mass close to the initial seed mass (though
the BH can still grow by BH-BH mergers). As the galaxy grows in mass
and its potential well deepens, the effectiveness of SN feedback
diminishes, and the BH grows quickly to higher masses, concentrating
at roughly $\Mbh\sim 10^{8.3}\msun$ where the kinetic mode feedback
becomes important and self-regulates the further growth of the
BH. \citet{McAlpine2018} found a similar process occurring in the
EAGLE simulations.

The picture for the kinetic mode is more complex: 60 per cent of all
galaxies in the \zeq{0} sample have experienced no kinetic mode
feedback at all throughout the history of the (simulated)
universe. These are invariably of low BH mass $\Mbh<10^8\msun$, or
$10^{10.5}\msun$ in stellar mass. In this mass range the threshold
$\chi$ to be in the kinetic mode is very low, implying that the
required accretion conditions are never satisfied. In order to show
these galaxies we have placed them `by hand' at the bottom of
the plot.

The two `branches' in the kinetic mode energy vs.\@ the BH mass
relation of \cref{fig:bhModel_bhMass} come about from the distribution of BH masses mentioned earlier, in which low-mass BHs are concentrated close to the seed mass and most of the high-mass BHs are found at masses of $10^{7.5\--8.5}\msun$. If the BH mass is low when it experiences its first kinetic mode injection event, it will progress on the shallow-slope left branch. If the BH experiences its first injection event only after it grows to the high-mass group, it will progress on the steeper, right branch.

Due to the scarcity of injection events at low masses, the energy
input is quite low, but rises steeply with BH mass as the likelihood
of being in the kinetic mode grows. A transitional mass scale appears
at $\Mbh=10^{8.3}\msun$ or $10^{10.5}\msun$ in stellar mass, where the
energy injected in the kinetic mode becomes comparable to that of
the thermal mode, with a typical ratio of $\sim 0.3$ (see inset in
\cref{fig:bhModel_bhMass}). Beyond this mass scale the relation with
$\Mbh$ is also linear, but with a larger scatter compared to the
thermal mode relation. In SMBHs above this mass scale, the gas
accretion is self-regulated by kinetic mode feedback, and the SMBHs
experience mostly epochs of kinetic mode feedback. BH-BH mergers
become an important mass growth channel for these objects
\citep{Weinberger2018}.

The transitional mass scale is set in part by the parameters of the AGN model (e.g.\@ \equnp{accRatio_thresh}), but is also influenced by other physical processes in the TNG model convolved with the coupled growth of galaxies and their BHs across cosmic epochs. For example, in a simulation run with the TNG model in which SN feedback was turned off, the stellar mass transitional mass scale occurs at a lower value. This most likely occurs since the mass growth of the BHs is no longer regulated by stellar feedback and thus massive BHs (with a propensity for kinetic mode feedback) are found in galaxies of lower stellar mass, when compared to the fiducial simulations. 

As we show in this study, this mass scale, marking the onset of the kinetic feedback is a major determinant of the properties of the galaxies and their surrounding atmospheres \citep[see also][]{Terrazas2020} and is also found in the distributions of $f_\mathrm{Edd}$ and AGN occupation fractions
\citep{Habouzit2019}, despite it being sub-dominant in terms of overall energy output compared to the thermal mode.

Although it supplies most of the feedback energy, the thermal mode appears to be relatively ineffective. This is likely due to the fact that the thermal energy is deposited in a region close to the BH which is usually comprised of dense, star-forming gas. The injected energy is then {\bf (1)} radiated away very efficiently, since cooling processes scale with \mbox{$\propto \rho^2$}, or {\bf (2)} if the gas is star-forming it is governed by en effective equation of state (see \cref{sec:SFmodel}) which will set an artificial temperature regardless of the amount of injected thermal energy \citep{Weinberger2018}. Since kinetic energy is agnostic to both radiative cooling and the equation of state of the gas, the kinetic mode can have a larger affect on the gas even if its total energy injection is considerably lower. %However, we do note that the simulations of \citet{Yuan2018}, carried out with a different physical model, also find that the `wind' feedback mode is more effective than the `radiative' mode despite being sub-dominant in terms of energy input. 

\subsection{Star formation and the effective EOS}\label{sec:SFmodel}
Star formation in the TNG physical model follows the
\citet{Springel2003} model for star formation in a pressurised,
multi-phase ISM. Gas elements whose number density is above $\sim 0.1\units{cm^{-3}}$ are designated as
`star-forming' and no longer follow the ideal-gas equation of state
(EOS), but are placed on an effective EOS, which mimics the pressure
support supplied by unresolved SN feedback. This star-forming gas is
converted to stars stochastically following the Kennicutt-Schmidt
relation and assuming a \citet{Chabrier2003} initial mass
function.

Since the thermodynamics of the star-forming gas are governed by this
effective EOS, which is not meant to reproduce the realistic physical
state of the gas, we will ignore this gas phase when extracting the
mean values of gas properties such as temperature, density, entropy,
and cooling time.

\subsection{The TNG galaxy sample}\label{sec:galSample}
Identification of bound objects in the TNG simulations is
obtained through a two-tier approach: the Friends-of-Friends (FOF)
method is used to identify distinct host haloes or `FOF Groups' . For
each of these haloes we define the `virial' parameters: The virial
mass, \mbox{$\Mv$}, and virial radius, \mbox{$\Rv$}, according to the relation
\begin{equation}\label{eq:virialDef} 
\frac{3 \Mv}{4 \pi \Rv^3}=200\rho_{\mathrm{crit}} .
\end{equation} 
We further define the virial temperature and virial velocity 
\begin{align}
\Vv &= \sqrt{\frac{G \Mv }{\Rv}} \label{eq:vvirDef}\\
\Tv &= \frac{\mu \mproton \Vv^2}{2\kboltz}, \label{eq:tvirDef}
\end{align}
where $\kboltz$ is the Boltzmann constant and $\mu \mproton$ is the
average mass per particle, with $\mproton$ being the proton mass, and
$\mu\simeq0.59$ being the mean molecular weight.
  
Within each host halo, the \textsc{subfind} algorithm
\citep{Springel2001,Dolag2009} is used to identify groups of
gravitationally bound resolution elements (dark matter particles,
stellar particles, gas cells, etc.), commonly known as
sub-haloes\footnotemark. With this approach every resolution element
of a sub-halo is naturally also part of the host halo, but some
resolution elements of the host halo are not bound to any of the
sub-haloes. We filter out objects from our sample which were not formed in the traditional cosmological framework of structure formation through gravitational collapse, but rather are baryonic clumps which fragmented out of galactic gas in already formed galaxies.  In detail, we follow the method described in \citet{Nelson2019a}.\footnotetext{Only one level of sub-halo identification is considered here -- sub-haloes of sub-haloes are typically below the limiting mass resolution.}

The centres of both host haloes and sub-haloes are set as the position
of the most bound resolution element. Sub-haloes typically contain a
stellar population, i.e.\@ a galaxy, and the centre of this stellar
distribution typically, but not always, corresponds to the sub-halo
centre. For each host halo, the most massive, most
gravitationally-bound sub-halo, whose centre invariably coincides with
the host halo centre, is defined as the central sub-halo, and in what follows we define the galaxy of this sub-halo as the central galaxy of host halo. In most cases, the central galaxy is the most
massive one in the halo; i.e.\@ in the group or cluster.

In this study we focus on the central galaxies found in the TNG100 and
TNG300 simulation volumes. We focus on the central galaxy population and
exclude satellite galaxies since the latter are affected by
environmental quenching processes. To ensure that our galaxy sample includes only well-resolved systems, we enforce a lower mass limit for the galactic
stellar mass of $10^9\msun$. In our analysis we examine various
properties of the gaseous medium in and around the galaxies; e.g.\@
temperature, entropy, density, etc. Obviously these properties can
be defined only for those galaxies which contain gas, further limiting
our sample to galaxies with $M_\mathrm{gas}>0$.% within $\rgal$.

There is an important point to be made about our chosen mass limit
with regards to the figures shown here. In many figures in
this paper, we plot the distribution of our galaxy sample in various
2D galaxy-property spaces, e.g.\@ sSFR--stellar mass and
entropy-stellar mass (\cref{fig:galPop_ssfr}). In these figures we use
contours to show the number of galaxies in different regions of the
plane, which taper off toward the enforced lower mass
limit. Naturally, if lower mass galaxies had been included in our
sample, the contours would have extended to include those as
well.
%\footnote{Especially considering that due to
%  the bottom-up nature of structure formation in the CDM paradigm, the
%  number of objects at a given mass grows as one considers lower
%  masses.}.

Finally, with the goal of examining the role of AGN in the quenching of
central galaxies, we wish to filter out those galaxies which, although
at the time of inspection are labelled as centrals, in the recent
past were identified as satellites; i.e.\@ those that may have undergone
environmental quenching processes specific to satellites and would otherwise bias our
description of the physical processes at play. These galaxies may be
`back-splash' galaxies- galaxies which recently travelled as
satellites through a host halo and came out on the other side
\citep{Gill2005}. To remove these galaxies, we follow the
main-progenitor branch of the \textsc{sublink} merger tree
\citep{RodriguezGomez2015} for each central galaxy (at \zeq{0} and remove it from our sample if \emph{both} of the following conditions are satisfied within the last 5 Gyr (since $z<0.5$):
\begin{itemize}
 \setlength\itemsep{1em}
    \item[-] Identified as a satellite by the \textsc{subfind} algorithm.
    \item[-] Found at a distance of more than $0.1\Rv$ from the centre of
      its host halo.
\end{itemize}

Our final sample includes $\sim 135,000$ central galaxies in
TNG300. While most of the results we present are based on the galaxy
population analysis of the TNG300 simulations, we have repeated our
analysis on the TNG100 simulation box, with a sample of $\sim 10,000$
galaxies, with identical results (see \cref{sec:resCompare}).

\subsection{Some basic definitions}\label{sec:define}
We here present the definitions of important quantities we use
throughout the paper: the subdivision of the gas into different
spatial regions, gas entropy, cooling time, and free-fall time. We
emphasise that when calculating both the entropy and cooling-time we
ignore the star-forming gas, as described in \cref{sec:SFmodel}. 

\subsubsection{The gas components}\label{sec:gasComponents}
Since we wish to study the effects of AGN feedback on the galaxy
itself as well as the surrounding CGM, we divide the gaseous medium of
each sub-halo into the following 3 components: galactic gas, inner CGM
and outer CGM. These components are determined by two radii: the
galactic radius is defined to be twice the 3D stellar half-mass radius
 in the sub-halo
\begin{equation}\label{eq:rgalDef}
  \rgal \equiv 2r_{1/2,\mathrm{stars}},
\end{equation}
and the half-mass radius of the gaseous distribution in the sub-halo
$\rhGas$. Throughout the paper, galaxy stellar masses are obtained by summing up the stellar particles within $\rgal$ and the star formation rates are, unless otherwise stated, obtained from the instantaneous SFRs of the star-forming gas cells within the same aperture.
\begin{figure}
  \centering
   \includegraphics[width=\columnwidth,keepaspectratio]{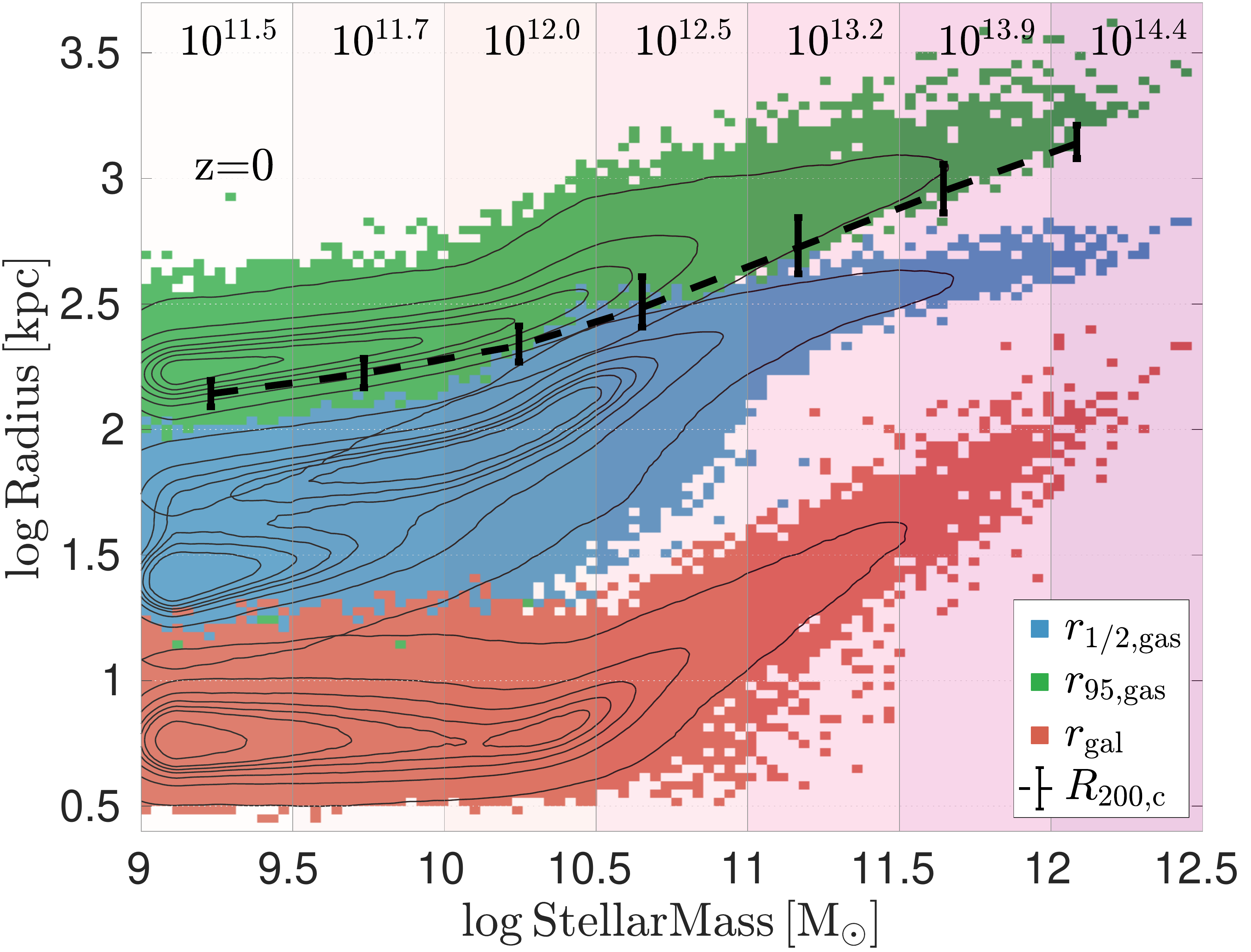}
  \caption{The galactic radius $\rgal$, the half gas radius $\rhGas$,
    and the radius enclosing 95 per cent of the gas mass,
    $r_\mathrm{95,gas}$ vs.\@ the stellar mass, of the \zeq{0} TNG300
    galaxies are shown in red, blue, and green, respectively. Overlaid
    contours show the galaxy population distribution. The sharp drop
    of the contours at the low-mass end is due to our enforced stellar
    mass limit. The black error-bars show the median value, and the
    $10\--90$ per cent quantiles of $\Rv$ of the host haloes of the
    galaxies in each of the coloured vertical bands. The numbers at
    the head of the coloured bands show the mean $\Mv$ of the host
    haloes. The gas which is gravitationally bound to the sub-halo
    extends to well beyond $\Rv$ in most cases. Throughout this paper,
    we call galactic gas, inner CGM, and outer CGM the mass and volumes
    of gravitationally-bound gas contained, respectively, within
    $\rgal$, between $\rgal$ and $\rhGas$, and beyond $\rhGas$.}
  \label{fig:radiiDist}
\end{figure}

In \cref{fig:radiiDist} we show the distribution of these radii for
our galaxy sample, as a function of galaxy stellar mass. We also
present the radius which encloses 95 per cent of the gas mass in the
central \textsc{subfind} object. The median virial radius
$\Rv$ of the host haloes is shown for comparison. As can be seen,
the gas that is gravitationally bound to the sub-haloes usually
extends to well beyond the `virial' radius, $\Rv$ defined by the total
mass distribution. In this sense, $\Rv$ should not be taken to be a
sharp `edge' of the system, but rather a typical length-scale of the
host halo \citep{Zinger2018}.

The three components are thus defined to be 
\begin{itemize}
  \setlength\itemsep{1em}
\item Galactic Gas -- all the gaseous material in the sub-halo enclosed
  within $\rgal$.
\item Inner CGM -- All the material found beyond $\rgal$ and within $\rhGas$. 
\item Outer CGM -- All the material found beyond $\rhGas$, yet still
  gravitationally bound to the sub-halo. This typically includes gas
  found beyond $\Rv$ of the host halo.
\end{itemize}

In the calculation of the thermodynamic properties: temperature, entropy and cooling times, we ignore the star-forming gas (which is placed on an effective EOS \cref{sec:SFmodel}), and find the
mass-weighted average over the non-star-forming gas cells. The
galactic gas component is typically comprised on average of several hundreds of
non-star-forming gas cells in galaxies of stellar masses of $<10^{11}\msun$. The inner and outer CGM components of these galaxies each contain several thousands to tens of thousands of non-star-forming gas cells. In higher mass galaxies, the number of cells in these component is even larger. In \cref{tab:TNGproperties} we give the average cell size for each of the three components, with the average taken over all non-star-forming gas cells in all objects of our sample (\cref{sec:galSample}). 

\subsubsection{Gas entropy}\label{sec:entropyDef}
The gas entropy is an important indicator of the thermodynamic state
of the gas. In what follows, we use the common definition of
`astrophysical' entropy for a gas element
\begin{equation}\label{eq:entropyDef}
S \equiv \log K = \log \left( \frac{\kboltz T}{n^{2/3}} \right),
\end{equation}
with $\kboltz$ the Boltzmann constant and $n$ the number density of
the gas cell. For a collection of gas cells we take the mass-weighted
average as a representative value. 

It is instructive to obtain an estimate of the `virial' entropy -- what
the entropy of the gas would be within the gravitational potential of
a host halo, in the adiabatic case, i.e.\@ when cooling/heating and
feedback processes are ignored. By using
\cref{eq:virialDef,eq:vvirDef,eq:tvirDef} we arrive at the relation
   \begin{equation}\label{eq:entropyVir}
      K_\mathrm{vir} \equiv\ \frac{\kboltz \Tv}{\left(200 \rho_{\mathrm{crit}}f_\mathrm{b}\right)^{2/3}} =14.8 \left(\frac{f_\mathrm{b}}{0.15} \right)^{-\frac{2}{3}} \left(\frac{\Mv}{10^{12}\msun}\right)^{\frac{2}{3}}\units{keV\,cm^2},     
     \end{equation}
%      \begin{align}\label{eq:entropyVir}
%     K_\mathrm{vir} &\equiv\ \frac{\kboltz \Tv}{\left(200 \rho_{\mathrm{crit}}f_\mathrm{b}\right)^{2/3}}  \nonumber \\
%&=14.78 \left(\frac{f_\mathrm{b}}{0.15} \right)^{-\frac{2}{3}} \left(\frac{\Mv}{10^{12}\msun} %\right)^{\frac{2}{3}}\units{keV\,cm^2},     
          %&=14.78 \left(\frac{\delvir}{200}\frac{\rho_\mathrm{ref}}{\rho_\mathrm{crit}} \right)^{-\frac{1}{3}}     \left(\frac{f_\mathrm{b}}{0.15} \right)^{-\frac{2}{3}} \left(\frac{M_\mathrm{vir}}{10^{12}\msun} \right)^{\frac{2}{3}},
    %\end{align}
where $f_\mathrm{b}$ is the baryon fraction within the halo. In the
purely adiabatic case the gas fraction within a halo should follow the
universal baryonic fraction
\mbox{$f_\mathrm{b}=\Omega_\mathrm{b}/\Omega_\mathrm{m}=0.157$}, and
we adopt this value to gain an estimate for the entropy of a halo of a
given $\Mv$ in the adiabatic case.

\subsubsection{Cooling time}\label{sec:tcoolDef}
The cooling time of a gas cell is calculated as
\begin{align}\label{eq:tcool1}
\tcool &\equiv \frac{U}{n_\mathrm{i}n_\mathrm{e}\Lambda}
=  \frac{2+3\epsilon}{1+\epsilon}  \frac{3}{2} \kboltz T \left(\frac{\chi}{\mu\mproton}\rho \Lambda\right)^{-1} 
\simeq 6 \frac{\kboltz T}{n\Lambda} \nonumber \\
&= 0.26\, T_6\, n_{-3}^{-1}\, \Lambda_{-22}^{-1}\units{Gyr}, 
%&= 0.26 \left(\frac{T}{10^6\units{K}}\right) \left(\frac{n}{10^{-3}\units{cm^{-3}}}\right)^{-1} \left(\frac{\Lambda}{10^{-22}\units{erg\,cm^3\,s^{-1}  }}\right)^{-1}\units{Gyr}
%=  \frac{1+2\epsilon}{1+\epsilon}  \frac{3}{2} \frac{\kboltz }{\chi^2} \frac{T }{ n \Lambda } 
\end{align}
where $U$ is the internal energy per unit volume in the cell,
$n_\mathrm{i},n_\mathrm{e}$ are the number density of ions and
electrons, and $\Lambda$ is the normalised cooling rate (in units of
$\mathrm{erg\, cm^3\, sec^{-1}}$). For a primordial composition, with a Hydrogen mass fraction of \mbox{$X_\mathrm{H}\simeq0.76$}, 
\mbox{$\epsilon\simeq0.08$} is the ratio of Hydrogen to Helium atoms,
$\mu\mproton$ is the average mass per particle \mbox{($\mu\simeq0.58$)}, and
\mbox{$\chi\simeq 0.52$} is the number of electrons per particle. To obtain a typical value of the cooling time (last line of \equnp{tcool1}) we define the scaled parameters \mbox{$T_6=T/10^6\units{K}$}, \mbox{$n_{-3}=n/10^{-3}\units{cm^{-3}}$}, and \mbox{$\Lambda_{-22}=\Lambda/10^{-22}\units{erg\,cm^3\,sec^{-1}}$}.   

In practice, the simulation code records the internal energy per unit
mass $u=U/\rho$, the net cooling rate $\Lambda$, and the gas mass density
$\rho$ for each cell in the simulation. The number density of the ions
and electrons are given by the relations
\begin{equation}
  n_\mathrm{i}=(1+\epsilon)\frac{X_\mathrm{H}}{\mproton}\rho\quad ; \quad  n_\mathrm{e}=\frac{\chi}{\mu\mproton}\rho , 
\end{equation}
and thus the cooling time for each cell is calculated with the following expression
\begin{equation}\label{eq:tcool2}
  \tcool= \frac{\mu \mproton^2}{(1+\epsilon) \chi X_\mathrm{H}} \frac{u}{\rho\Lambda}\simeq1.37 \mproton^2 \frac{u}{\rho\Lambda}. 
\end{equation}

The cooling time is set by the \emph{local} thermodynamic conditions
of a given gas element. When calculating the cooling time of a given
gaseous region we take the mass-weighted average over all gas cells in
that region. In addition, we note that some cells have a negative cooling rate, or rather a heating rate -- we ignore these cells when calculating the mass-weighted average.
% \footnotetext{For temperatures $T \gtrsim10^{6.5}$ in the
% Bremsstrahlung regime $\Lambda\sim T^{1/2}$ and thus the cooling
% time $\tcool \sim n^{-1}T^{1/2}$.}

\subsubsection{Free-fall time}\label{sec:tffDef}
The free-fall time sets a timescale for a gas element to reach the
centre of the potential well, and is set by the total mean-density of
the FoF halo enclosed within the radial position of a gas element,
$\overline{\rho}(\le r)$,
\begin{equation}\label{eq:tff}
t_\mathrm{ff}\equiv\sqrt{\frac{3\pi}{32G\overline{\rho}(\le r)}}.
\end{equation}
We note that the free-fall time is a monotonic function of the radial
position within the halo. Regardless of the mass profile of any
individual haloes, the definition of the virial quantities,
\cref{eq:virialDef}, sets the free-fall time at $\Rv$ to be
$1.6\units{Gyr}$.

%trim=0.7cm 5cm 0.7cm 0.7cm, clip=true
\begin{figure*}
  \centering
    \includegraphics[width=17.7cm,keepaspectratio]{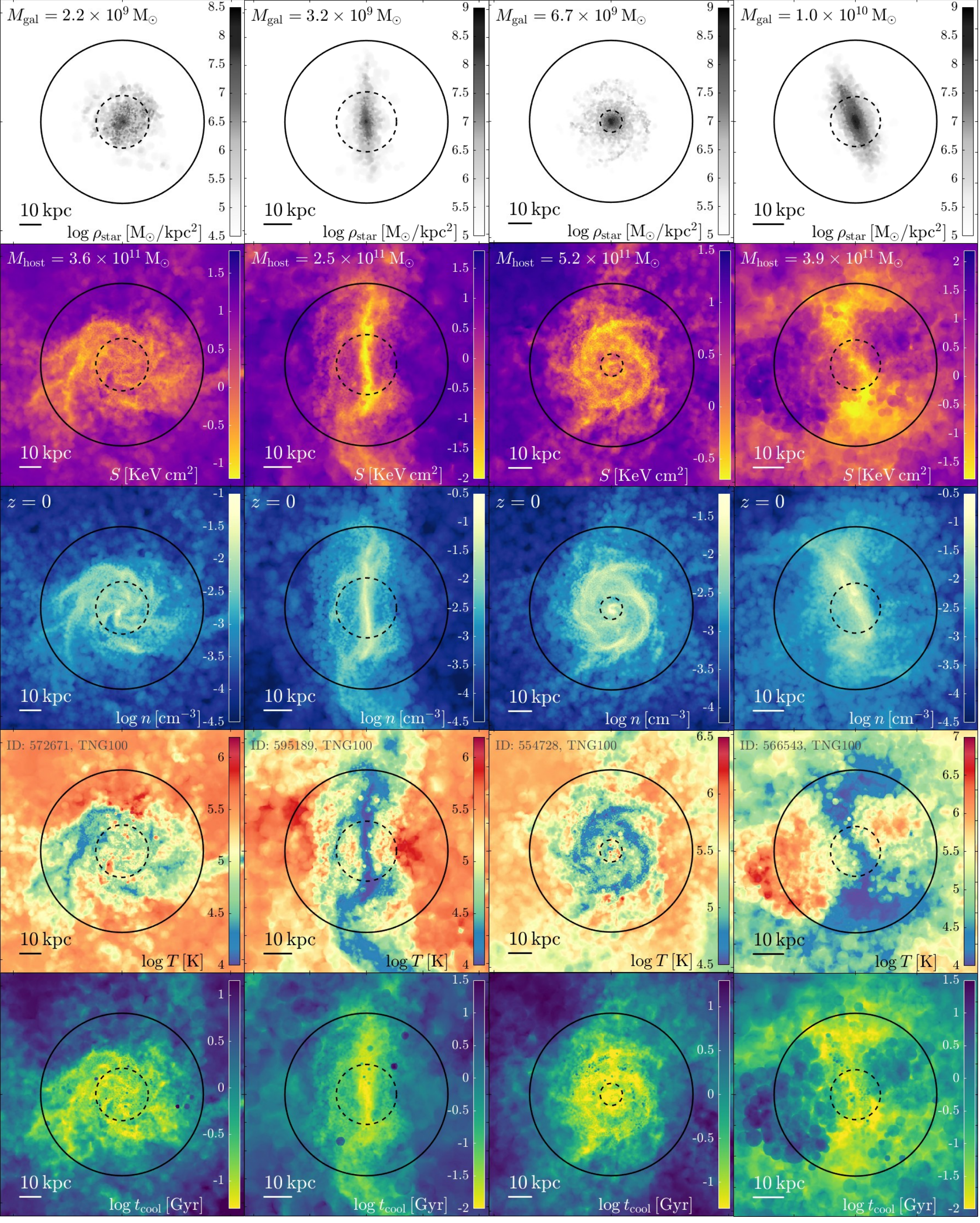}
          \caption{Central galaxies from TNG100 at \zeq{0} arranged in
            columns of ascending stellar mass (left to right). From
            top to bottom we show the stellar surface density
            (projected along the slice), and the entropy, gas density,
            temperature, and cooling time. The averages of the
            temperature, entropy and cooling time only include gas
            which is \emph{non}-star-forming and are mass weighted
            averages over the simulation slice. The number density
            depicts \emph{all} the gas and is a simple average over
            the cell densities along the simulation slice. Inner
            dashed circle demarks $\rgal$ and solid outer circle
            demarks $\rhGas$ for each galaxy. The slice thickness is
            twice $\rhGas$ (diameter of outer circle). Stellar mass
            and FoF host mass are shown for each galaxy in the top two
            rows.}
  \label{fig:Maps1}
\end{figure*}

\begin{figure*}
\ContinuedFloat
\captionsetup{list=off,format=cont}
  \centering
    \includegraphics[width=18cm,keepaspectratio,]{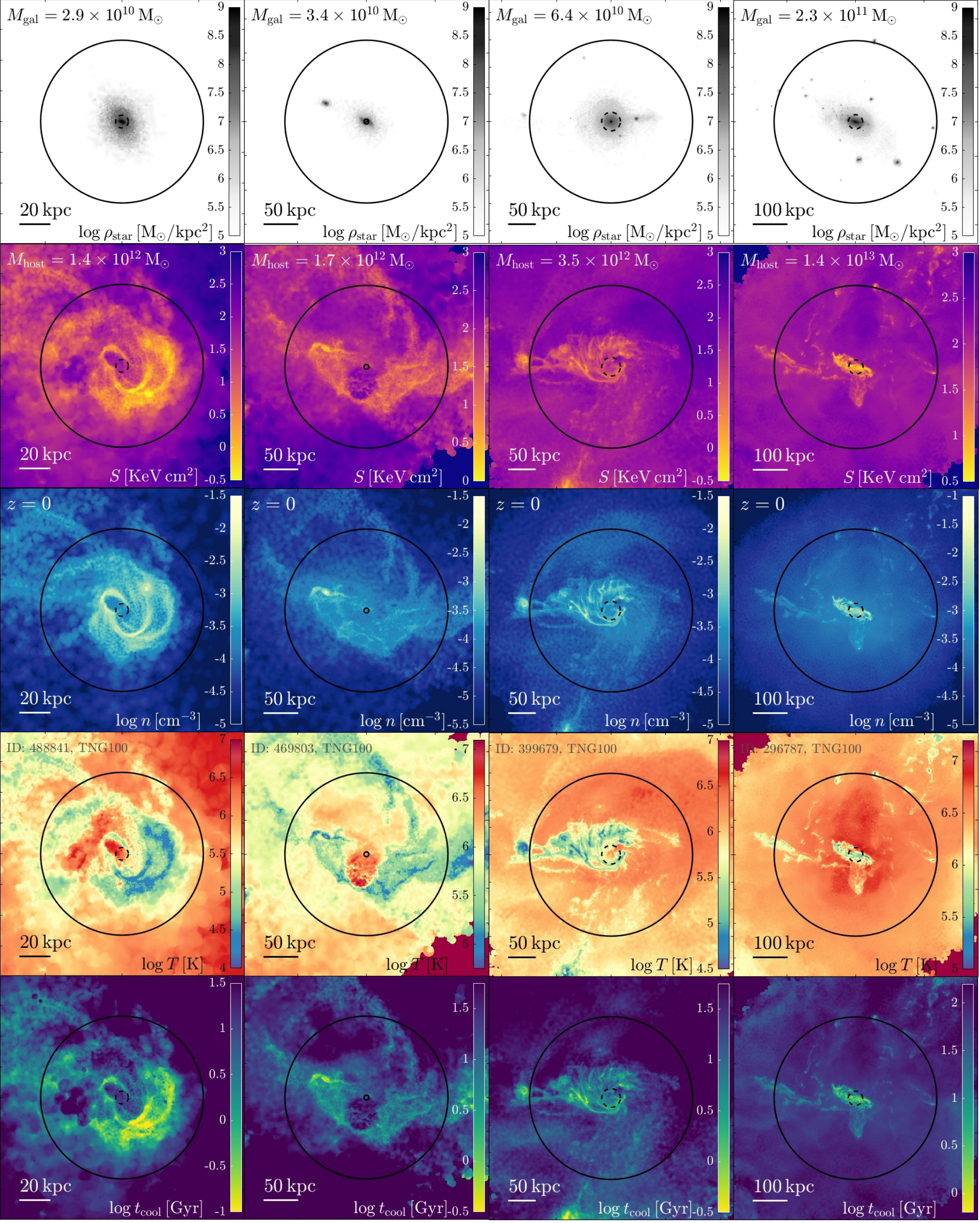}
          \caption{Continued. }
  \label{fig:Maps2}
\end{figure*}

\section{The galactic gas and gaseous haloes of TNG galaxies}\label{sec:results}
\subsection{Visualisation of the gas within and around galaxies}\label{sec:galaxyMaps}
In \cref{fig:Maps1} we show 8 representative central galaxies at
\zeq{0} extracted from the TNG100 simulation that span the mass range
explored in this study. The images focus on the inner CGM and
galactic gas regions of the galaxy. For each galaxy we plot, from top
to bottom, the projected stellar surface density, entropy
(\equnp{entropyDef}), gas number density, temperature, and cooling
time (\equnp{tcool1}). As in every calculation of these quantities,
only gas which is \emph{not} star-forming is used to calculate the
entropy, temperature, and cooling times shown in these images. The
number density of the gas, however, depicts all the gas, including the
star-forming gas cells.

Since the host galaxy and host mass span several orders of
magnitude, the typical values of these properties can vary greatly
over the galaxy selection, and thus the colour-bars of these images
are \emph{not} consistent across images, and have been set to enhance
the important features in each image.

The inner dashed circle, of radius $\rgal$, denotes the `galactic gas'
region, and the outer solid circle is of radius $\rhGas$. The region
between the two circles is the `inner CGM' region, with the `outer
CGM' found beyond the circle, as defined in \cref{sec:gasComponents}.

These maps are plotted from slices taken from the TNG100 simulation,
and contain all the material from the FoF host and thus also include
stellar particles and gas cells which do not belong to the central
\textsc{subfind} halo. The slice thickness for each galaxy is equal to
twice the gas half-mass radius, $\rhGas$, and thus is equal to the
diameter of the outer circle in each plot. The stellar surface density
is projected along the slice. For the other quantities we show the
average along the slice -- a simple average for the gas number density
and mass-weighted averages for the other quantities. The stellar mass
and host mass, $\Mv$, of each system are shown in the first and second
rows, respectively (the galaxy ID in the TNG catalogues is
shown in the second row from the bottom).

Examining the stellar surface density plots (top row) we see that the
stellar distribution often extends beyond our chosen galaxy size
$\rgal$ which is meant as a typical length-scale rather than a sharp
border.
%beyond the galaxy limit we set
%$\rgal$ -- defining an `edge' for a galaxy is always a somewhat
%subjective matter and is often constrained by the available
%observational/computational resolution limits. We are quite happy with
%the our choice of galaxy radius, but we urge the reader to treat it as
%a typical length-scale, rather than a sharp border.

In the low-mass galaxies (part 1 of the figure), the galactic disk is
embedded within a much larger gas disc, which can seamlessly connect
to the large scale filaments of the cosmic web, as seen in the second
system from the left. Here too, it is difficult to determine where the
galactic gas ends and the CGM begins. In the higher mass systems (shown in part 2), the gas distribution beyond the galaxy is less `ordered' and shows more signatures of past and ongoing merger events. In all cases, the conditions of the CGM are \emph{not} uniform -- areas of hot and cold gas are found side by side and the cooling times of
different gas regions can vary by several orders of magnitude. We
address the multi-phase nature of the CGM, as it pertains to our
findings in \cref{sec:multi-phase}.

The right-most system in the lower-mass, first part of the figure,
captures a bipolar high-entropy outflow originating from the central
region of the galaxy. The outflow is hotter than the surrounding gas, but of similar density to its surroundings. The source of this outflow may be stellar/SN feedback, thermal mode or kinetic mode AGN feedback, or some combination of all three. The stellar mass of this system is found below the transitional mass scale of $10^{10.5}\msun$, and its BH mass, $\Mbh=10^{7.6}\msun$, is likewise below the threshold where the kinetic mode has become comparable to the thermal mode. However, we have confirmed that this BH has indeed experienced kinetic feedback injection events in its history, with a cumulative injected kinetic energy of $\sim 10^{58}\units{erg}$, which is relatively high compared to most BHs of this mass (see \cref{fig:bh_stuff}).

\citet{Nelson2019} carried out an extensive study of such outflows in the TNG50 simulation (the smaller volume/higher-resolution run of the TNG project)
showing that these outflows also originate from the action of the kinetic mode AGN feedback. Even though each kinetic feedback energy injection event is assigned a random direction, the resulting outflows follow the path of least resistance and form collimated outflows perpendicular to the galactic disc. While the cooling time in the outflow is quite long, $1\--10\units{Gyr}$ (see bottom panel), the cooling time in the extended gas disk is of only of order $10\units{Myr}$. So even as the AGN is pushing gas out of the galaxy, more gas can be accreted to take its place. The continuous feeding of gas into the galaxy and the BH can generate a continuous outflow. 

We can see the AGN feedback at work in the two systems on the left of
part 2 of the figure. In both systems a hot, dilute (and therefore
high-entropy) region can be seen originating from the galaxy. In the
more massive of these two, the high-entropy `bubble' has a diameter of
$\sim 50\units{kpc}$.

In the most massive system, right-most in the second part of the
figure, one can make out a succession of shock fronts, especially
visible in the temperature map (second panel from bottom). These
shocks may originate from gas pushed out by kinetic mode feedback
and interacting with its surroundings. In high-mass systems such as
this, there is little new gas accretion, which in turn makes the
kinetic feedback more episodic. The individual shock fronts may
correspond to individual injection events.

Many of the physical processes described above such as gas outflows, shock formation etc.\@, can be seen in action in a movie which follows the evolution of a central galaxy in the TNG100 simulation. We invite the reader to view the movie here: \href{www.tng-project.org/movies/tng/tng100_sb0_194946_zinger20.mp4}{www.tng-project.org/movies/tng/tng100\_sb0\_194946\_zinger20.mp4}.
%An additional interesting feature we find in this system is the
%satellite galaxy found just below the main galaxy. Comparing the
%stellar and gaseous distribution around this satellite, we find an
%extensive gas tail trailing behind the galaxy. This is an example of
%a `Jellyfish' galaxy \citep{Poggianti2017} -- a satellite undergoing
%ram-pressure stripping by its interaction with the gaseous environment
%of its host halo. The stripped gas mixes with the CGM and may create
%regions of lower temperature and higher density (and higher
%metallicities) within the CGM. A study of these satellites in the
%TNG100 simulation is found in \citet{Yun2019}.

\subsection{The gas entropy of star-forming and quenched galaxies in TNG300}\label{sec:quenching}
The important role that AGN feedback plays in quenching the
star formation of its host galaxy within the framework of the TNG
physical model has been studied previously. For example,
\citet{Nelson2018a} showed that the TNG model reproduces the galaxy
colour bi-modality, with the transition stellar mass scale found at
$\sim10^{10.5}\msun$ as in observations. \citet{Weinberger2018}
demonstrated that the onset of quenching occurs thanks to the kinetic
mode feedback, and \citet{Terrazas2020} showed that the 
quenching of star formation follows once the cumulative energy injection of the
kinetic mode exceeds the binding energy of the gas in a galaxy,
leading to gas depletion. In what follows, we consider the quenching
process as it relates to the conditions found in the gaseous
components within the galaxy and beyond, in the CGM. We begin by
characterizing the entropy of the gas.

\begin{figure}
  \centering
  \subfloat[Entropy of galactic gas on the sSFR-stellar mass plane]{\label{fig:galPop_ssfr_ent}
      \includegraphics[width=\columnwidth,keepaspectratio]{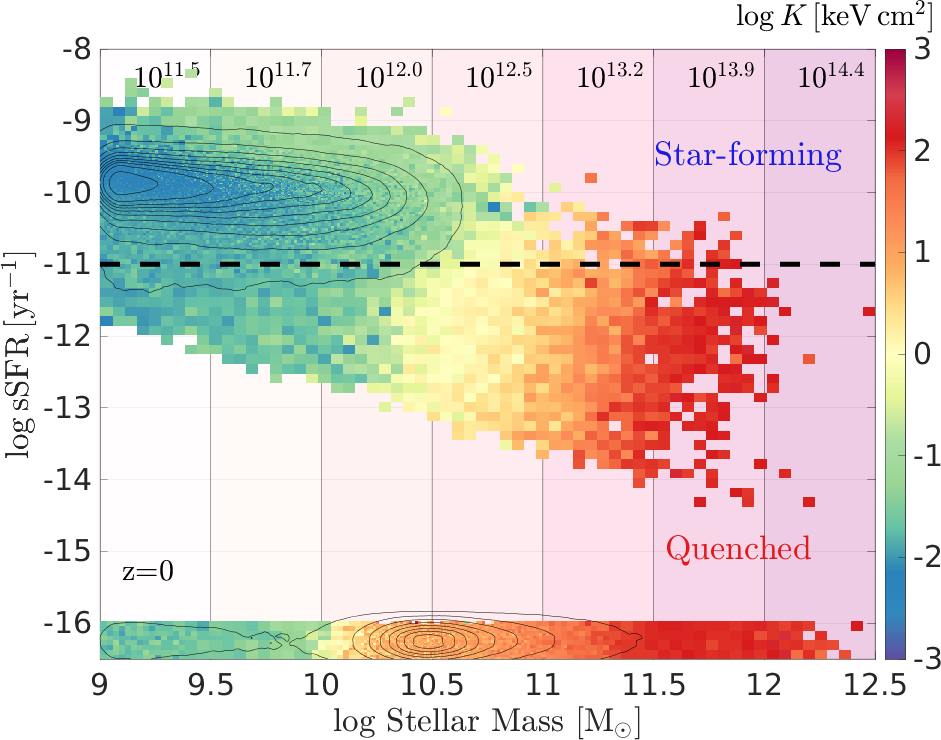}}\\
         \subfloat[Galaxy sSFR  on the entropy-stellar mass plane]{\label{fig:galPop_ent_ssfr}
     \includegraphics[width=\columnwidth,keepaspectratio]{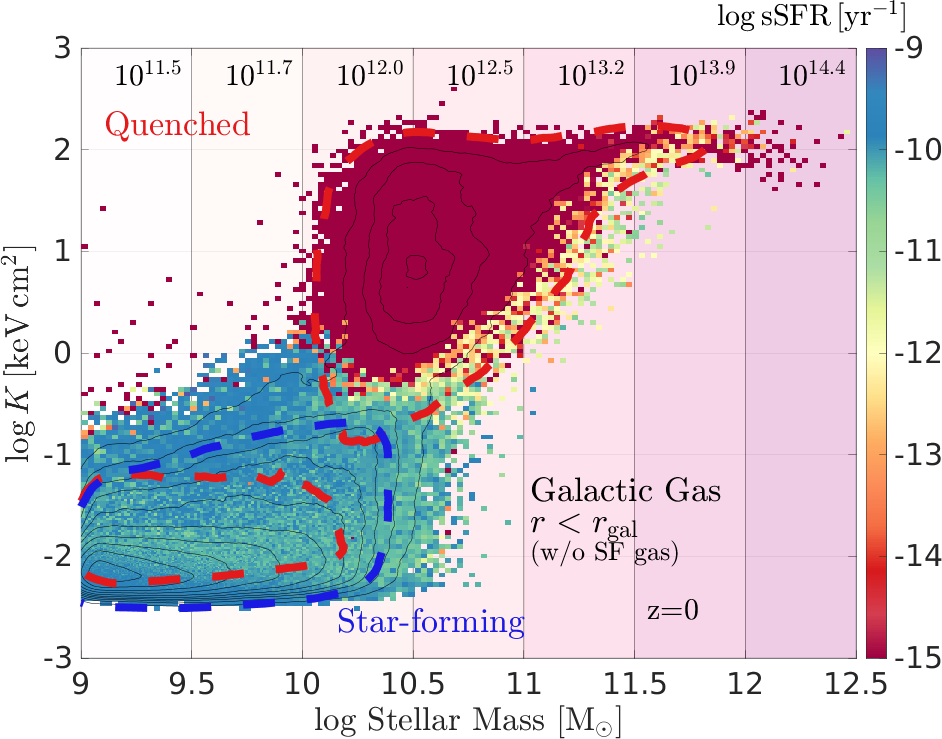}}
     \caption{The \zeq{0} TNG300 central galaxy population in the
       sSFR-stellar mass plane, with pixel colours corresponding to
       galactic gas entropy in \subrfig{galPop_ssfr_ent} and
       the complementary distribution in the entropy-stellar mass
       plane coloured by sSFR in \subrfig{galPop_ent_ssfr},
       illustrating the relation between the thermodynamic state of
       the galactic gas and the SFR (or lack thereof). The
       star-forming sequence, clearly seen in the low-mass end is
       dominated by galaxies of low-entropy galactic gas, while galaxies above the \mbox{$10^{10.5}\msun$} mass scale are predominantly quenched, and characterised by an increase in gas entropy.
       Galaxies with unresolved star formation, $\mathrm{SFR}=0$, are
       placed `by-hand' in the band on the bottom of the plot. The
       pixels in the plot vary in size to include no more than 30
       galaxies, with the colour corresponding to the median of the
       entropy or sSFR. Black contours show the distribution in
       the plane for all galaxies. The dashed line in \subrfig{galPop_ssfr_ent} marks
       our fiducial sSFR threshold of $10^{-11}\units{yr^{-1}}$ used
       to label star-forming vs.\@ quenched galaxies. The blue and red
       contours in \subrfig{galPop_ent_ssfr} each enclose 90 per-cent
       of the SF/Quenched population. The low-mass/low-entropy
       quenched population comprises roughly 15 per cent of the total
       quenched population and is discussed in \cref{sec:lowK_quenched}.}
  \label{fig:galPop_ssfr}
\end{figure}

In \cref{fig:galPop_ssfr_ent} we show the distribution of the TNG300
central galaxy population on the stellar mass-sSFR plane. In this
plot, and in all similar plots that follow, we group the galaxies on
the plane into bins which are adaptively sized to ensure that no more
than 30 galaxies are found within each bin. The colour of each bin is
set by the median value of the property under consideration, in this
case the entropy of the galactic gas.

It is important to note that owing to the mass resolution limit of the
simulation, the SFR in some galaxies falls below a certain resolved
value and so it is in practice equal to zero \citep[see][for a
  discussion]{Donnari2019}. Throughout this paper, such galaxies are
given an arbitrary sSFR value `by hand' in the range $5\times10^{-17}
\-- 10^{-16}$ so that they will appear in the plots -- in
\cref{fig:galPop_ssfr_ent} this is the flat horizontal band in the
bottom of the top panel. In addition, black contours denote the
distribution of the galaxy population on the plane: they enclose the 
5, 15, 25, 35, etc.\@ percentiles of the entire galaxy population in the
depicted plane, with the outermost contour enclosing 95 per cent
of the population. The values at the top of the coloured bands in the
figures show the mean $\Mv$ of the host haloes of the galaxies within
the mass range defined by the bands.

\begin{figure*}
   \centering
   \subfloat[sSFR on entropy-$\Mbh$ plane]{\label{fig:mbh_ssfr}
     \includegraphics[width=\columnwidth,keepaspectratio]{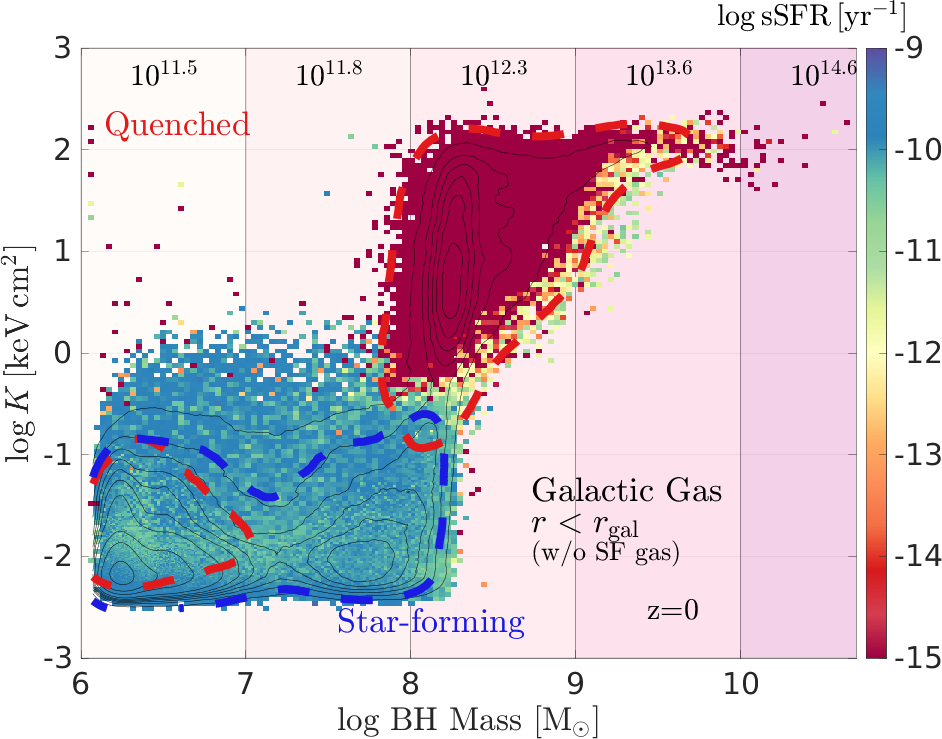}}
     \subfloat[sSFR on the kinetic/thermal mode energy ratio vs.\@ stellar mass plane]{\label{fig:mbh_bhrat}
     \includegraphics[width=\columnwidth,keepaspectratio]{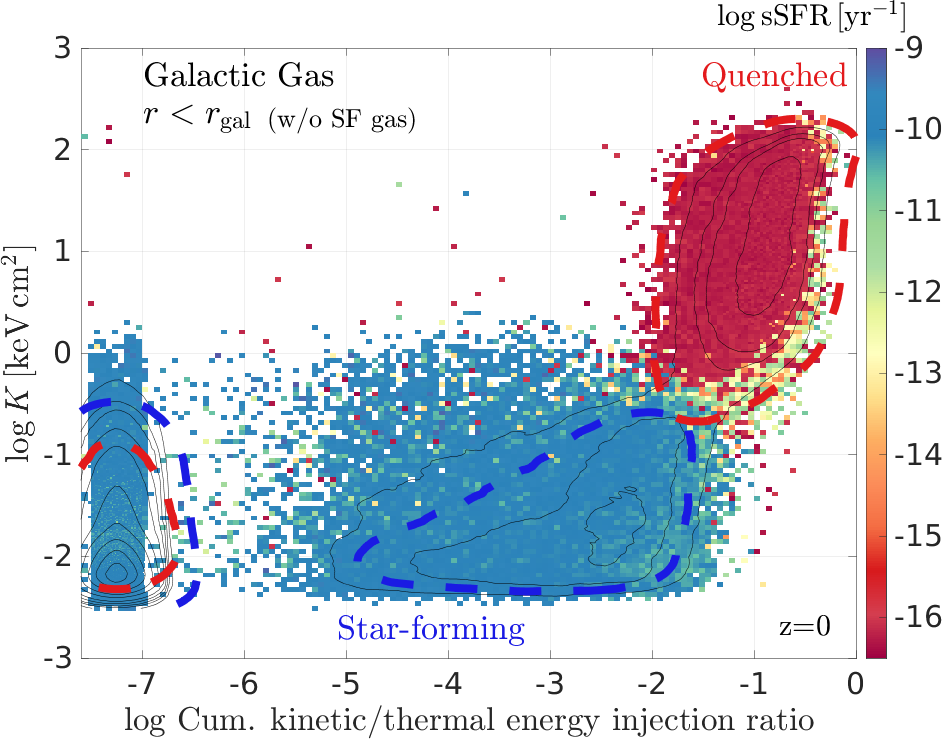}}
   \caption{The role of the AGN feedback in setting the entropy and
     SFR of the galaxy. The distribution of \zeq{0} TNG300 galaxies in
     the entropy-$\Mbh$ plane, with sSFR shown by colour
     \subrfig{mbh_ssfr}, shows that the jump in galactic gas entropy
     and SF quenching occur at a sharp BH mass scale, where the energy injection of 
     kinetic mode feedback becomes comparable to the thermal mode. This can be seen in
     \subrfig{mbh_bhrat} where the entropy is shown vs.\@ the ratio
     between the cumulative energy injected by the two accretion
     modes, coloured by sSFR as before. The vertical band of points on
     the left of the plot are galaxies which have not experienced a
     single kinetic mode injection event, and are given a random value
     \mbox{$<10^{-7}$} for the ratio. Red and blue contours enclose 90
     per cent of the SF/Quenched populations respectively. The onset
     of quenching occurs when the kinetic mode energy becomes
     appreciable ($>1\%$ of the thermal mode).}
   \label{fig:galPop_mbh}
\end{figure*}

The star-forming sequence of galaxies is clearly evident in the low
mass end. Above a stellar mass of $\sim 10^{10.5}\msun$, most galaxies
are `quenched', i.e.\@ fall below the locus of the star-forming main
sequence (or rather, below its extrapolation towards the high-mass
end). In fact, most of them possess so low star-formation rates that
they are unresolved by the simulation. We note that there are many
galaxies, especially in the high-mass end, which are forming stars at
rates of $0.1\--10 \units{M_\odot\,yr^{-1}}$, but would still be
considered quenched due to their high stellar mass, and thus low
sSFR. In terms of numbers, the quenched population comprises 23 per
cent of the total galaxy sample in TNG300 and 15 per cent of the
TNG100 galaxy sample for the same stellar mass minimum of
$10^{9}\msun$ (see \cref{sec:galSample}).

We divide the \zeq{0} galaxy population into star-forming vs.\@
quenched galaxies by using an sSFR threshold of
$10^{-11}\units{yr^{-1}}$ to separate the two groups. This threshold
is based on the bi-modal distributions of galaxies at \zeq{0} in
observations \citep[e.g.\@][]{Brinchmann2004,Kauffmann2004,Wetzel2012}, as well as simulations \citep[e.g.][for the TNG suite]{Donnari2019}.

The transition between the star-forming and quenched population is
seen to correspond to a jump in the entropy of the galactic gas. This
transition is more striking when displayed in the entropy-stellar
mass plane (\cref{fig:galPop_ent_ssfr}) where the colour-coding now
shows the sSFR of the galaxy. The star-forming and quenched population
each inhabit a distinct region in this plane, with the transition
between the two occurring at the transitional stellar mass scale of
$10^{10\--11}\msun$, where the kinetic mode feedback becomes a
significant source of energy, as seen in
\cref{fig:bhModel_stellarMass}. Because the entropy of the gas scales
as $ K \propto T n^{-2/3}$, this separation may indicate that the gas
within galaxies is depleted or heated up, or a combination of the two.

\begin{figure*}
  \centering
  \subfloat[Temperature of the galactic gas]{\label{fig:gal_temp}
    \includegraphics[width=6cm,keepaspectratio]{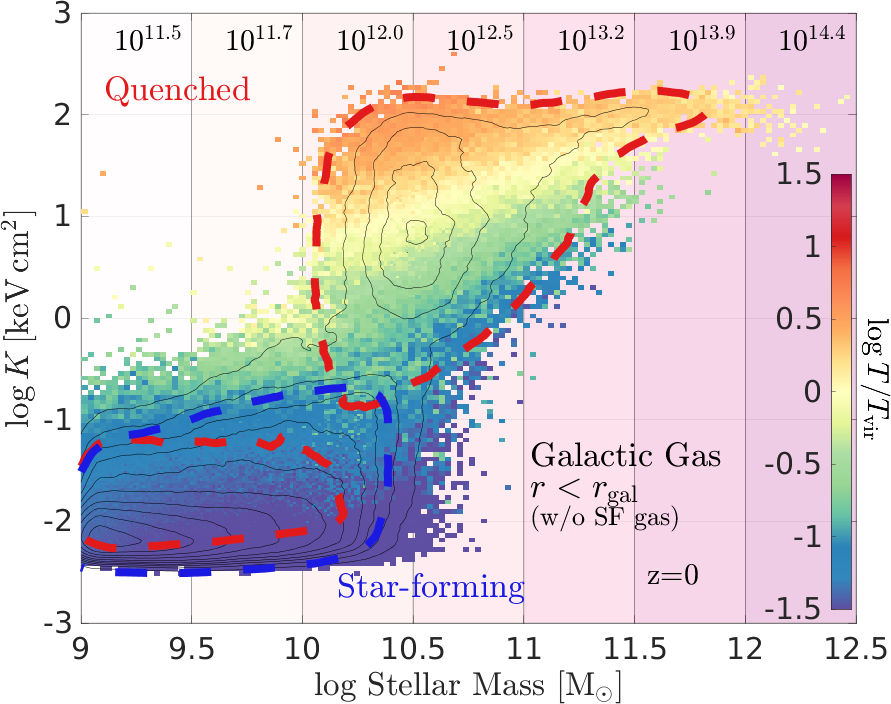}}
  \subfloat[Temperature of the inner CGM]{\label{fig:cgm_temp}
    \includegraphics[width=6cm]{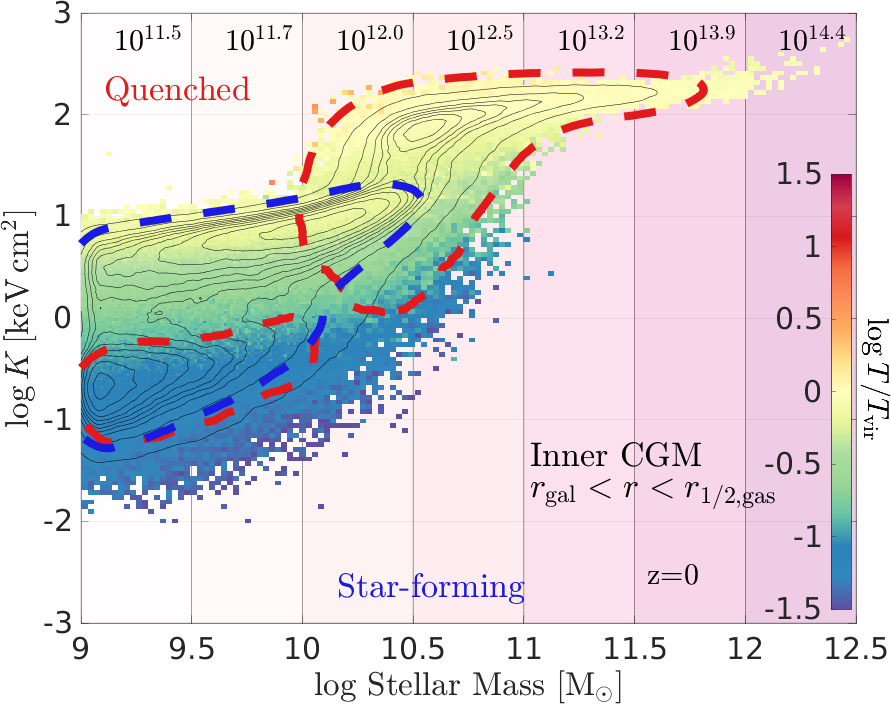}}
  \subfloat[Temperature of the outer CGM]{\label{fig:out_temp}
    \includegraphics[width=6cm,keepaspectratio]{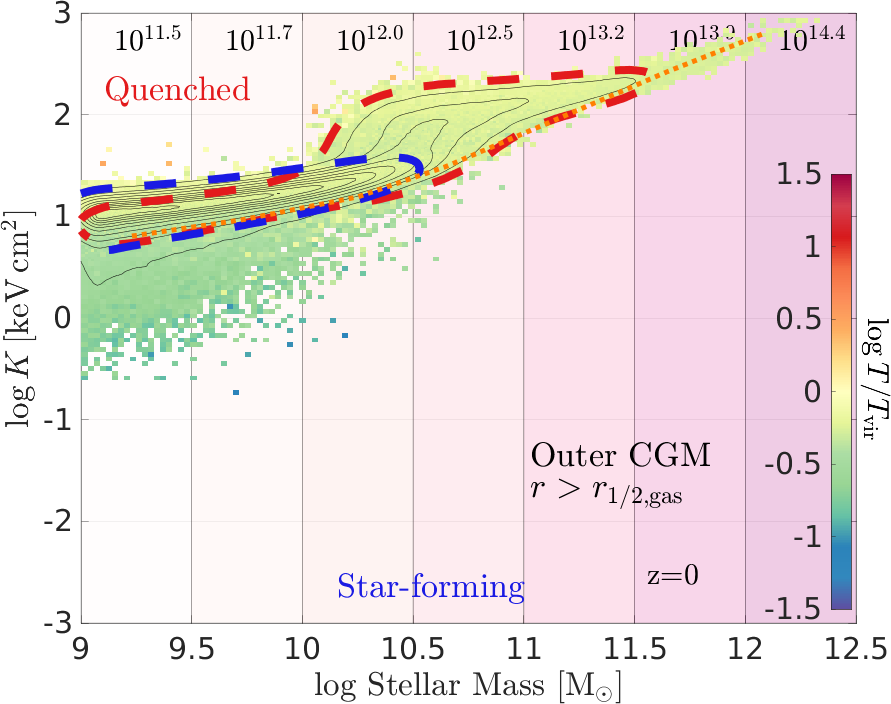}}\\
  \subfloat[Gas fraction in the Galaxy]{\label{fig:gal_massFrac}
    \includegraphics[width=6cm,keepaspectratio]{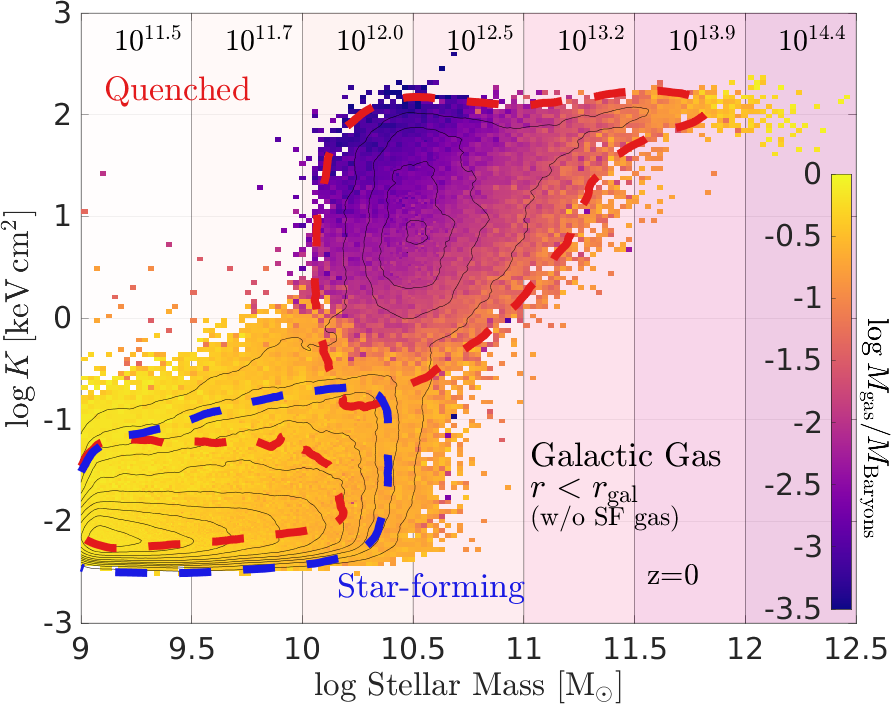}}
  \subfloat[Density of the inner CGM]{\label{fig:cgm_dens}
    \includegraphics[width=6cm,keepaspectratio]{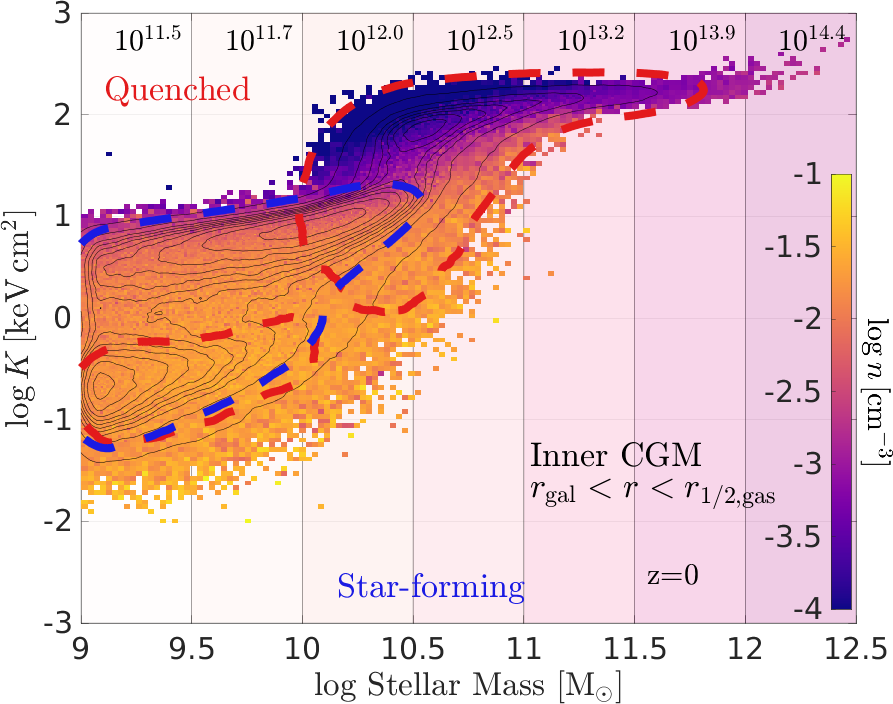}}
  \subfloat[Density of the outer CGM]{\label{fig:out_dens}
    \includegraphics[width=6cm,keepaspectratio]{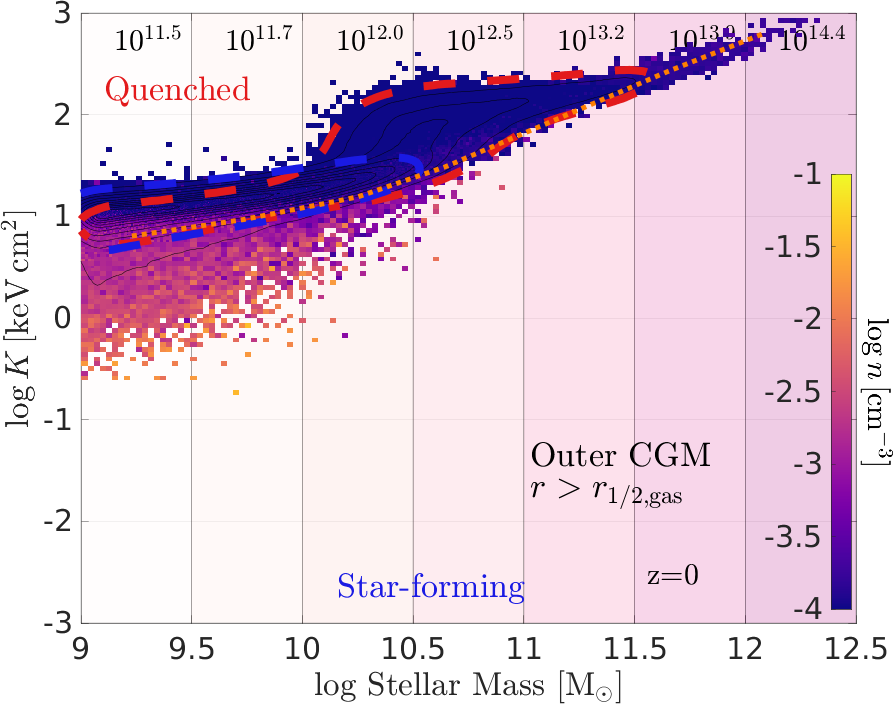}}\\
  \subfloat[Cooling time of the galactic gas]{\label{fig:gal_tcool}
    \includegraphics[width=6cm,keepaspectratio]{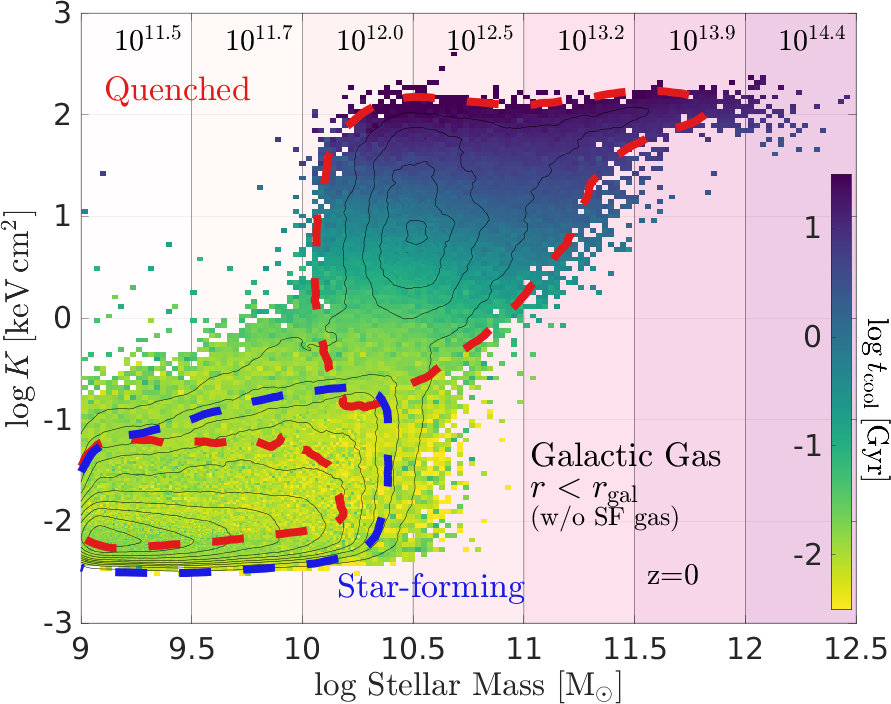}}
  \subfloat[Cooling time on the inner CGM]{\label{fig:cgm_tcool}
    \includegraphics[width=6cm,keepaspectratio]{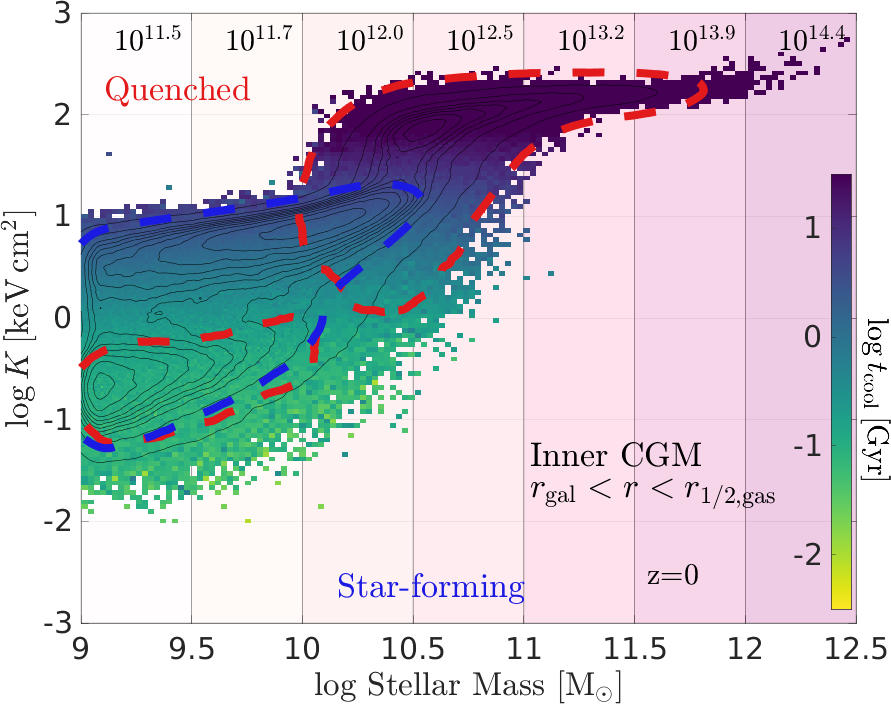}}
  \subfloat[Cooling time of the outer CGM]{\label{fig:out_tcool}
    \includegraphics[width=6cm,keepaspectratio]{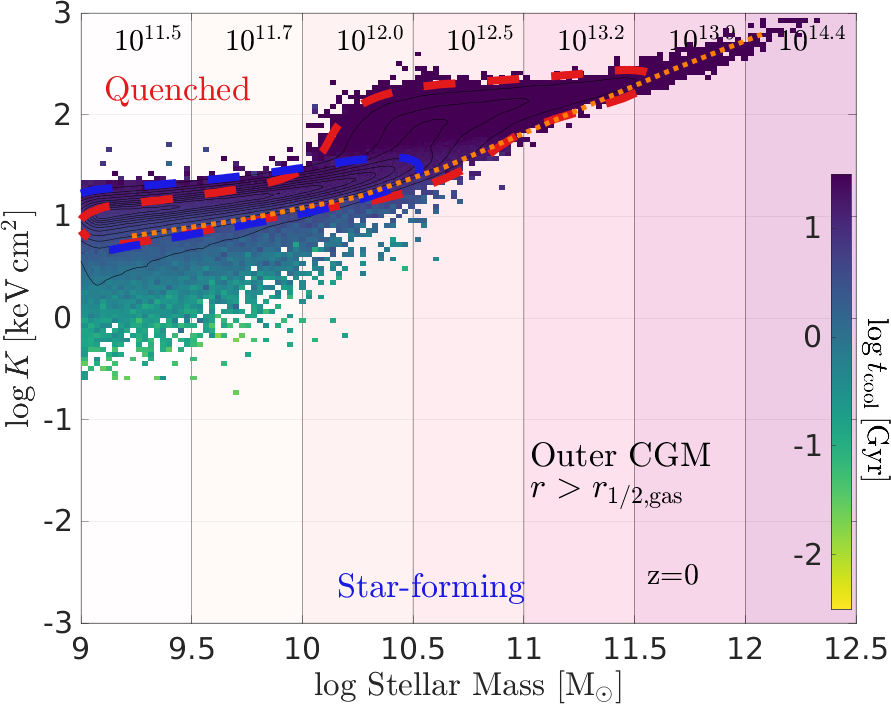}}\\
  \caption{The thermodynamic properties of the \zeq{0} central
    galaxies of TNG300 separated into the galactic gas (left column),
    inner CGM (middle column), and outer CGM (right column), shown as
    the distribution in the entropy-stellar mass plane, with the
    colours corresponding to the gas temperature in units of $\Tv$ (top
    row) and the cooling time (bottom row). In the middle row, the
    leftmost panel \subrfig{gal_massFrac} shows the gas fraction in
    the galaxy and the other panels show the mean gas density in the
    inner and outer CGM. Red and blue contours enclose 90 percent of
    the SF/Quenched populations respectively. The orange dotted line
    in the panel for the outer CGM (right column) show the
    entropy-halo mass relation in the adiabatic case
    (\equnp{entropyVir}), based on the host halo masses shown in
    purple bands.}
  \label{fig:galPop_temp_dens_tcool}
\end{figure*}

An interesting feature of \cref{fig:galPop_ent_ssfr} is the appearance
of an entropy `floor' for the low-mass galaxies and an entropy
`ceiling' seen in the high-mass galaxies. The existence of the entropy
`floor' is set by the physical model of the TNG simulation: in the
simulation gas cannot cool below $10^4\units{K}$, and as specified in
\cref{sec:SFmodel}, cold gas with a number density above
$\sim 0.1\units{cm^{-3}}$ is deemed star-forming and placed on an
effective EOS and thus is no longer part of our analysis. Combining
these two values in \cref{eq:entropyDef} results in a minimal entropy
value for the gas of \mbox{$\log\,K/(\units{keV\,cm^2}) \simeq -2.4$}. There is
no straightforward physical reason for the existence of a limiting
entropy value, and indeed, when examining the entropy of the inner and
outer CGM (see \cref{fig:galPop_temp_dens_tcool}) we see that there is no maximal value for the entropy of the gas. In the outer CGM of high-mass galaxies, where the gravitational
potential of the host halo determines the gas entropy, it is seen to
rise with host mass. The entropy limit seen in the galactic gas may be
a matter of buoyancy -- when the entropy of a gas element is higher
than its surrounding, a buoyant force is exerted on it, accelerating
it outward \citep{Keller2020}, until it is no longer found within our
prescribed radial distance limit $\rgal$.
%\ez{ but why is it the same limit for all masses?}

The classification of star-forming vs.\@ quenched galaxies is shown in
\cref{fig:galPop_ent_ssfr}, and in all subsequent figures, by the blue
and red contours, each of which encloses 90 per cent of the
star-forming/quenched population respectively. While there are some
galaxies at the high-mass end which are considered to be star-forming,
their numbers are very small. However, we do find a
non-negligible group of low-mass, low-entropy galaxies which are
quenched, as marked by the red contour in the bottom-left
corner. Though the size of the contour is \emph{not} an indication of
the size of this galaxy sub-sample, but rather their spread in stellar
mass and entropy, we find that these objects comprise roughly 3.5 percent of the total galaxy population and 15 percent of the quenched
population in TNG300 for stellar mass $\geq 10^9\msun$ (in TNG100, a
similar population comprises 0.5 percent of the total galaxy
population and 3.4 percent of the quenched population). We reiterate
that these numbers reflect the population \emph{after} removing
satellites and `back-splash' galaxies from the sample. We discuss this
curious sub-sample of quenched galaxies in \cref{sec:lowK_quenched}.

The decisive role of the kinetic mode feedback in this transition is
all the more evident when the galaxy entropy is displayed vs.\@ the mass
of the central black hole, as shown in \cref{fig:mbh_ssfr}. The
transition from star-forming to quenched, and the corresponding jump
in entropy occurs very sharply at the transitional black hole mass of
$\Mbh\sim 10^{8.3}\msun$ found in \cref{sec:BHpops} (\cref{fig:bhModel_bhMass}). We speculate that the bi-modality in BH masses, seen earlier in the inset of \cref{fig:bhModel_stellarMass}, is also evident in this figure. The cause for the bi-modality is most-likely the regulation of the BH growth by SN feedback in low mass galaxies (see \cref{sec:BHpops} for details).

The relative role of the two feedback modes is further studied in
\cref{fig:mbh_bhrat} where we show the ratio between the cumulative
energy injected by the central BH in the kinetic mode and the energy
injected by the thermal mode\footnotemark. As shown in
\cref{sec:BHpops}, roughly 60 percent of all galaxies, all of low
mass, have not experienced a single kinetic energy feedback event. We
have shown these in the vertical band at the left end of the plot, by
assigning them arbitrary $x$-axis values below $10^{-7}$. It is
interesting to note that most (78 percent) of low-mass/low-entropy
quenched galaxies are of this group.  \footnotetext{The value of the
  cumulative energy in both the kinetic mode and thermal mode for a BH
  particle includes the energy injected by all the progenitors of the
  BH. Thus, some of this energy may have been injected in the smaller progenitors which built up the central sub-halo, before they were accreted.}

When the kinetic mode mode is responsible for more than a few percent
of the total energy output of the AGN, the entropy of the galactic gas
jumps to much higher values and the host galaxy becomes quenched. The
contribution of the kinetic mode to the total energy output is always
sub-dominant, yet even this small contribution has an appreciable
effect on the galaxy properties, which can extend even to the
surrounding gas and leave its mark on the entire sub-halo CGM (see
\cref{fig:galPop_temp_dens_tcool} and \cref{sec:cgm}).

The jump in entropy which follows the onset the kinetic mode mode
feedback indicates that the gas has been heated, depleted or most
likely both, by the kinetic energy supplied by the AGN (see
\equnp{entropyDef}). We observe this process in action in
\cref{fig:Maps1}: in the two left-most galaxies in the second part, of
masses $2.9 \times 10^{10}\msun$ and $3.4\times 10^{10}\msun$, we see
a hot, dilute `bubble' emanating from the central region.

\subsection{The thermodynamic state of the gas within galaxies}
In \cref{fig:galPop_temp_dens_tcool}, left column, we characterize the
temperature, gas mass fraction, and cooling time of the gas within the
galaxies \mbox{($<\rgal$)}, from top to bottom.  As a reminder, the
color-coded values are averages of all the galaxies in each pixel,
where values for individual galaxies are themselves an average of the gas
within the galactic component.

In the top left panel (\cref{fig:gal_temp}), we show the mean
temperature of the galactic gas, normalised by the virial temperature
$\Tv$ of the host halo, for the distribution of the galaxies in the
entropy-stellar mass plane. As mentioned above and throughout the
paper, we only consider the gas within galaxies and haloes that is
\emph{not} star-forming when deriving an average gas temperature. While
the (non-star-forming) gas within star-forming galaxies is cool, and
close to the lower limit of $10^4\units{K}$ (as seen in the low-mass
galaxies shown in \cref{fig:Maps1}), in the majority of quenched
population the average gas temperature is as hot, or as much as 10 times
hotter than, the virial temperature of the host halo. This shows that
feedback processes inject additional energy into the gas beyond what is
available through gravitational infall into the host halo.

In \cref{fig:gal_massFrac} we show the fraction of gas to total baryon
mass, $M_\mathrm{gas}/\left(M_\mathrm{gas}+M_\mathrm{stars}\right)$,
within the galaxy ($<\rgal$). Here too we see high gas content in
star-forming galaxies, whereas in most of the quenched galaxies, less
than a few percent of the mass is comprised of gas. 

The kinetic mode feedback therefore not only heats the gas to beyond
the virial temperature, it can remove it altogether to beyond
$\rgal$. This is especially true for the galaxies just at the
$10^{10.5}\msun$ mass scale. Of all galaxies affected by the kinetic
mode feedback, these have the shallowest potential wells and their gas
is most susceptible to the feedback. This result is complemented by
the findings of \citet{Terrazas2020}, who analysed the TNG100
simulation, and found that the shutdown of star formation occurred in
galaxies in which the cumulative energy injected by the kinetic mode
of the central BH exceeded the gravitational binding energy of the gas
in the galaxy.

Indeed, when examining galaxies of higher mass, we see that they
still contain large amounts of gas, which is kept hot by the AGN and thus
maintaining their quenched state. This can be seen in the most massive
galaxy shown in \cref{fig:Maps1}: a gaseous disk of density $\gtrsim
0.01\units{cm^{-3}}$ is found in the centre of the galaxy, with a
temperature of $\sim 10^6\units{K}$, and cooling times of order
several Gyrs.

In \cref{fig:gal_tcool} we examine the cooling time of the galactic
gas, which naturally reflects the star-forming efficiency of the
galactic gas. The value of the cooling time in the gas here and in \cref{fig:cgm_tcool,fig:out_tcool} appears to correlate strongly with the entropy \citep[see also][]{Davies2019}. In star-forming galaxies the cooling time is of order
$10\--100\units{Myr}$. For the quenched population, galaxies with high
entropies of $\log K/(\units{keV\,cm^2})\geq 1$ exhibit cooling times of
$1\--10\units{Gyr}$ and above. This implies that the quenched state of
these galaxies will be long-lived since the gas cannot cool
efficiently to form stars. Gas repletion from the CGM of these
galaxies is even less likely, since the cooling time in the CGM is
of order $30\units{Gyr}$ and above for these galaxies, as can be seen
in \cref{fig:cgm_tcool}.

For quenched galaxies in the entropy range
\mbox{$1\--10\units{keV\,cm^2}$} the cooling times are of order
several hundred Myrs, but, as shown above, they contain very little
gas. For a galaxy found at the peak of the quenched distribution, of
mass $10^{10.5}\msun$ and entropy of $\sim 10\units{keV\,cm^2}$, the gas content is of order $\sim 10^{8.5}\msun$ (\cref{fig:gal_massFrac}) with typical cooling times of order $\sim 300\units{Myr}$. Even if 10 percent of the available gas would form stars over a single cooling time \citep[with actual star-formation efficiencies being of order $\sim 1$ per cent per free-fall time,]{Krumholz2019}, the fractional stellar mass increase would be $\sim 10^{-3}$. In comparison, the fractional stellar mass increase for a galaxy on the sSFR threshold over the same period would be roughly 3 times this value -- thus even under these very favourable conditions, the galaxies are expected to remain quenched. The cooling times in the CGM of these galaxies is $\gtrsim 10\units{Gyr}$ (see \cref{fig:galPop_temp_dens_tcool} and next Section) and is therefore not expected to be a significant source of gas accretion onto the galaxy. We revisit this line of reasoning in the Discussion section.
% This shows that the mode of energy injection, namely kinetic vs.\@ thermal, is more important in setting the gas properties than the actual amount of energy. With growing $\Mbh$ the energy injected by the LAM is higher leading to higher entropies in the gas. This results in an `entropy floor' -- a minimum value for galactic entropy which grows with increasing $\Mbh$. 
% \ez{But we still need to understand what are the parameters which set the scatter above this floor}.

\begin{figure*}
  \centering
   \subfloat[BH masses]{\label{fig:cgm_bhMass}
    \includegraphics[width=\columnwidth,keepaspectratio]{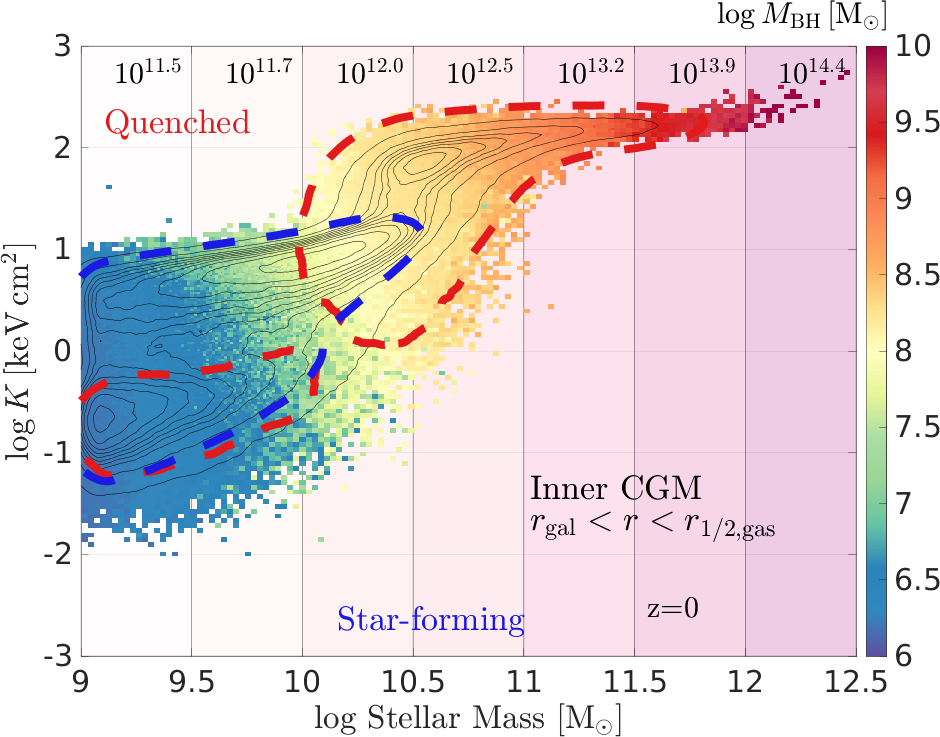}}
  \subfloat[kinetic mode/thermal mode energy injection ratio]{\label{fig:cgm_bhrat}
    \includegraphics[width=\columnwidth,keepaspectratio]{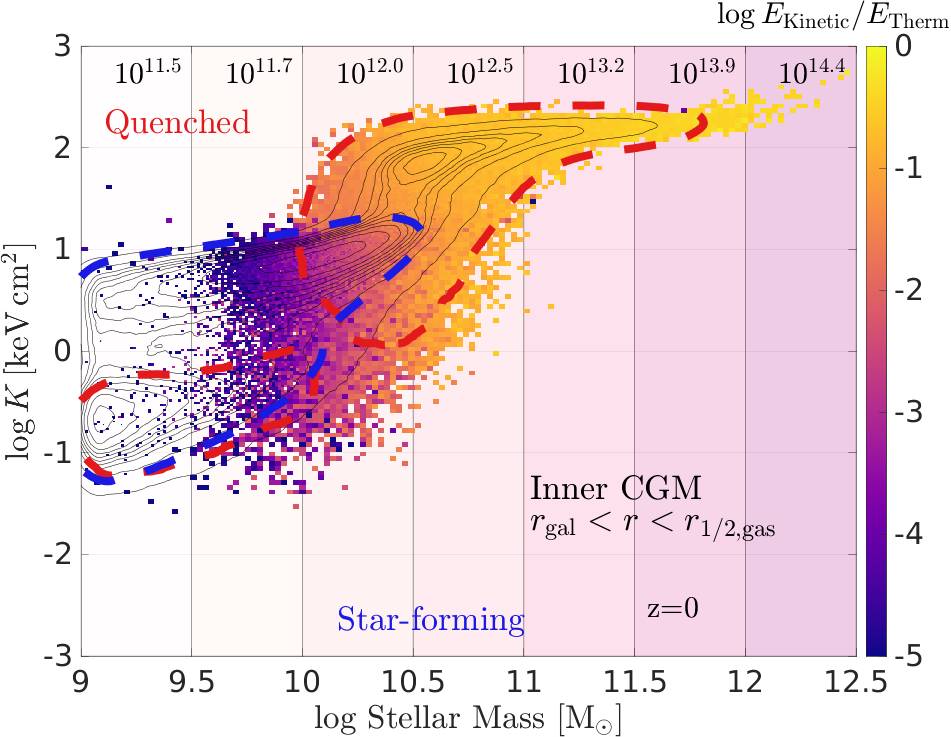}}
   \caption{The relation between AGN feedback and the state of the
     non-star-forming gas in the inner CGM as shown by the
     distribution in the entropy-stellar mass plane of the \zeq{0}
     TNG300 central galaxies, colour coded by the BH mass
     \subrfig{cgm_bhMass} and the cumulative energy ratio of the
     kinetic mode to thermal mode feedback. Red and blue contours each
     enclose 90 percent of the quenched and star-forming populations,
     respectively. The transition to substantial kinetic feedback at
     the transitional mass scale is found in an entropy jump in the
     inner CGM. In the low-mass regime, the concentration of galaxies
     at low entropies corresponds to galaxies in which the kinetic
     mode feedback has not operated at all, while in the concentration
     at intermediate entropies, although kinetic feedback events have
     occurred, they have contributed a very small fraction of the
     total feedback energy, which nevertheless has a noticeable effect
     on the CGM entropy.}
  \label{fig:galPop_cgm_bh}
\end{figure*}

\subsection{AGN effects on the gas properties in the CGM}\label{sec:cgm}
The effects of the AGN are not limited to the gas within the galaxy
but extend out to the surrounding CGM. We see this explicitly in
\cref{fig:Maps1}, in the left-most galaxy of the first part, and the
two right-most galaxies in the second part. In these galaxies,
outflows originating from the central AGN create extended regions and
`bubbles' of hot and dilute gas in the CGM region, in which the
cooling time is considerably longer than in unaffected regions.

A more comprehensive and quantitative analysis is shown in
\cref{fig:galPop_temp_dens_tcool}, middle and rightmost columns, where
we study the distribution in the entropy-stellar mass plane not only
of the gas within galaxies (see previous Section), but also of that
within the inner and outer CGM. Colour-maps show the gas
temperature in units of the host halo $\Tv$ (top row), the mean gas
density (middle row), and cooling time of the gas (bottom row). Again,
the depicted values are averages across galaxy samples of average gas
properties within the different CGM components.

We find that the onset of the kinetic mode feedback at the
transitional mass scale of $\sim 10^{10.5}\msun$ leaves a distinct
imprint in the gas entropy in the inner CGM and even in the outer
regions of the gas halo, up to distances of several hundreds of kiloparsecs
(\cref{fig:radiiDist}). In the distribution of the outer CGM
(\cref{fig:out_tcool,fig:out_temp,fig:out_dens}), we find that most
galaxies lie on a relatively tight relation between stellar mass and
entropy, set by the properties of the host halo. The orange dotted
line shows the global entropy-halo mass relation one obtains in the
adiabatic case (see \equnp{entropyVir}). In essence, this line shows
what the average entropy of the gas would be for a halo of a given
mass and a gas fraction equal to the universal baryon fraction,
affected only by the gravitational potential of the host halo. The
distribution of galaxies roughly follows this relation, except in the
$10^{10.5}\msun$ region where a prominent `bump' of higher entropy
values is found, where kinetic mode feedback becomes important. This
shows that the effects of the AGN feedback can leave an imprint that
extends to the gaseous atmospheres around galaxies.

Furthermore, it is interesting to note that in the low-mass range
(below the transitional mass scale), the average entropy of the outer
CGM is found to be above the adiabatic prediction, whereas in the high-mass
range, the entropy of the outer CGM lies closer to the adiabatic
relation. An excess in entropy above the adiabatic relation indicates
that additional energy, beyond the gravitational potential of the
halo, is imparted into the gas. The injected energy affects both the
thermodynamic properties of the gas, and the gas fraction in the
halo by removing some of it.

In the low-mass range, the energy injection is likely due to SN
and thermal mode AGN feedback. When the kinetic mode becomes
important, the excess in entropy grows considerably, indicating that
the injection mode (kinetic vs.\@ thermal) is of extreme importance in
coupling the energy to the gas. As we examine haloes of even higher
mass, the amount of energy which can be injected by SN and AGN becomes
sub-dominant with respect to the gravitational energy of the host halo
and the global entropy is set largely by the gravitational potential.

The average cooling time in the CGM of galaxies found at the
transitional mass scale is the same as in the central galaxies of
massive clusters, and is of order 10--30 Gyr. With average cooling
times exceeding the Hubble time, the gas in the CGM is not expected to
cool in any significant quantity, ensuring that these galaxies
remain quenched in the future. In this way, the AGN feedback not only
shuts off star formation in the galaxy but also maintains the
quenched state for extended times. Of course, if gas is funnelled to
the galaxy in a different way, such as in a merger with a gas-rich
companion or via smooth accretion of pristine or feedback gas, the
galaxy may acquire more gas and star-formation may resume.

The kinetic mode feedback heats the CGM of the central galaxies, and,
if the potential well of the halo is shallow enough, may actually
evacuate gas from the inner CGM as well. This can be seen in
\cref{fig:cgm_dens} where the mean density of the CGM gas is
colour-coded. The extremely low densities in the lower mass quenched
galaxies, at masses of $\gtrsim 10^{10}\msun$, show that the energy
supplied by the AGN is enough to unbind the gas not only from the
galaxy itself but to push out significant amounts of gas from the
surrounding medium as well. This has also been seen in the EAGLE simulation \citep{Davies2019a}, even though the implementation of the BH feedback in the EAGLE model does not include a kinetic mode. 

When examining the distribution of galaxies in the CGM entropy-stellar
mass plane, as represented by the black contours of
\cref{fig:cgm_temp,fig:galPop_cgm_bh} for example, we find three
distinct galaxy concentrations: a low entropy concentration at low
stellar masses (\mbox{$< 10^{9.5}\msun$} and \mbox{$\log\,K < 0 $}),
an intermediate-entropy concentration at stellar masses of in the
range \mbox{$ 10^{9.5}\--10^{10.5} \msun$} and entropies of \mbox{$ 0
  < \log\,K \lesssim 1$}, and a high entropy concentration with masses
of $>10^{10.5}\msun$ and entropies of \mbox{$ \log\,K>1.5$}. Both the
low- and intermediate-entropy concentrations are comprised of
star-forming galaxies, while the high-entropy concentration is
composed of quenched galaxies.

The cooling time in the low entropy concentration is very short, of
order $\sim 100\units{Myr}$, thus the galactic gas in these galaxies
can be easily replenished by CGM gas cooling. For the intermediate
mass concentration, comprised of galaxies of mass $\sim10^{10}\msun$,
the CGM cooling time is of order $1\units{Gyr}$. The gas depletion
timescale due to star-formation is also of order several Gyrs
\citep{Kong2004,Bigiel2008,Leroy2008,PflammAltenburg2009,Bauermeister2013},
thus these galaxies can remain star-forming since the gas cooling in
the CGM can balance the gas depletion from star-formation. In the
high-entropy concentration, above a stellar mass of $10^{10.5}\msun$
the cooling time is $\gtrsim 10\units{Gyr}$, thus these galaxies are
expected to remain quenched since they have been depleted of cold
galactic gas which cannot be replenished on timescales shorter than
that. The caveat to this statement is the induced cooling of the CGM by cold, high-metallicity gas ejected by SN winds from the galaxy itself \citep{Marinacci2010,Fraternali2013,Armillotta2016}, though this effect is more important in star-forming galaxies rather than in galaxies which have already been quenched.

In \cref{fig:galPop_cgm_bh} we can find clues as to what causes the groupings of galaxies in CGM entropies: the typical BH masses found in these galaxies
(\cref{fig:cgm_bhMass}) and more importantly, the amount of kinetic
feedback released by the BH, or lack thereof (\cref{fig:cgm_bhrat}). In the
low-entropy concentration BH masses are small and there has been no
kinetic mode feedback at all. Once the kinetic mode feedback
starts to occur, even at very low levels (below the $\Mbh\sim
10^8\msun$ mass scale), the entropy of the inner CGM is seen to
increase by two order of magnitudes, and the typical temperature rises
to $\Tv$ (\cref{fig:cgm_temp}), though the gas density remains high
(\cref{fig:cgm_dens}). Once the BH mass scale is crossed and the
kinetic feedback jumps to higher values, the entropy of the inner CGM
grows even further, reaching values identical to those of haloes of
masses 100 times larger. The temperature in the high entropy
concentration is still in the vicinity of $\Tv$, but the mean density
drops to very low values. Thus, small amounts of kinetic energy input heat the CGM gas, but only when the kinetic mode becomes the prevalent mode does the gas become diluted. 

A question naturally arises -- Why do we not see a similar distinction
in entropy in the galactic gas of star-forming galaxies between those
galaxies with no kinetic feedback at all and those with some kinetic
feedback (e.g.\@ \cref{fig:gal_tcool})? We think that once the kinetic
feedback begins to be active, even with low levels of energy injection, some portion of the gas will be affected, and its entropy will increase. This gas would be hydrodynamically unstable and thus rise outwards to beyond our fiducial $\rgal$, thus becoming part of the CGM. The rest of the gas is unaffected and
retains its low entropy and short cooling time.

\begin{figure*}
  \centering
    \subfloat[Density ]{\label{fig:comp3_dens}
    \includegraphics[width=7.5cm,keepaspectratio]{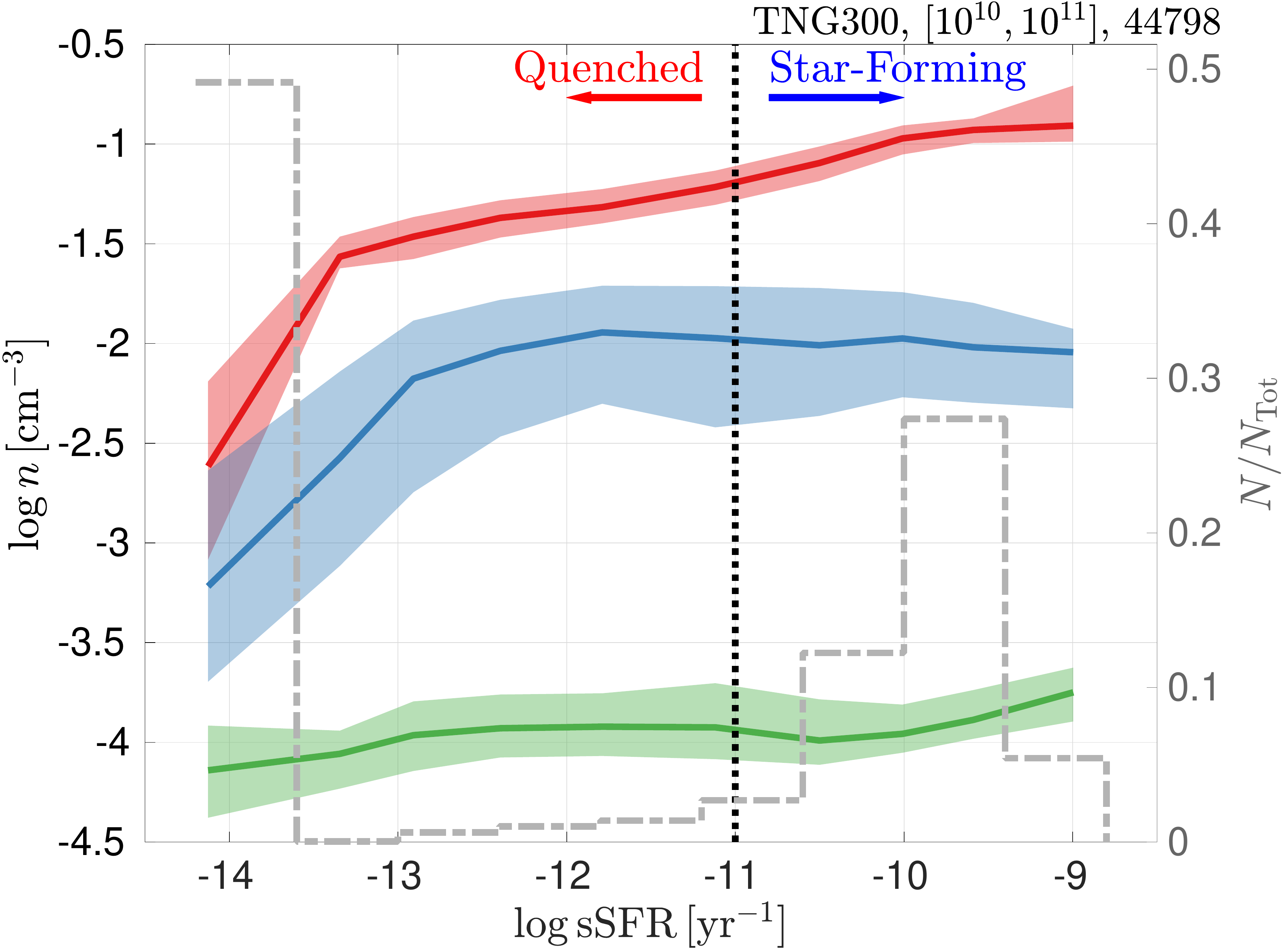}}
  \subfloat[Temperature]{\label{fig:comp3_temp}
    \includegraphics[width=7.5cm,keepaspectratio]{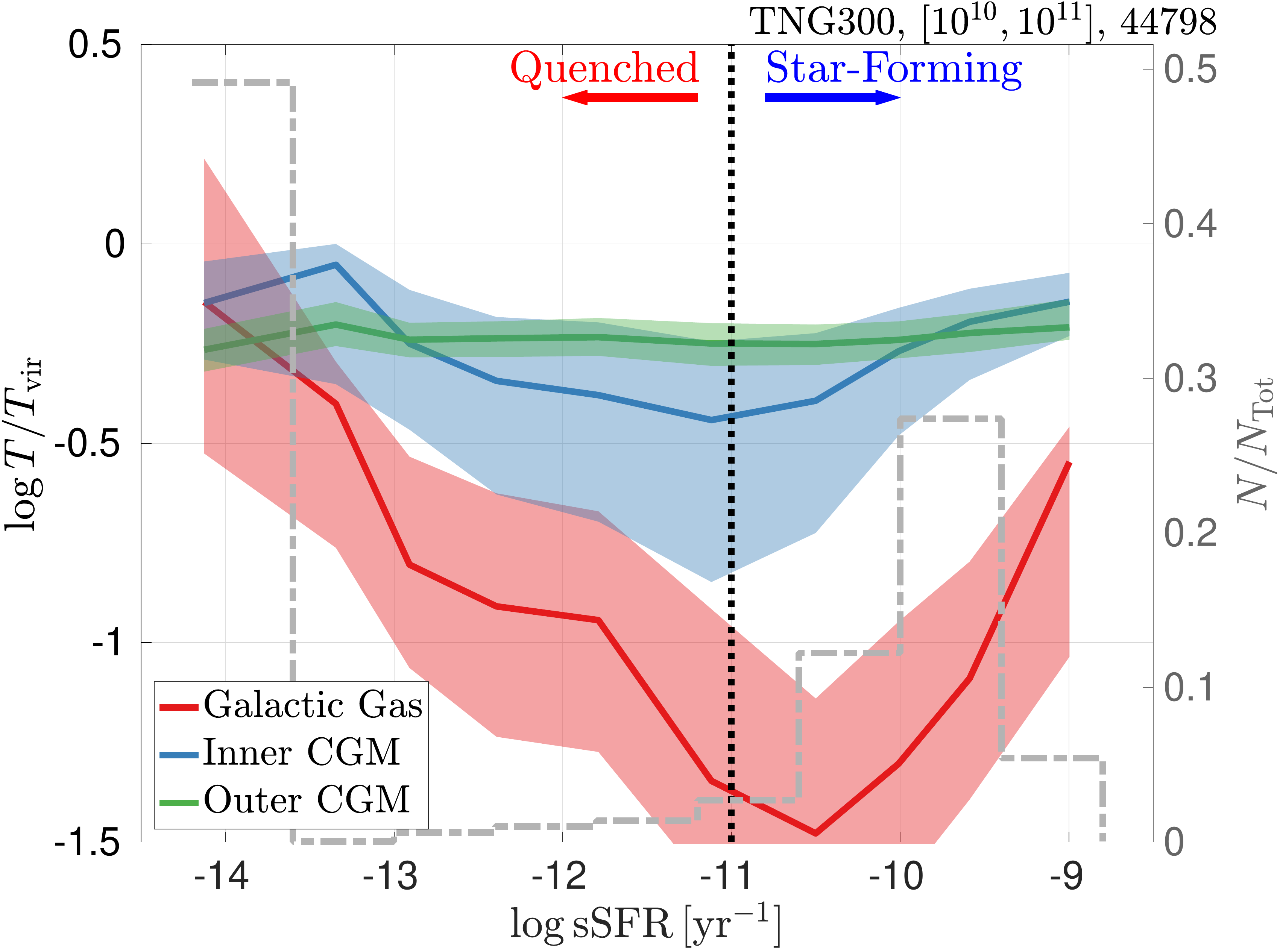}} \\
  \subfloat[Entropy]{\label{fig:comp3_ent}
    \includegraphics[width=7.5cm,keepaspectratio]{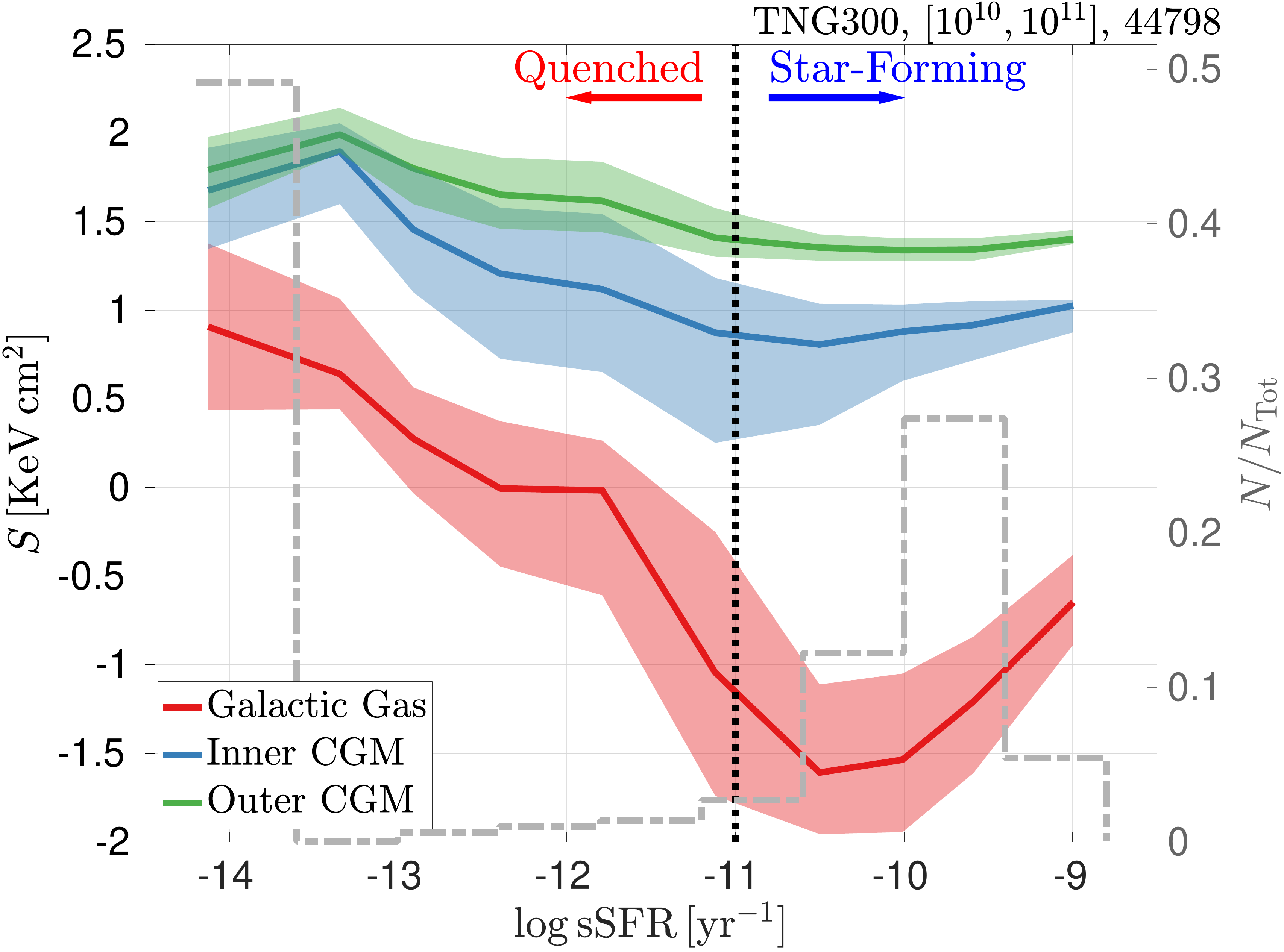}}
  \subfloat[Cooling time]{\label{fig:comp3_tc}
    \includegraphics[width=7.5cm,keepaspectratio]{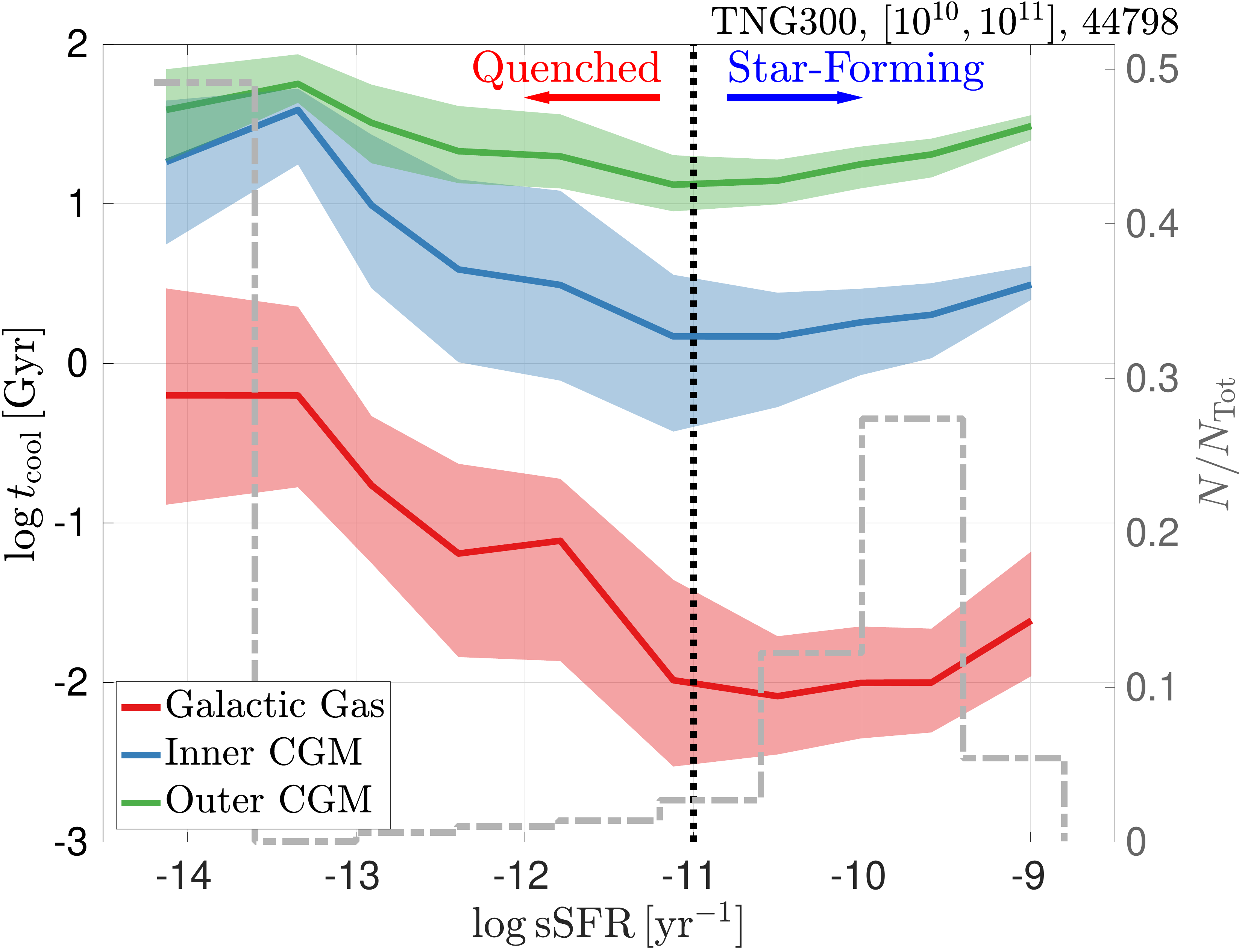}}
  \caption{Thermodynamic properties of the non-star-forming galactic
    gas (red), inner CGM (blue), and outer CGM (green), shown as a
    function of the sSFR of the galaxies in the transitional mass
    range $10^{10}\--10^{11}\msun$ in the \zeq{0} TNG300 simulation
    (44798 galaxies). The median values within each bin of the number
    density and (top left), temperature, normalised by $\Tv$ for each
    galaxy (top right), entropy (bottom left) and cooling time (bottom
    right) are shown by solid lines with the shaded region indicating
    the 25--75 percentile range. The right axis and corresponding
    brown histogram show the fraction of galaxies in the sSFR
    bins. The sSFR threshold between the star-forming and quenched population is marked by the vertical dotted line. Changes in the galactic gas, which reflect the transition
    from star-forming to quiescence, are mirrored in the CGM, showing
    that the driving force for this transition, the kinetic mode
    feedback, also affects the gas beyond the galaxy.}
  \label{fig:compTransition}
\end{figure*}

\subsection{Thermodynamics of the CGM for star-forming vs.\@ quiescent galaxies at the transitional mass scale}
We explore what happens to the central galaxies as they transition
from star-forming to quenched by examining the galaxies in the
transitional mass range $10^{10}\--10^{11}\msun$ in the TNG300 simulation, where the kinetic mode becomes effective. In this mass range
we find 44,798 galaxies with roughly equal numbers of star-forming and
quenched galaxies. As shown in \cref{fig:galPop_mbh}, the transition
between star-forming and quenched in this mass range is tied to the
onset of the kinetic mode feedback.

In \cref{fig:compTransition} we bin these galaxies by their sSFR, and
plot the thermodynamic conditions: density, temperature (normalised
by $\Tv$), entropy, and cooling time, within the 3 gas components: the
galactic gas, inner CGM, and outer CGM (red, blue, and green curves,
respectively). The values plotted are the median values over all
galaxies in a given sSFR bin, with the shaded region showing the
25--75 percentile range. Also shown is a histogram of the number of
galaxies within each bin.

As seen earlier, the transition from star-forming to quenched is
marked by a rise in the temperature -- the galactic gas is very cool in
the star-forming galaxies, but in the quenched galaxies it is as hot
as the inner and outer CGM with values of order $\Tv$. The inner and
outer CGM temperatures remain fairly constant. In all three gas
components we see a rise in entropy, with the most significant change
in the galactic gas. The entropy change is very similar to the changes
in cooling times, showing once again (see
\cref{fig:galPop_temp_dens_tcool}) that the entropy is a good
indicator of the cooling time in the gas.

The cooling time in star-forming galaxies is quite short, of order
$\sim 10\units{Myr}$, but rises to values of $\sim 1\units{Gyr}$ in
the quenched population. The density of the gas also drops
considerably, so while the cooling time is not very long, the small
amount of gas that remains in these galaxies (see also
\cref{fig:gal_massFrac}) is of very low density and is not likely to
be a source of considerable star formation in the future.

This behaviour is mirrored in the inner CGM component -- the average
cooling time grows from several Gyrs in the star-forming galaxies to
exceeding the Hubble time, and the density drops by more than an order
of magnitude. In this way we see that the onset of the kinetic mode
feedback greatly affects the inner CGM as well as the gas within the quenched
galaxies and thus, left to itself, the CGM is not expected to be a substantial source for gas accretion.

The most star-forming galaxies in this mass range, with
\mbox{$\mathrm{sSFR}\gtrsim 10^{-9.5}\units{yr^{-1}}$}, actually show
a rise in the normalised temperature and entropy of the galactic gas
compared to the rest of the star-forming population
(\cref{fig:comp3_temp,fig:comp3_ent}). We speculate that this may be
due to heating of the galactic gas by the action of SN feedback
and/or the thermal mode of the AGN.

 \begin{figure*}
  \centering
     \subfloat[Inner CGM in low-mass galaxies]{\label{fig:gasHist_cgm_low}
    \includegraphics[width=5.5cm,keepaspectratio]{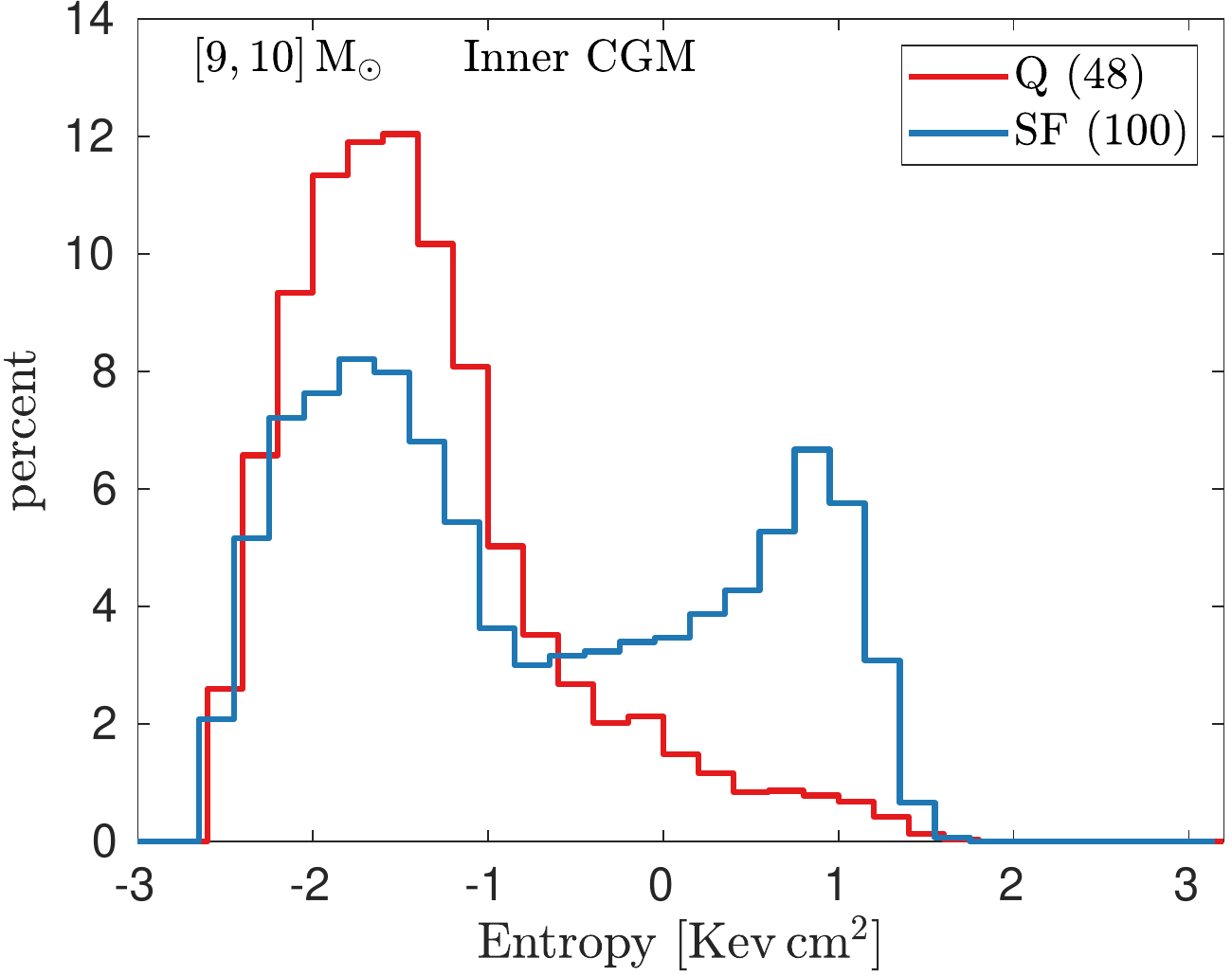}}
        \subfloat[Inner CGM of intermediate-mass galaxies]{\label{fig:gasHist_cgm_trans}
    \includegraphics[width=5.5cm,keepaspectratio]{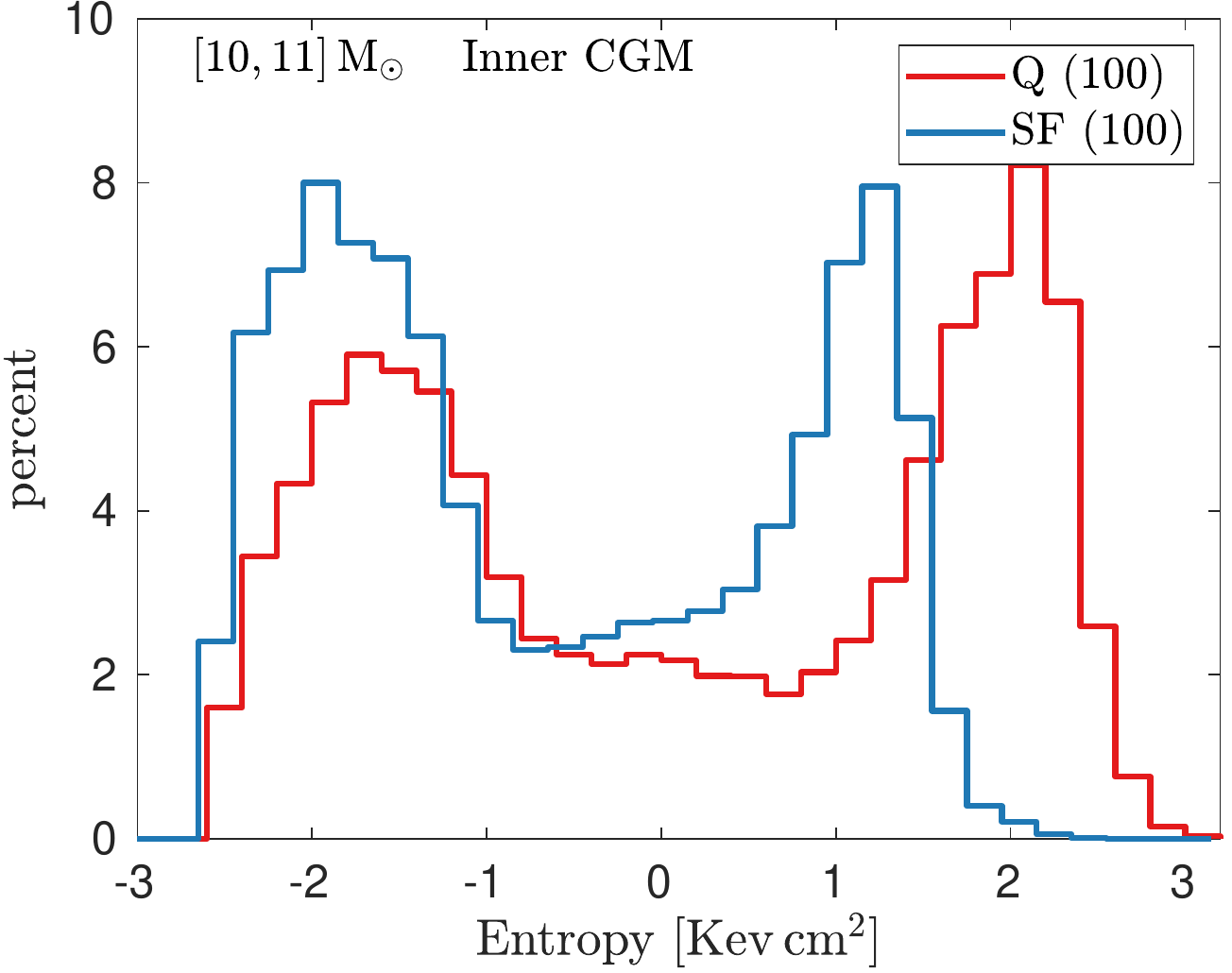}}
        \subfloat[Inner CGM of high-mass galaxies]{\label{fig:gasHist_cgm_high}
          \includegraphics[width=5.5cm,keepaspectratio]{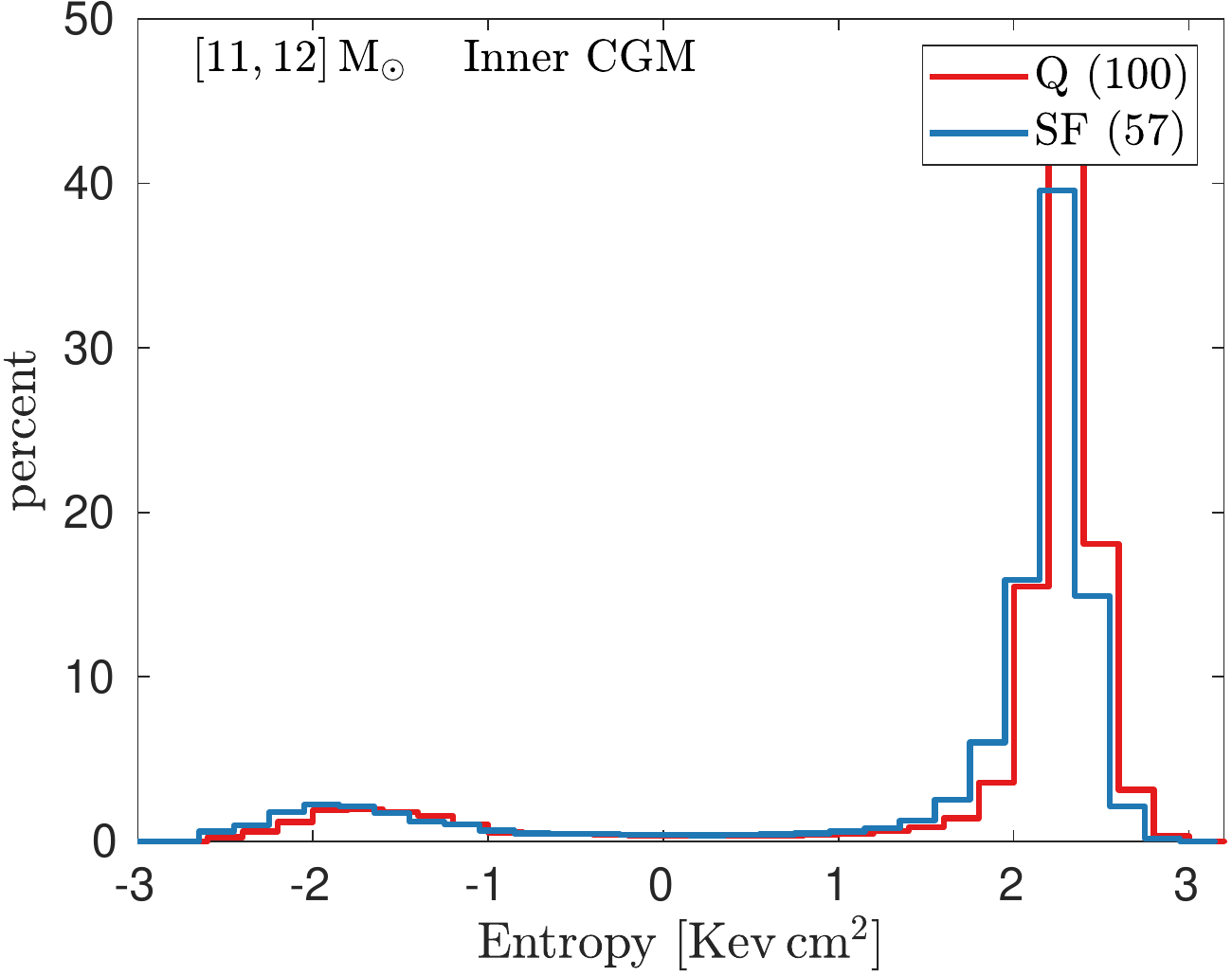}} \\
          \subfloat[Outer CGM of low-mass galaxies]{\label{fig:gasHist_out_low}
    \includegraphics[width=5.5cm,keepaspectratio]{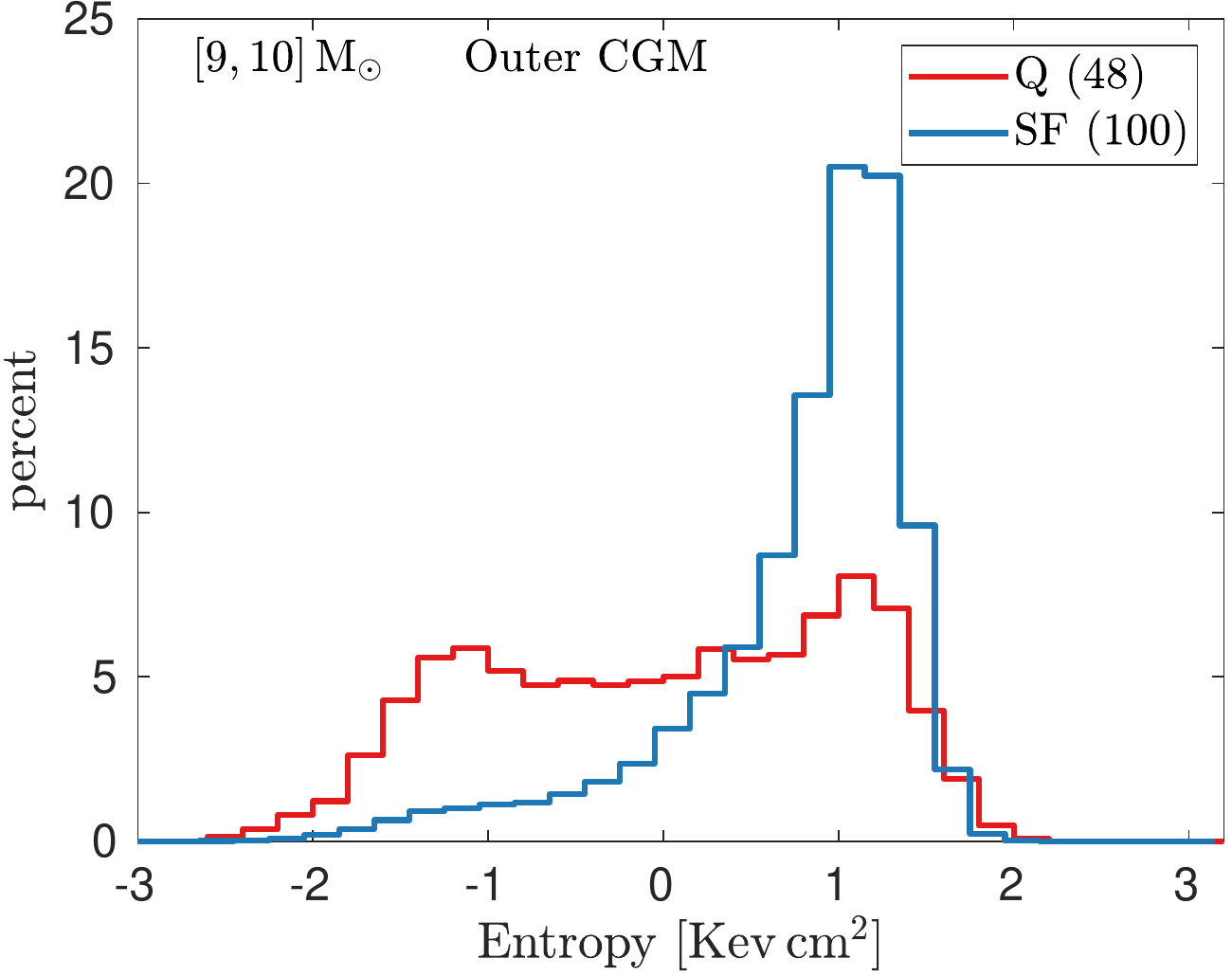}}
        \subfloat[Outer CGM of intermediate-mass galaxies]{\label{fig:gasHist_out_trans}
    \includegraphics[width=5.5cm,keepaspectratio]{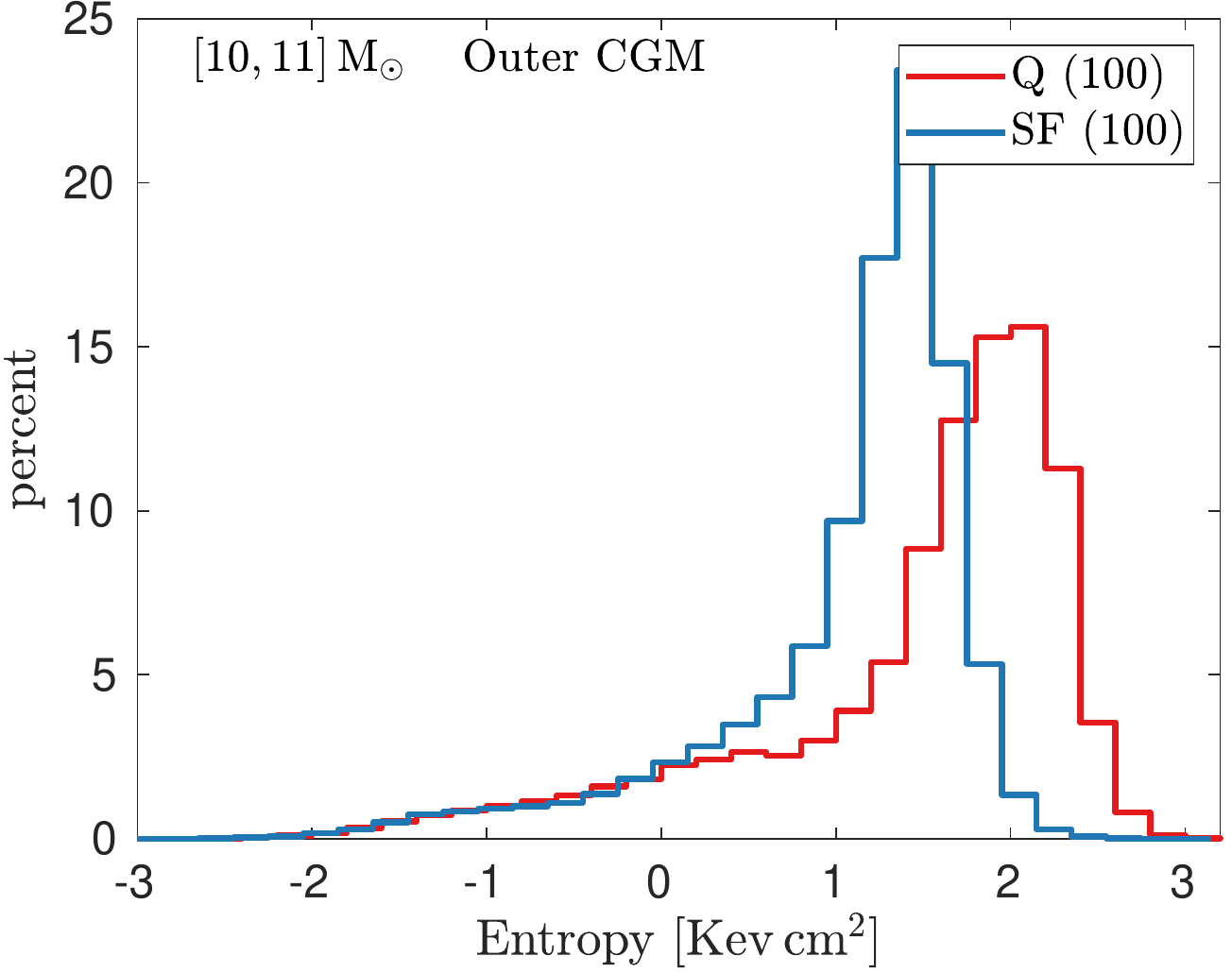}}
    \subfloat[Outer CGM of high-mass galaxies]{\label{fig:gasHist_out_high}
    \includegraphics[width=5.5cm,keepaspectratio]{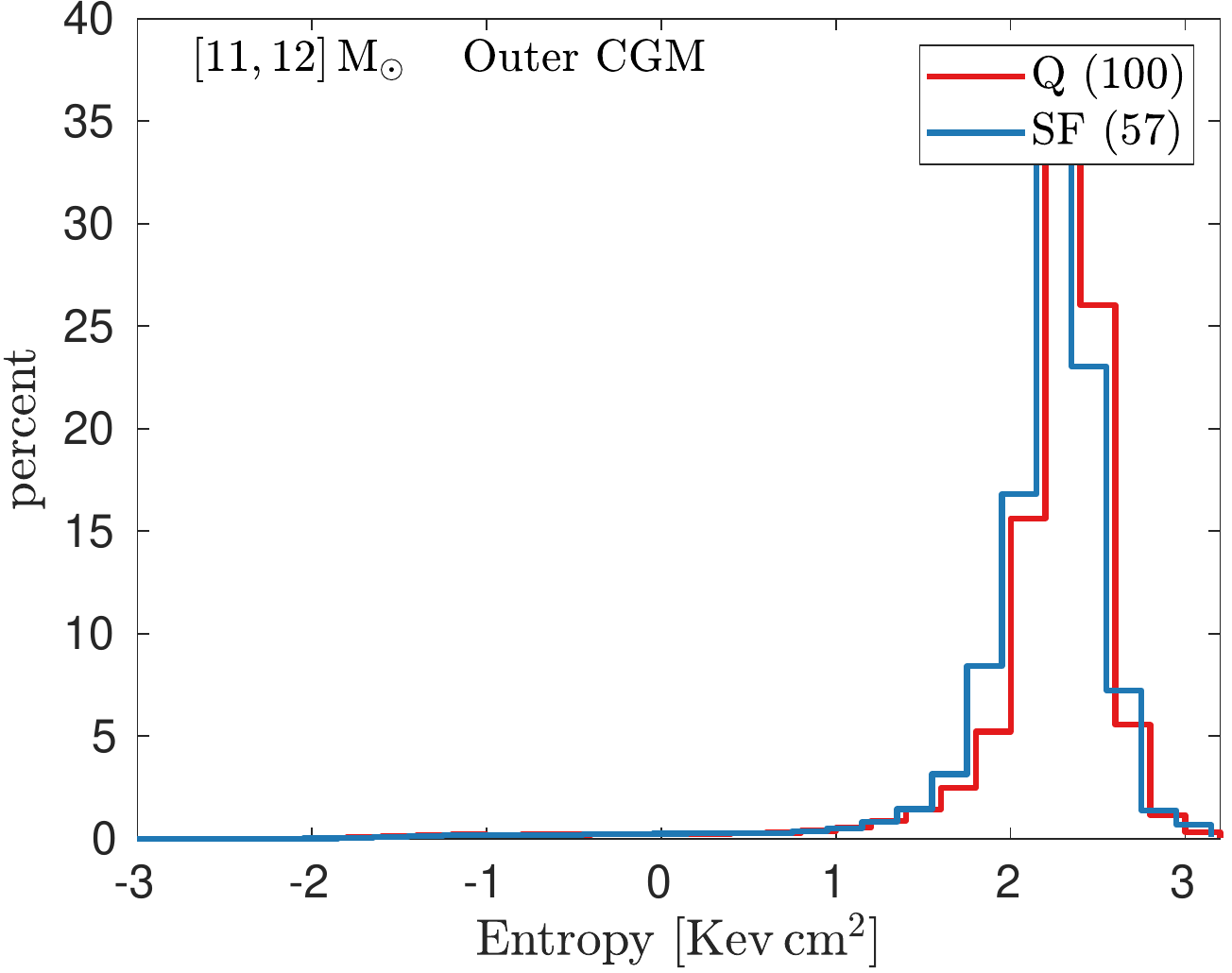}} 
    \caption{The multi-phase nature of the inner and outer CGM (top and
      bottom, respectively) is shown by stacked mass-histograms,
      counting only the non-star-forming gas, of the CGM entropy for
      star-forming (blue) and quenched (red) galaxies in three stellar mass ranges of the TNG100 simulation: \mbox{$10^9\--10^{10}\msun$}  (left), \mbox{$10^{10}\--10^{11}\msun$}(middle), and \mbox{$10^{11}\--10^{12}\msun$ } (right). In each mass bin, the individual mass-histograms of the CGM entropy of 100 (or less) galaxies of each group were stacked. The number of stacked histograms for each group and mass bin is shown in the legend. In all cases, most (but not all) of the individual stacked galaxies have qualitatively similar distributions to the stacked histogram shown here. The bi-modal distribution in the inner CGM shows that the gas is not monolithic in its properties. The shift to higher values in the high-entropy peak of the quenched, intermediate-mass galaxy compared to similar mass star-forming ones shows that the
      transition between being star-forming and quenched involves a
      change in the CGM properties as well. The quenched galaxies in \subrfig{gasHist_cgm_low} and \subrfig{gasHist_out_low} are the low-mass/low-entropy quenched population discussed in detail in \cref{sec:lowK_quenched}. }
  \label{fig:gasHistograms}
 \end{figure*}
 
 \subsection{The multi-phase nature of the CGM and the role of AGN feedback}\label{sec:multi-phase}
Up to this point we have treated the galactic gas and the CGM as
monolithic components whose thermodynamic state is defined by single
average values of density, entropy, temperature, and cooling time. In
reality, observations show that the gas within galaxies and in the
CGM is multi-phase in nature and comprised of regions which vary in
density and temperature leading to greatly varying cooling times \citep[][and references therein]{Tumlinson2017}. This multi-phase nature of the CGM gas is realised also in the TNG simulations, as demonstrated to some extent already by
\citealt{Nelson2019} with the TNG50 run.

This can also be seen in the maps of \cref{fig:Maps1} -- the CGM of
low-mass galaxies contain disc-like structures of low entropy (dense and
cool) gas. Especially striking is the right-most galaxy in the first
part of the figure, in which the inner CGM, found within the solid
black circle, is comprised of a cold and dense disk (seen edge-on) with
two lobes of hot outflows emanating from the galaxy at the centre.

In the higher mass galaxies, seen in the second part of the figure, we
still find regions of varying entropy in the CGM, though they are not
as structured as in the large gas discs seen in the low-mass
examples. In the three most massive galaxies of \cref{fig:Maps1} we
also find satellite galaxies, best seen in the stellar mass map on the
top row. These satellite galaxies can also be a source of lower
entropy gas in the CGM -- `Jellyfish' galaxies which lose gas due ram-pressure as they travel through the CGM \citep{Poggianti2017,Yun2019}. The cold, low-entropy stripped gas joins the much hotter medium, and may even lead to cooling and entrainment of gas from the CGM \citep{Marinacci2010,Gronke2018,Gronke2020}.

A quantitative characterization of the multi-phase nature of the CGM is
shown in \cref{fig:gasHistograms} where we separately stack histograms
of the CGM entropy of star-forming and quenched galaxies selected from
the TNG100 simulation, in three galaxy stellar mass ranges. In each
mass range, $10^9\--10^{10}\msun$, $10^{10}\--10^{11}\msun$, and
$10^{11}\--10^{12}\msun$, we randomly select 100 galaxies from each
group, star-forming or quenched. In the low (high) mass range we find
less than 100 quenched (star-forming) galaxies, and in those cases
collect all the available galaxies of that group. For each galaxy, a
mass histogram of the CGM entropy is created, i.e.\@ how much gas mass
is found at a given entropy value, and these histograms are stacked
together. As in the analysis presented thus far, here we include only central galaxies and we account only for the gas that is gravitationally bound to each central, i.e.\@ excising the gas contribution from satellites. We choose to present here the results from the TNG100
galaxies since the higher resolution of the simulation better captures
the multi-phase nature of the CGM, but repeating the analysis with the
TNG300 sample gives very similar results.

In the inner CGM, shown on the top row of \cref{fig:gasHistograms},
the entropy distribution is bi-modal for (nearly) all galaxy types and
mass ranges, indicating that the CGM contains both low and high
entropy gas, although the amount of low-entropy gas in the high-mass
regime is found to be very small. The high amount of low-entropy gas
found in the quenched galaxies of the transitional mass range
(top-middle panel) is especially striking: these are mass histograms
and thus the top middle panel in \cref{fig:gasHistograms} shows that
there can be as much (in mass) low-entropy as high-entropy gas in the
CGM of the depicted galaxies. Though the AGN feedback has already
affected some of the CGM gas, (seen in shift of the high-entropy peak
compared to that of the star-forming galaxies), there are still
considerable amounts of low-entropy gas. This points to a localised
effect of the AGN, as can be achieved with collimated outflows or
bubbles (the mid-range galaxy images of \cref{fig:Maps1} show this
nicely). We note that a similar bi-modal distribution is found in the 
mass histograms of the temperature and density separately (not shown here), 
with the high-temperature/low-density peak of the quenched galaxy groups 
shifted to higher/lower values compared to the star-forming group. 

The quenched galaxies at the low-mass end (red histograms in \cref{fig:gasHist_cgm_low}) deviate from the general description proposed so far, as in fact they are almost completely lacking in high-entropy gas in their inner CGM. There are only 48 such objects in TNG100: they are the low-mass, low-entropy galaxies already seen in \cref{fig:galPop_ssfr}. We discuss this interesting group of galaxies in \cref{sec:lowK_quenched}.

In the outer CGM the distribution of entropy is no longer bi-modal,
with the single peak moving to higher values in higher mass ranges as
expected. The exception to this statement is again the low-mass quenched
population, which curiously seems to be characterised by a lower
entropy distribution compared to the star-forming counterparts (see \cref{sec:lowK_quenched}).

When stacking a large number of different distributions, a bi-modal
distribution can occur naturally if all (or most) of the individual
distributions are of a similar bi-modal shape. However, a similar
distribution will be achieved if the sample contains only single-peak
distributions, some with the peaks in the low entropy regime and some
in the high entropy regime. We examined the individual distributions
and found that for the inner CGM, in both groups and all mass ranges,
\emph{most} ($\gtrsim 60$ percent) of the individual galaxies exhibit
distributions similar to the final stacking, that is, most of the
individual galaxy distributions are also bi-modal. The rest of the
distributions usually had a single peak, either in low entropy regime
($\log\,K<0$) or in the high-entropy regime ($\log\,K>0$). In the
star-forming groups, there were more low entropy single-peak
distributions and the opposite was true for the quenched group: most
single-peaked distributions were in the high-entropy regime.

In the transitional mass range, $10^{10}\--10^{11}\msun$, while both
groups show a similar bi-modal distribution of entropy in the inner
CGM, the quenched group distribution is shifted to higher values, even
for the low entropy peak, as a result of the energy injected by the
kinetic mode feedback of the AGN (see also \cref{fig:comp3_ent}). This
is also the case for the outer CGM. In the high-mass end, both
quenched and star-forming groups are practically identical, both in
the inner and outer CGM. In these galaxies, the deep potential of the
massive host halo is the dominant factor in setting the thermodynamic
state of the gas.

\begin{figure*}
  \centering
    \includegraphics[width=18cm,keepaspectratio,]{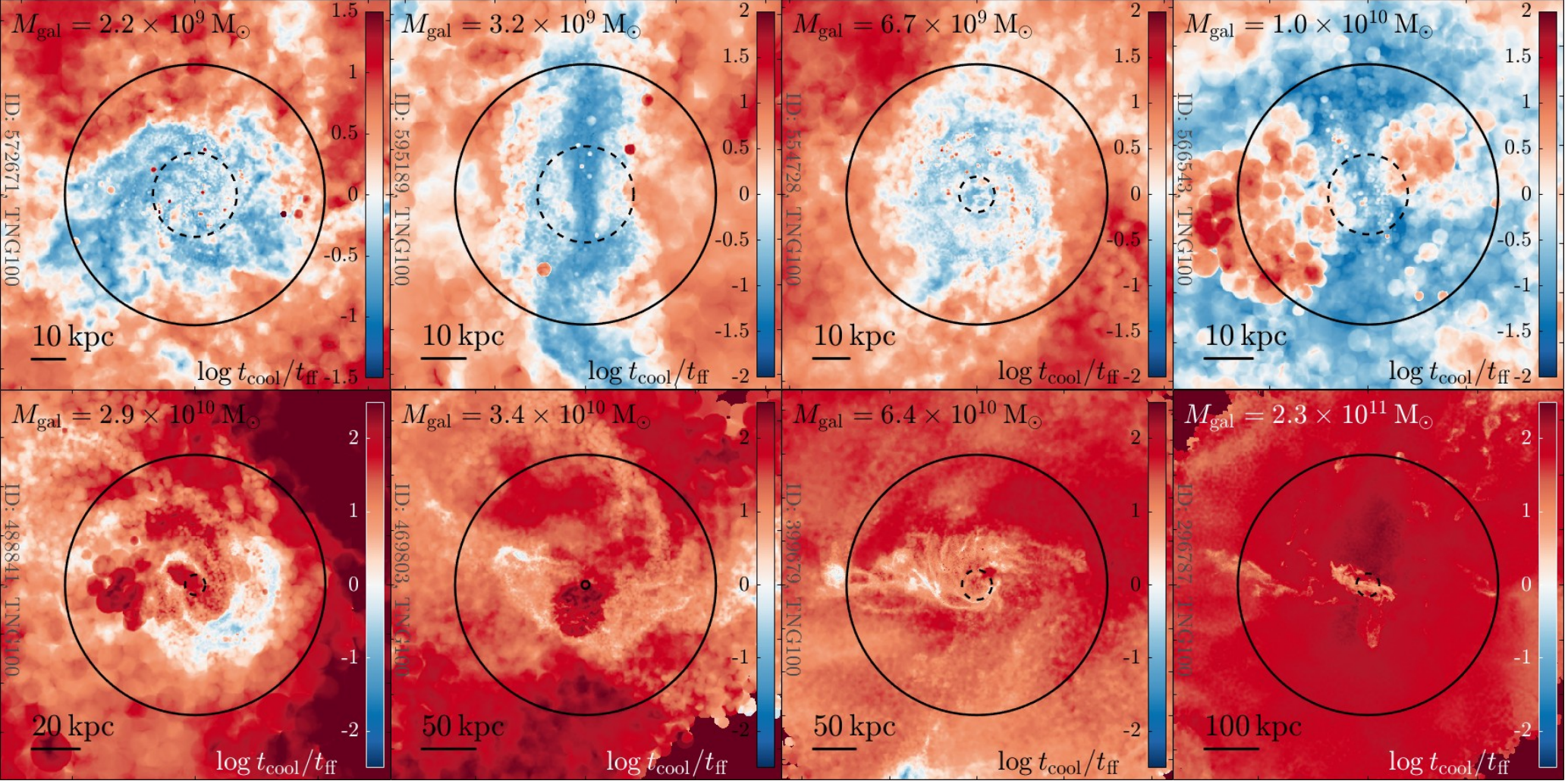}
    \caption{Maps of the same 8 representative galaxies shown in
      \cref{fig:Maps1}. The colour map corresponds to the ratio of the
      cooling time to free-fall time, $\log\, \tcool/\tff$. If
      $\tcool>\tff$ the gas cannot cool quickly enough and will most
      likely not be able to contribute to star formation. Outflows and `bubbles' originating from the central
      galaxy, found in all 3 panels in the middle row, are
      characterised by high $\tcool/\tff$ values.}
  \label{fig:tctffMaps}
\end{figure*} 

\section{Discussion and implications}\label{sec:discuss} 
From both the TNG simulations studied here, we can trace a relationship between the star-forming and thermodynamic states of the galaxy and its CGM with the stellar mass of the galaxy and its SMBH mass, which can be summarised as follows. In low-mass galaxies, $\lesssim10^{9.5}\msun$, the (\emph{non}-star-forming) gas in the galaxy is cold and dense with very low entropy values and short cooling times of order \mbox{$\sim 10\units{Myr}$} -- within the main stellar body of galaxies. The BH masses are also low, around the initial seed mass of $10^6\msun$. In most of these galaxies, AGN feedback is limited to the high accretion mode alone (\cref{fig:mbh_bhrat,fig:cgm_bhrat}). Under these conditions, the sSFR of these galaxies is high and they are typically found on the
star-forming sequence. Star formation in these galaxies is expected to
continue unabated due to the conditions in the gas immediately
surrounding the galaxy, which we have dubbed the inner CGM. The
cooling time in the inner CGM is of order $\sim 100\units{Myr}$
ensuring that any gas used up through star formation will be
replenished by gas cooling from the CGM. In these galaxies the BH
masses remain low due to low accretion rates. 
%As the galaxies increase their overall mass, the SN feedback can no
%longer push gas out and the BH mass can grow more quickly.

In more massive galaxies, $\sim 10^{10}\msun$, the BH mass is
$10^{7-8}\msun$. At these BH masses the kinetic feedback is present
and weak, but its effects can already be seen. Part of the galactic
gas is affected by the AGN and thus has elevated entropy and is
pushed out of the galaxies to become part of the CGM, however most of
the galactic gas is unaffected and the galaxy remains
star-forming. The entropy of the inner CGM is higher and thus the
cooling time is $\sim1\units{Gyr}$. In these galaxies the cooling of
the CGM can still balance gas depletion due to star formation
since both have roughly the same timescales.

Once the BH mass crosses the transitional mass scale of \mbox{$\log
  \Mbh \sim 8.3$}, energy injected by the kinetic mode kinetic
feedback becomes comparable ($\sim 1/3$) to that supplied by the
thermal mode (and therefore $1/4$ of the total energy output), and as
a result the galaxy entropy jumps by $2\--3$ orders of magnitude, and the
galaxy transitions from star-forming to quenched. Naturally, this transition will also manifest in observable properties of the galaxy such as colour \citep{Nelson2018a}. 

Throughout this paper (and elsewhere) the transition in galaxy and CGM properties - e.g.\@ sSFR, entropy, cooling time, halo gas content \citep{Davies2019,Terrazas2020}, color \citep{Nelson2018a}, size \citep{Genel2018}, and others - is seen to occur at specific values of stellar or BH mass as a result of the way the AGN feedback is implemented in the TNG model. Whether or not such a sharp transition is supported by observations is still an open question, but we would suggest that entropy, due to its dramatic increase at the transition, constitutes a powerful diagnostic for future studies.  

In galaxies close to the transition mass of $10^{10.5}\msun$ the kinetic
feedback overcomes the binding energy of the gas in the galaxy
\citep{Terrazas2020} and can deplete much of the galactic gas. In some
cases, even the inner CGM becomes extremely dilute
(\cref{fig:cgm_dens}): the kinetic feedback in the TNG model is
therefore `ejective' and star-formation quenching is triggered, in
effect, by gas removal from the star-forming regions.

However, the onset of kinetic mode feedback also leads to a change in
the CGM of the galaxies: the CGM is hotter, more dilute and of higher
entropy. Therefore, in a nutshell, the TNG AGN feedback ejects and
heats up the gas within and around galaxies. The change in conditions
may be observable, as the high-entropy CGM becomes a detectable X-ray
source, albeit strongly susceptible to the underlying gas density
\citep{Truong2020,Oppenheimer2020}.

The cooling time, both in the galactic gas as well as the CGM is now
quite long, of order $10\units{Gyr}$. These galaxies are quenched, and
expected to remain so, since the cooling from the CGM, even when
taking into account the multi-phase nature of the CGM, will not
replenish the gas in the galaxy. Only if gas is accreted into the galaxy
in a different way, by a merger for example, can star formation be
ignited in the galaxy.

It is important to note that a similar physical occurrence is found in the EAGLE simulations \citep{Davies2019,Oppenheimer2019}, a consensus which lends credibility to the outcome at the galaxy and CGM levels of from two very different galaxy formation models. 
In fact, in the details, this is achieved by a different implementation of BH feedback: the very effective kinetic mode in the TNG model operates at accretion rates well below the Eddington rate whereas in the EAGLE model, in which the AGN feedback efficiency is constant, the BH feedback is most effective at accretions rates close to the Eddington rate. This in turn highlights the challenges and uncertainties present in the modelling of this important process, as well as the need for detailed studies of the physical outcomes of its implementation. 

In the inner CGM we found the imprints of the kinetic mode feedback at
two mass scales -- when the first event of kinetic feedback injection occurs (even when the galaxy is still star-forming), and when the kinetic to thermal energy ratio becomes roughly constant. In both, the typical entropy of the CGM increases noticeably. In the outer CGM we only see the imprint of the latter. This is probably
due to two processes which work in tandem at this mass scale: only
when the energy injected by the kinetic mode feedback is high enough
can it influence the outer regions of the CGM. In addition, for galaxies
at this mass scale the dark matter host halo becomes massive enough to
sustain stable shocks emanating from its centre
\citep{Birnboim2003,Dekel2006} allowing them to expand and encompass
the entire host halo. It may be that the shocks being driven by the
kinetic feedback in lower mass systems cannot expand far enough to affect
the outer CGM.

As shown by \citet{Nelson2019}, the kinetic mode of AGN feedback
can drive strong bi-polar outflows of gas, which emerge perpendicular
to the galaxy disk (`path of least resistance'). Examples of this can
be seen in \cref{fig:Maps1}. In addition to gas removal from the
galaxy itself, the high velocity gas in the outflow
generates shocks as it encounters the ambient gas \citep[see idealized
  tests in][]{Weinberger2017}. These shocks, in turn, convert the
kinetic energy of the outflows into thermal energy and
lead to the heating of the CGM gas and a subsequent rise in entropy
and cooling times. Regions of high-entropy gas (`bubbles') embedded in
lower entropy gas (see \cref{fig:Maps1}) which rise due to their
buoyancy \citep{Bower2017,Keller2020} can also transport energy from
the central regions to the outer halo gas.
%In our analysis we identified a transitional mass scale in which the
%kinetic mode AGN feedback transforms the galaxy population from
%star-forming to quenched. This mass scale is roughly the same as the
%corresponding halo mass scale where one expects the CGM to become hot
%as shown by \citet{Dekel2006}. While this should not affect the
%galactic gas, the development of a hot atmosphere as the halo becomes massive
%enough to support large-scale accretion shocks \citep{Birnboim2003} would also result in a %rise in entropy in the outer CGM. It is very difficult to disentangle the relative %importance of the two processes in setting the state of the gas, especially considering that %there is most likely some synergy at play: the AGN feedback 

\subsection{On the preventative nature of the TNG BH feedback}\label{sec:futureSF}
We have argued that in galaxies in which the kinetic mode feedback
from the AGN is active, the thermodynamic conditions of the CGM (in
an averaged sense), namely high entropy and long cooling times, will
inhibit this gas from accreting and so prolong the quenched state of a galaxy.
However, we have also demonstrated that the CGM is comprised of both low and high entropy components, which therefore
have very different cooling timescales.

To assess whether the TNG black hole feedback model is indeed
preventative, we calculate an estimate for the total mass of (inner
and outer) CGM gas which may turn into stars in the near future for
each galaxy, and try to determine whether the galaxy would be
classified as quenched or not based on this value. To do so, we assume
that a gas cell in the CGM can only be accreted onto the galaxy and
form stars if it can cool in the time it would take it to reach the
centre, namely that its cooling time (\equnp{tcool1}), must be less
than the free-fall time (\equnp{tff}) at its position, $\tcool<\tff$.

This approach is inspired by the work of \citet{McCourt2012} who find
that the condition for the development of thermal instabilities in the
gas haloes surrounding galaxies, which allows gas to condense and cool
from the medium (and potentially from stars) to be $\tcool/\tff\lesssim 1$. Additional studies of the hot, dilute plasmas which populate the intra-cluster medium estimate the condition for condensation to be $\tcool/\tff\lesssim 10$ \citep{Sharma2012,Voit2015b,Voit2015c}

In \cref{fig:tctffMaps} we show the same 8 galaxies presented earlier
in \cref{fig:Maps1}, with the colour-maps showing the ratio of the
cooling time to the free-fall time, $\tcool/\tff$. In regions where
the cooling time is longer than the free-fall time (in red), the
gas cannot cool quickly enough and therefore cannot contribute to
star formation. We posit that only gas with $\tcool/\tff<1$
(white to blue pixels) can potentially fuel star formation in the
future. Low-mass galaxies contain large regions in which this ratio is
below 1, whereas in the CGM of high-mass galaxies the ratio
predominantly above 1 -- note the logarithmic scale of the color
bar. The regions in which outflows and hot and dilute `bubbles' were
formed (see panels in the middle row) are characterised by high values
of the ratio.

In each galaxy in our sample, for a given timescale $\ts=1\units{Gyr}$, we sum up the mass of all gas cells for which $\tcool\le \tff$ and $\tff\le \ts$
to estimate how much gas is available to fuel star formation, by assuming that
only gas cells which a) can reach the galaxy within the given timescale \emph{and} b) can cool within that time can in turn form stars. Studies of star-forming gas clouds show that only a small fraction of the cloud mass actually forms stars, with star formation efficiencies of order \mbox{$\sim 1$} percent per cloud free-fall time \citep[see][and references therein]{Krumholz2019}. We likewise assume that not all the available gas cooled from the CGM will form stars and assume a high fraction of \mbox{$\epsilon_\mathrm{SF}=5$} percent, so as to obtain a high estimate of the total amount of gas which can be transformed into stars\footnotemark. Thus the total amount by which the stellar mass can increase within a timescale
of $\ts$ is
\begin{equation}
    \Delta M_\mathrm{SF}(\ts)=\epsilon_\mathrm{SF} \sum m_\mathrm{g}\left(\tcool\le \tff\, \&\, \tff\le \ts\right),
\end{equation}
where we take $\ts=1\units{Gyr}$. 
\footnotetext{When taking a value of \mbox{$\epsilon_\mathrm{SF}=0.01$} per free-fall time convolved with the number of free-fall times found in the time span of $\ts=1\units{Gyr}$ for different parcels of gas we find that the integrated star-formation efficiency over all the gas is $\approx 0.04$.}

This estimate takes into account the \emph{local} thermodynamic
conditions in the CGM and therefore reflects the star-forming
potential of the multi-phase CGM surrounding each galaxy. An additional
assumption we make in this estimate is that the actual process of
forming stars from gas occurs on timescales much shorter than
$\ts$. Of course, in making this estimate of the stellar mass formed
in the future, we are deliberately ignoring the effects of feedback, SN and AGN, that in practice would continue to act on galaxies. As such, this estimate reflects the conditions within the CGM and should \emph{not} be considered as an attempt to predict the stellar mass in the future of a galaxy. Rather, these are very strong conditions to assess the `maintenance' i.e.\@ `preventative' nature of the AGN feedback we are seeking to prove.

\begin{figure}
  \centering
    %  \subfloat[$\ts=1\units{Gyr}$, $\epsilon_\mathrm{SF}=0.01$]{\label{fig:massProj_1G_1}
    \includegraphics[width=\columnwidth,keepaspectratio]{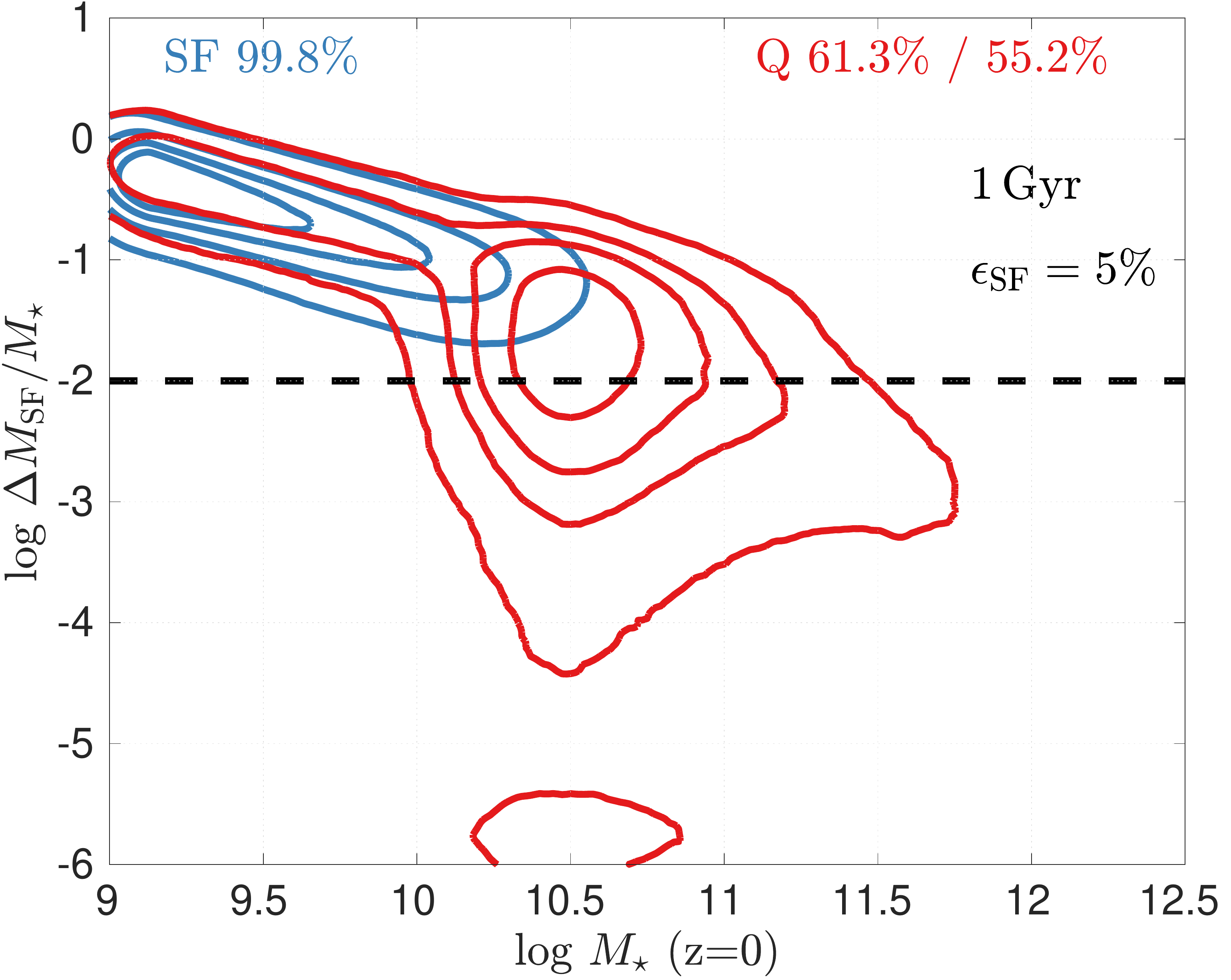}
    %  \subfloat[$\ts=1\units{Gyr}$, $\epsilon_\mathrm{SF}=0.05$]{\label{fig:massProj_1G_5}
    % \includegraphics[width=\columnwidth,keepaspectratio]{smassFuture2_cgm_sfe005_tf1_ts1G_TNG300.pdf}}
     \caption{An estimate for the fractional increase in
       stellar mass, for our TNG300 central galaxy sample (at \zeq{0}),
       over the next $\ts=1\units{Gyr}$ in the future. The estimate is
       based on the total amount of gas for which $\tcool<\tff$ and
       $\tff<\ts$, modulated by a high star formation efficiency of 5 percent. Galaxies that are star-forming or quenched at \zeq{0} are coloured in blue and
       red, respectively, with contours showing the 25, 50, 75 and 95
       percentiles of each population. The dashed line corresponds to
       the expected fractional mass increase of a galaxy with
       $\mathrm{sSFR}=10^{-11}\units{yr^{-1}}$ (our SF/Quenched
       threshold) over a period of $\ts=1\units{Gyr}$. The blue and
       red numbers at the top of each panel show the percentage of the
       star-forming and quenched \zeq{0} populations that are found
       above the black dashed lines after $1\units{Gyr}$. For the quenched population, we also
       quote the percentage of galaxies of stellar mass higher
       $10^{10}\msun$ found above this line. With a lower choice of star-formation efficiency of 1 percent (not shown), over 80 percent of the galaxies that are quenched at \zeq{0}, will remain quenched.}
  \label{fig:sfMassProj}
 \end{figure}

In \cref{fig:sfMassProj} we show the projected (in time) stellar mass increase, normalised by the current stellar mass $\Delta M_\mathrm{SF}/M_\mathrm{star}$, for all the galaxies in our TNG300 sample. We separate the galaxy population based on their star-forming state at \zeq{0}: star-forming galaxies in blue and quenched galaxies in red, and show the contours enclosing the 25,50,75
and 95 percentiles of the each population. The results shown assume a 
star formation efficiency of of \mbox{$\epsilon_\mathrm{SF}=0.05$}, for a representative timescale of $\ts=1\units{Gyr}$.

To asses whether the projected increase in stellar mass is typical of
a star-forming or quenched galaxy, we calculate what would be the
(normalised) increase in stellar mass, for a galaxy with exactly the
threshold value of $\mathrm{sSFR}=10^{-11}\units{yr^{-1}}$, within the
timescale $\ts$. This value is shown as the horizontal dashed line in
\cref{fig:sfMassProj}. Thus, galaxies found above this line will have
formed \emph{more} stars than a galaxy which has been marginally
quenched; i.e.\@ just on the sSFR threshold, for the entire time
$\ts$.

The percentages shown on the top of each panel indicate the fraction
of galaxies found above the dashed lines for the two populations;
namely, the percentages give the fraction of galaxies that, selected
to be e.g.\@ quiescent at \zeq{0} (red annotations), would become star-forming within the next \mbox{$1\units{Gyr}$}. For the quenched population at \zeq{0} (red annotations), we also show the percentage of galaxies with
stellar mass above $10^{10}\msun$ that are then found above the dashed
line after the chosen time span (here $1\units{Gyr}$). For longer timescales (not shown here) the galaxy distribution in the figure remains unchanged -- the radial extent corresponding to $\tff\gtrsim1\units{Gyr}$ encompasses nearly all of the gravitationally bound gas in the CGM (see \cref{sec:tffDef}). With the high value of the star-formation efficiency of 5 percent (and no energy feedback of any kind), nearly half of quenched galaxies above $10^{10}\msun$ (55 per cent), will still be considered quenched over the next $\units{Gyr}$, with the gas in the CGM in a thermodynamic state which is disadvantageous for future star formation. For values of \mbox{$\epsilon_\mathrm{SF}=0.01$} the percentage of massive quenched galaxies found above the line exceeds 80 percent.  

When taking the higher threshold of \mbox{$\tcool/\tff<10$} for the cooling condition to see how much star-forming gas will be available within $\tff<1\units{Gyr}$, the above results do not change by much: the fraction of quenched $10^{10}\msun$ and higher galaxies found above the SF-threshold line
is now $\sim 66$ percent for $\epsilon_\mathrm{SF}=0.05$ and $\sim 28
$ percent for $\epsilon_\mathrm{SF}=0.01$.

The line of arguments provided above demonstrates that the BH feedback
in the TNG model is not only ejective, but also preventative in the sense that it sets the conditions within the CGM to be non-conducive for fuelling future star formation. In fact, within the TNG implementation, both the quenching of star formation and the maintenance of a quenched state are accomplished by the same mode of BH feedback, namely the kinetic BH-driven winds at low-accretion rates.
%We could have proved the maintenance nature of the TNG BH feedback
%also by inspecting the star-formation histories or the thermodynamical
%state of the gas on a galaxy-by-galaxy basis starting from populations
%of TNG galaxies at some earlier time.

\subsection{On the state of the gas and star formation activity at earlier epochs}\label{sec:past}
We also inspected the thermodynamic state of galaxy and halo gas at
earlier epochs: by and large, most of the results of this study are already
in place at $z\sim 0.5$ (\mbox{$ \sim 5\units{Gyr}$} ago), with long cooling times in all gaseous components, for example. As expected, the star-forming sequence is found at higher values of sSFR \citep{Donnari2019}, yet the shape of the distribution in the entropy-stellar mass plane is remarkably similar to that found at \zeq{0} (\cref{fig:galPop_ssfr}), with the transitional stellar mass scale where the entropy increases and galaxies become predominately quenched located at the typical $10^{10.5}\msun$ scale as for more recent epochs. The increase in entropy at the transitional mass scale is seen in the inner and outer CGM components as well, and the cooling times in the CGM of galaxies above the transitional mass scale is of order $10\units{Gyr}$, showing that preventative aspect of AGN feedback is effective since this time to the present.

A complementary, in-depth study of the evolution over the history of individual galaxies may still be necessary for a complete understanding of the role the AGN plays in the evolution of the gas conditions in tandem with the star-forming
state of the galaxy.

\subsection{Addressing issues of resolution}\label{sec:resolution}
The analysis presented above rests on the results of both the TNG300 simulation, with its extremely large sample of galaxies, as well as the higher-resolution TNG100 simulation. In this way we could better explore the distribution of galaxy properties over a large sample with TNG300 (especially in the high-mass end), as was done in \cref{fig:galPop_ssfr,fig:galPop_mbh,fig:galPop_temp_dens_tcool,fig:galPop_cgm_bh,fig:compTransition,fig:sfMassProj}, and also study important small-scale features which are better resolved in TNG100 (\cref{fig:Maps1,fig:gasHistograms,fig:tctffMaps}). 

To study the effects of the simulation resolution, we carried out our population analysis on the TNG100 simulation. We find that the salient features remain unchanged when examining the better-resolved population: the transitional stellar mass scale is identical and the typical average values of entropy, temperature, density, and cooling times shown in \cref{fig:galPop_temp_dens_tcool} are by and large unchanged as well. We show a few examples of these comparisons in \cref{sec:resCompare} and \cref{fig:ssfr_100_300,fig:tcool_100_300}. The results presented in \cref{sec:futureSF} and \cref{fig:sfMassProj} also remain the same when conducted on TNG100 galaxies (see \cref{fig:sfMassProj_100_300}). In examining the multi-phase aspect of the CGM, \cref{sec:multi-phase} and \cref{fig:gasHistograms}, we repeated our analysis with the TNG300 galaxies and found similar trends in the properties of the CGM.

Thus, our conclusions do not depend on the resolution of our simulations, though we stress that this is also due to our studying averaged values both within the galaxies and over the galaxy populations. Simulations with increased resolution would undoubtedly reveal features and properties of the CGM, particularly pertaining to regions of low entropy, which the simulations studied here cannot capture \citep[e.g.\@][]{Voort2019,Suresh2019}. To gain insight as to how these small-scale differences may alter the way BH feedback affects the CGM, it would be interesting to carry out a detailed comparison between individual, similar-mass galactic systems in the TNG300, TNG100, as well as the higher resolution TNG50 simulation \citep{Pillepich2019,Nelson2019}, an endeavour we defer to  a later date. 

\section{Summary}\label{sec:summary}
In this paper we have quantified the effects of AGN feedback, as
realised in the TNG model, on the evolution of galaxies by specifically focusing on the physical conditions of the \emph{non}-star-forming gas within the galaxy and beyond, i.e.\@ in the surrounding circumgalactic medium. To that end, we have examined the population of central galaxies in the stellar mass range
$10^{9-12.5}\msun$ from the TNG100 and TNG300 simulations at \zeq{0}. We have drawn our conclusions from global properties of the entire population under the assumption that the galaxy histories are imprinted on their present day properties -- an assumption which may not hold in all cases. We have relied on a numerical framework whereby the feedback from SMBHs is implemented as thermal energy injection at high BH accretion rates and kinetic BH-driven winds at low
accretion rates capable of driving large-scale outflows.

In particular, we have quantified the temperature, density, entropy,
and cooling times of the \emph{non}-star-forming, gravitationally-bound
gas of tens of thousands of galaxies, by broadly distinguishing among
galactic gas (within twice the stellar half-mass radius), the inner
CGM (between the latter and the gas half-mass radius), and the outer
CGM (beyond the gas half-mass radius). We have connected the gas
thermodynamics to the star-forming state of the galaxies
(e.g.\@ star-forming vs.\@ quiescent) and hence to the current and
time-integrated properties and activity of their central SMBHs.

Our key conclusions are the following:
\begin{itemize} 
\setlength\itemsep{1em}
\item[*] Gas entropy is a sensitive diagnostic of the effects of feedback energy injection and is a strong indicator of the cooling time in the gas and, by extension, of the quenched vs.\@ non-quenched state of galaxies;
\item[*] The same AGN feedback channel can simultaneously function as `ejective' and `preventative'.
\end{itemize}

More specifically, our findings can be summarised as follows:
\begin{itemize} 
\setlength\itemsep{1em}
\item The population of \zeq{0} central galaxies in the TNG
  simulations (TNG100 and TNG300) transitions between a largely
  star-forming population at low masses, which forms a tight
  `star-forming sequence' in the sSFR-stellar mass plane, to a mostly
  quiescent population at higher masses (\cref{fig:galPop_ssfr_ent}).  The
  transition between the two populations comes about when the kinetic
  mode feedback of the AGN model (BH-driven winds) accounts for
  roughly 1/4 of the total energy output over the lifetime of the BH
  (\cref{fig:bh_stuff,fig:galPop_mbh}). This occurs at a BH mass scale of
  $\sim10^{8.3}\msun$, which corresponds to a stellar mass scale of
  $\sim 10^{10.5}\msun$.

\item The transition between star-forming and quiescent galaxies
  manifests itself in the gas properties of the galaxy in the form of
  a rapid increase in the average entropy of the galactic gas (\cref{fig:galPop_ent_ssfr},
  excluding star-forming gas) which corresponds to a jump in the
  typical gas cooling times from \mbox{$10\--100\units{Myr}$} to
  \mbox{$1\--10\units{Gyr}$} (\cref{fig:gal_tcool}).
  
\item The changes in galactic gas properties are mirrored in the CGM,
  showing that the kinetic mode feedback also affects the gas beyond
  the galaxy. Even very small amounts of energy injected via the
  kinetic mode feedback result in a noticeable jump in the CGM average
  entropy. In other words, in quenching galaxies, the effects of the
  kinetic mode AGN Feedback leave an imprint also on the outer reaches
  of the gas halo, up to distances of several hundreds of kiloparsecs
  (\cref{fig:galPop_temp_dens_tcool}). This separation in the typical
  thermodynamic properties of the gas within and around star-forming
  vs.\@ quiescent galaxies can be tested with X-ray observations
  \citep[see][]{Truong2020,Oppenheimer2020}.
  
 \item Within the framework of the TNG model, AGN
   feedback, and particularly the kinetic mode feedback at low BH
   accretion rates, ejects and heats up the gas within and around
   galaxies (\cref{fig:compTransition}). Namely, it acts as both an
   `ejective' feedback -- by expelling star-forming gas from the
   galaxy and triggering quenching -- as well as a
   `preventative' feedback -- by increasing the average CGM entropy and
   cooling time and making them non-conducive for fueling future star
   formation (\cref{fig:tctffMaps,fig:sfMassProj}). This remains true
   even though the CGM itself is multi-phase, with both high and low
   entropy (diffuse and dense) components (\cref{fig:gasHistograms}). While for some systems the 'ejective' part of the feedback is enough to ensure long-term quiescence, a significant sub-population would start to form stars again unless preventative AGN feedback is operating.
\end{itemize}

In conclusion, with this work we have examined how the three
components -- the SMBH, the galaxy, and the CGM -- are all
interconnected in their evolution within the context of a numerical
model for galaxy formation that yields galaxy populations in
reasonable agreement with observations. While the latter may not be
fully realistic in detail owing to limited resolution 
\citep[see e.g.\@][]{Voort2019,Suresh2019}, it still provides us with a rich and plausible laboratory to get insights and inspirations for what may happen in the real Universe. 
%Particularly, we have shown that the same AGN feedback
%channel (in the TNG case, the kinetic BH-driven winds that act at low
%BH accretion rates) can be simultaneously ejective and preventative and
%can have far-reaching effects, not only on the star-formation activity
%of galaxies, but also on the thermodynamical state of the gas at tens
%or hundreds of kiloparsecs from their centers.

Importantly, global evolutionary properties, such as a long-lasting quenched state, can be accompanied by a diversity of halo gas states, spanning orders of
magnitude in entropy values within the same galaxy or gaseous
halo. Finally and more generally, we hope we have inspired future
studies of the formation and evolution of galaxies to treat the entire galactic ecosystem -- from the small scales of the central BH, to the
outer reaches of the CGM -- as a complex, inter-dependent structure,
with each component having its own part to play in determining the
future of the system.

\section*{Data Availability}
\begin{itemize}
    \item The simulation data used for this analysis is publicly available through the Illustris TNG project website \citep[see][for details]{Nelson2019a}: \href{http://www.tng-project.org/data/}{http://www.tng-project.org/data} .

\item The scripts used to analyse the data and generate the figures can be found here: \href{https://github.com/eladzing/matlab_scripts.git}{https://github.com/eladzing/matlab\_scripts.git}. Please contact the corresponding author for assistance at \href{mailto:elad.zinger@mail.huji.ac.il}{elad.zinger@mail.huji.ac.il}. 
\end{itemize}

%\footnotetext{\href{http://www.tng-project.org/}{http://www.tng-project.org/}}\%citep[see][for details]{Nelson2019a}.

\section*{Acknowledgements}
The authors thank the anonymous referee for their time and effort in providing
constructive comments and suggestions which improved the paper. The authors would like to thank Martina Donnari for input on the
star formation rates of TNG galaxies and for many fruitful discussions. FM acknowledges support through the Program ``Rita Levi Montalcini'' of the Italian MIUR. MV acknowledges support through an MIT RSC award, a Kavli Research Investment Fund, NASA ATP grant NNX17AG29G, and NSF grants AST-1814053, AST-1814259 and AST-1909831. The flagship simulations of the IllustrisTNG project used in this work (TNG100 and TNG300) have been run on the HazelHen Cray XC40-system at the High Performance Computing Center Stuttgart as part of project GCS-ILLU of the Gauss centres for Supercomputing (GCS, PI: Springel).

\bibliographystyle{mnras}
\bibliography{centralQ.bib}
%\input{bibList.tex}

%\newpage

\appendix 
\section{Comparison between the TNG100 and TNG300 galaxy samples - resolution effects} \label{sec:resCompare}
\begin{figure}
  \centering
   \subfloat[Galaxy sSFR on the entropy-stellar mass plane - TNG100]{\label{fig:ssfr100}
    \includegraphics[height=6.6cm,keepaspectratio]{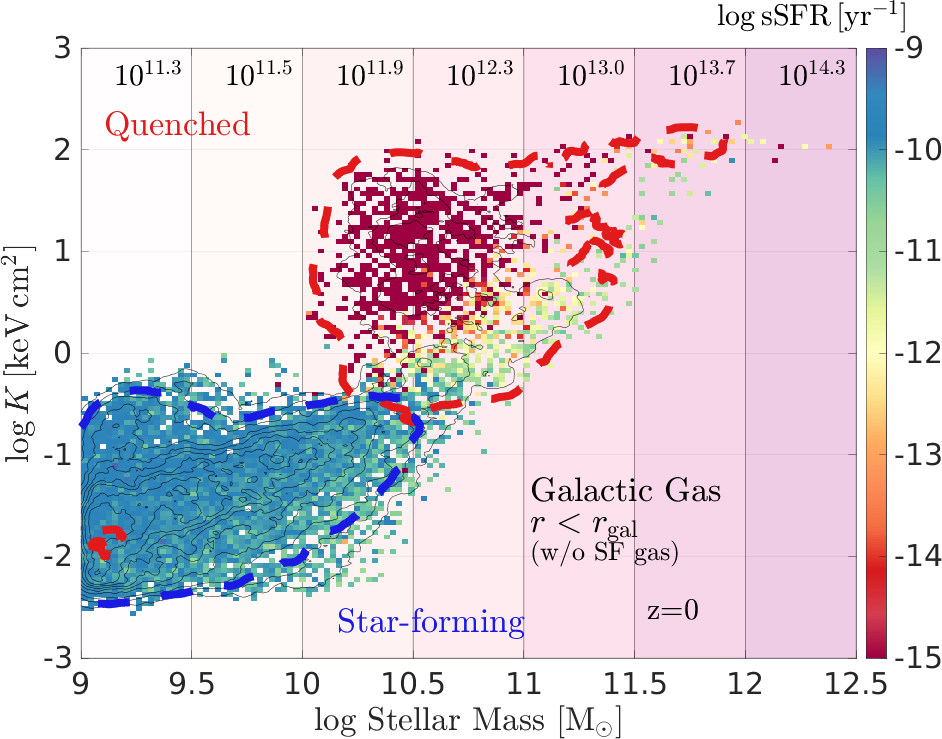}} \\
  \subfloat[Galaxy sSFR on the entropy-stellar mass plane - TNG300]{\label{fig:ssfr300}
    \includegraphics[height=6.6cm,keepaspectratio]{massEnt/massEnt_ssfr_Gal_centrals_snp99_TNG300.png}}
   \caption{The \zeq{0} central galaxy population on the entropy-stellar mass 
       plane, coloured by sSFR, shown for the TNG100 galaxy sample \subrfig{ssfr100} and the TNG300 galaxy sample \subrfig{ssfr300} (shown earlier in \cref{fig:galPop_ent_ssfr}). Black contours show the distribution of all galaxies in the plane and the blue and red contours each enclose 90 per-cent of the SF/Quenched population. The distribution of the galaxy population is similar between the two simulation boxes, as are the typical entropy values for the star-forming and quenched population. }
  \label{fig:ssfr_100_300}
\end{figure}
\begin{figure*}
  \centering
   \subfloat[Cooling time of the galactic gas - TNG100]{\label{fig:tcool_gal_100}
    \includegraphics[width=\columnwidth,keepaspectratio]{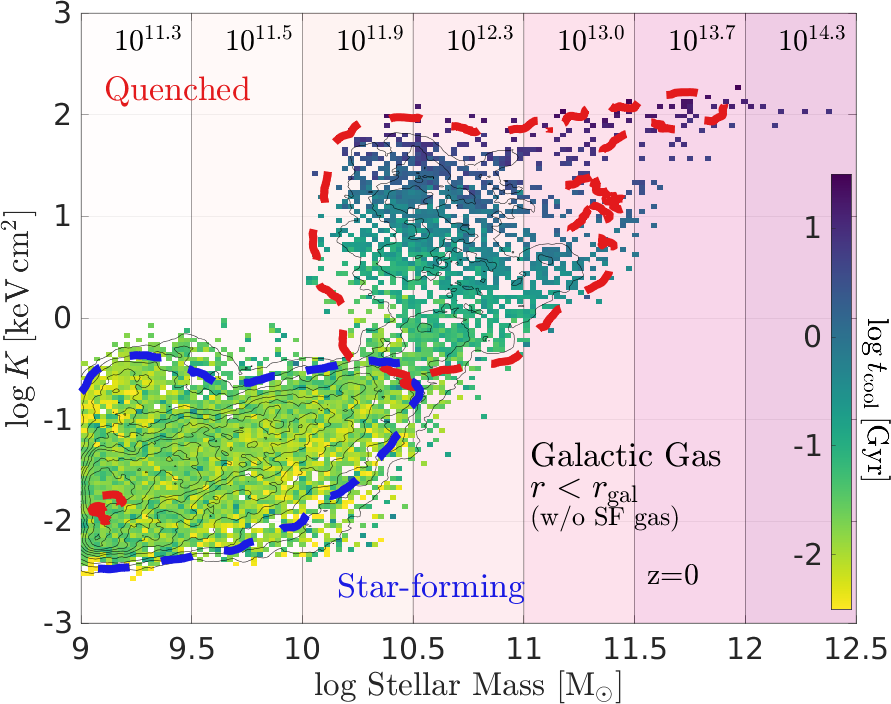}}
  \subfloat[Cooling time of the galactic gas - TNG300]{\label{fig:tcool_gal_300}
    \includegraphics[width=\columnwidth,keepaspectratio]{massEnt/massEnt_tcool_Gal_centrals_snp99_TNG300.png}} \\
    \subfloat[Cooling time of the inner CGM - TNG100]{\label{fig:tcool_cgm_100}
    \includegraphics[width=\columnwidth,keepaspectratio]{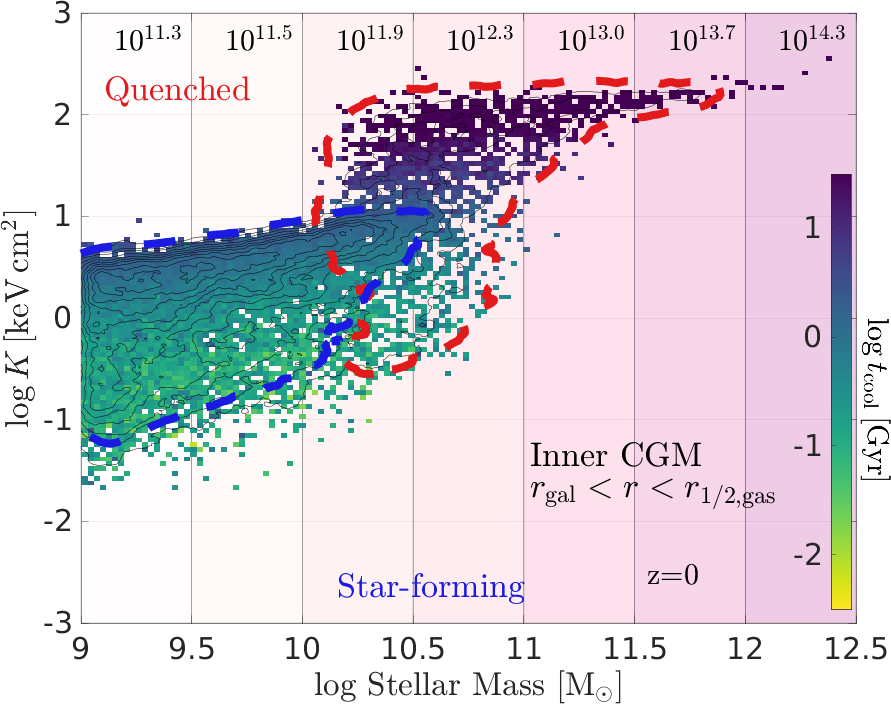}}
  \subfloat[Cooling time of the inner CGM - TNG300]{\label{fig:tcool_cgm_300}
    \includegraphics[width=\columnwidth,keepaspectratio]{massEnt/massEnt_tcool_CGM_centrals_snp99_TNG300.png}} 
   \caption{Comparison between the TNG100 sample (left column) and the TNG300 sample (right column) of the distribution in the entropy-stellar mass plane of the galactic gas (top) and inner CGM (bottom), with the colours corresponding to the cooling times (Panels \subrfig{tcool_gal_300} and \subrfig{tcool_cgm_300} shown earlier in \cref{fig:gal_tcool,fig:cgm_tcool}). The average entropy vales of the inner CGM component are very similar across the two samples and the values of the cooling times also do not appear to vary due to the different resolution.}
  \label{fig:tcool_100_300}
\end{figure*}
\begin{figure*}
  \centering
    \subfloat[TNG100]{\label{fig:sfmassProj_100}
    \includegraphics[width=\columnwidth,keepaspectratio]{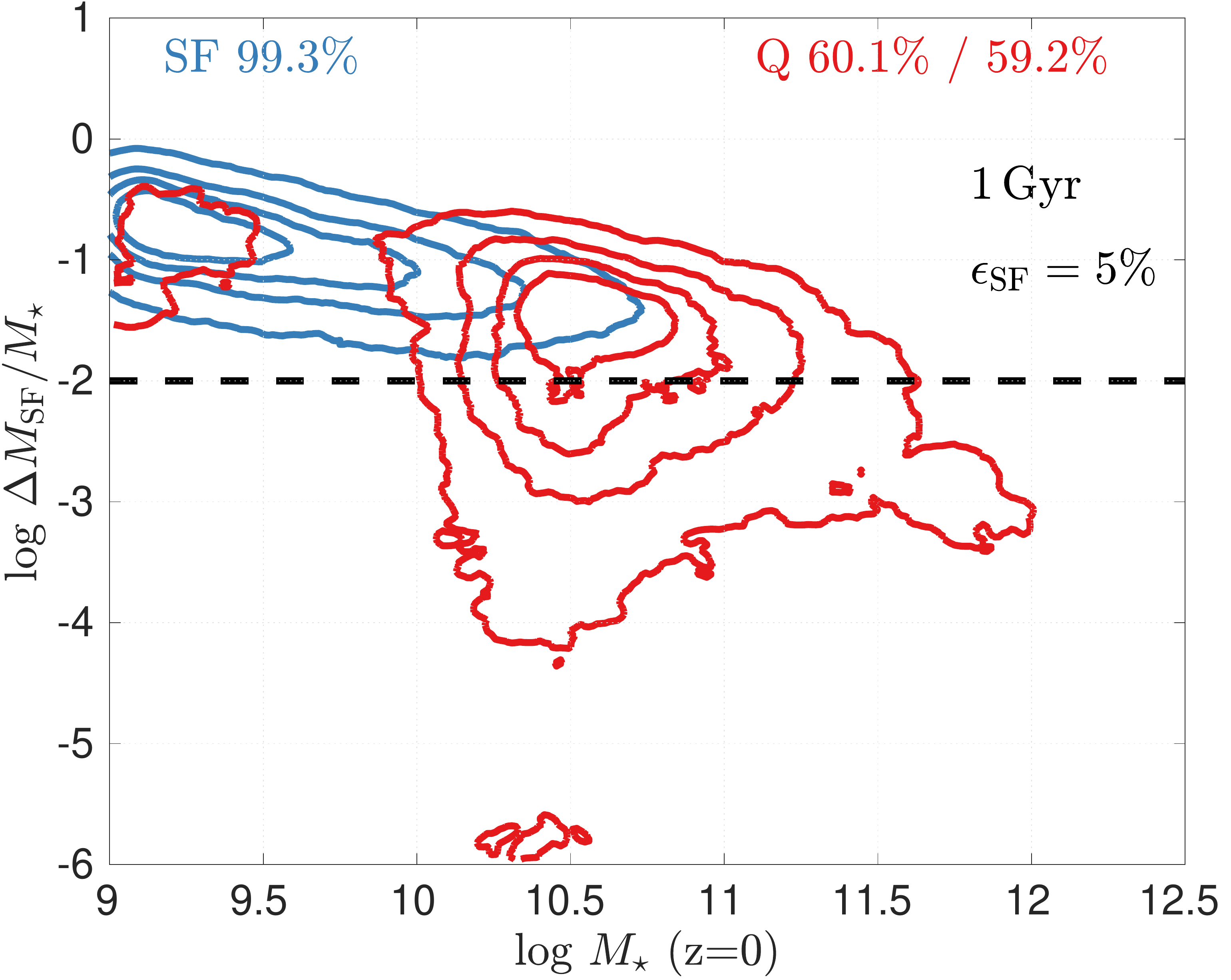}}
    \subfloat[TNG300]{\label{fig:sfmassProj_300}
    \includegraphics[width=\columnwidth,keepaspectratio]{smassFuture2_cgm_sfe005_tf1_ts1G_TNG300.pdf}}
     \caption{Comparison of the estimated fractional increase in
       stellar mass, between the TNG100 \subrfig{sfmassProj_100} and TNG300 \subrfig{sfmassProj_300} central galaxy sample (at \zeq{0}),
       over the next $\ts=1\units{Gyr}$ in the future. Panel \subrfig{sfmassProj_300} is identical to \cref{fig:sfMassProj}. Galaxies that are star-forming or quenched at \zeq{0} are coloured in blue and
       red, respectively, with contours showing the 25, 50, 75 and 95
       percentiles of each population. The dashed line corresponds to
       the expected fractional mass increase of a galaxy with
       $\mathrm{sSFR}=10^{-11}\units{yr^{-1}}$ (our SF/Quenched
       threshold) over a period of $\ts=1\units{Gyr}$. The blue and
       red numbers at the top of each panel show the percentage of the
       star-forming and quenched \zeq{0} populations that are found
       above the black dashed lines after $1\units{Gyr}$. For the quenched population, we also quote the percentage of galaxies of stellar mass higher $10^{10}\msun$ found above this line. The results for the two samples are very similar, and thus are not resolution dependant.}
  \label{fig:sfMassProj_100_300}
 \end{figure*}

In this section we present several comparisons of our results carried out between the galaxy populations of TNG100 and TNG300 simulations, in an effort to probe the effects of resultion on our findings. The TNG100 and TNG300 galaxy samples number \mbox{$\sim10,000$} and \mbox{$\sim 135,000$} respectively (see \cref{sec:galSample}).  

In \cref{fig:ssfr_100_300} we show the distribution of the galactic gas of central galaxies at \zeq{0} for the two simulation boxes on the entropy-stellar mass plane, with the colors corresponding to the sSFR of the galaxies (shown earlier in \cref{fig:galPop_ent_ssfr}. The advantages of using the much larger TNG300 sample are clear as the distribution is much more `fleshed-out' with this sample, especially at the high-mass end. 

The transition between star-forming and quenched occurs at the same mass scale, as expected, since this threshold is set by the physical model and the transition between the thermal and kinetic mode (\cref{sec:AGNmodel}). In addition, the entropy values for the star-forming and quenched population are by and large identical, with the same entropy `floor' (as set by the physical model) and `ceiling', and the same typical values for the peak of the distribution. 

In \cref{fig:tcool_100_300} we show the distribution for the galactic gas and inner CGM, top and bottom row, respectively, with the colors showing the cooling time in the gas, for the TNG100 sample on the left and TNG300 sample on the right (\cref{fig:tcool_gal_300,fig:tcool_cgm_300} are identical to \cref{fig:gal_tcool,fig:cgm_tcool}, respectively). The distribution of galaxies for inner CGM entropy values is also similar in both samples. Even though the gas is better resolved in the TNG100 sample, the average values of the cooling time in both the galactic and CGM gas is also practically identical between the two simulations. Higher resolution usually entails a more multi-phase medium, but it appears that this has little effect when considering median value of the (mass-weighted average) cooling time for the galaxies in a given region of the plane. 

In \cref{fig:sfMassProj_100_300} we repeat the analysis of \cref{sec:futureSF} for estimating the potential increase in stellar mass due to accretion from the CGM. In \cref{fig:sfmassProj_100} we show the results for the TNG100 sample, with the TNG300 sample results shown for comparison in \cref{fig:sfmassProj_300} (previously shown in \cref{fig:sfMassProj}). Here too we see that despite the difference in resolution, the quenched and star-forming populations inhabit similar regions in the plot, and the estimated fractional increase in stellar mass is by and large the same. 

\section{Low-mass/low-entropy quenched population}\label{sec:lowK_quenched}
In \cref{fig:galPop_ent_ssfr} we encountered a population of
low-mass, low-entropy galaxies, marked in the figure by the red
contour found in the lower-left corner. These galaxies comprise
roughly 3.5 percent of the TNG300 galaxies in our sample (with
stellar mass above $10^9\msun$), and about 15 percent of the quenched
galaxy population. It should be noted that we find these galaxies
\emph{after} removing `back-splash' galaxies as described in \cref{sec:galSample}. 

In TNG100, a similar population is found (see \cref{fig:ssfr100}), but of much lower relative numbers: only 0.5 percent of the total population and only
3.4 percent of the quenched population. The fact that the relative
size of this population is different between TNG100 and TNG300
points to either a resolution or volume dependent effect. We note that
when lowering the sSFR threshold for what should be considered a
quenched galaxy by a factor of two, this galaxy population shrinks by
roughly 25 percent, but still comprises 11 percent of the quenched
galaxy population.

In \cref{fig:mbh_ssfr} we learn that the BHs in all these galaxies
have low mass, and all have not yet experienced a single kinetic mode
injection event, as can be seen in \cref{fig:mbh_bhrat}. As such, the
inner CGM of these galaxies is characterised by relatively low
entropies (\cref{fig:cgm_temp}). The outer CGM of these galaxies is no
different overall from most other star-forming galaxies of similar
masses (\cref{fig:out_temp,fig:out_dens,fig:out_tcool}). We cannot
rely on \cref{fig:galPop_temp_dens_tcool} to learn about the
temperatures, densities, and cooling times of these galaxies directly
since the values of these quantities are averaged over all galaxies
within a pixel, most of which are likely star-forming. However, based on the entropy values for these galaxies one can confidently surmise that their cooling times are most likely similar to those of the star-forming population of order
$10\units{Myr}$ in the galactic gas and $100\units{Myr}$ in the inner CGM.

When examining the conditions in the CGM of similar galaxies in the
TNG100 simulation, by studying the red histogram of
\cref{fig:gasHist_cgm_low,fig:gasHist_out_low}, we see that these
galaxies have entropy distributions which are skewed to lower values in
the inner CGM and possess a more massive low-entropy component than their
star-forming counterparts.

All this points to the fact that whatever process is quenching these
galaxies, it is most likely \emph{not} energy injected by either
SN or AGN kinetic feedback -- the conditions within the gas of these
galaxies is conducive to star formation at a rate similar to that of
galaxies on the star-forming sequence. Indeed, when examining the
future star-forming potential of these galaxies, as was done in
\cref{fig:sfMassProj}, most of them occupy the same region as similar
mass star-forming galaxies. In light of this, we are confident that
the existence of this small group does not affect our conclusions as
to the role of AGN feedback in instigating the quenching of the
general population of galaxies.

We defer an in-depth analysis as to the reason this specific galaxy
population is quenched to a later date, but offer several conjectures
as to why these galaxies are found in this state. It is possible that
these are `back-splash' galaxies which evaded our filtering
procedure, i.e.\@ galaxies who are centrals at \zeq{0} but were identified as satellites at some point in the past $5\units{Gyr}$. When we further remove all galaxies which were identified as satellites within the past $12\units{Gyr}$, the low-mass, low-entropy galaxies now comprise 3.3 percent of total TNG300 sample and roughly 13 percent of the quenched sub-sample. In TNG100, most galaxies from this group were found to be
`back-splash' galaxies and only a negligible number of quenched low-mass
galaxies remain after their removal, based on the criteria set out in \cref{sec:galSample}.

Another possibility is that the quenched state of these galaxies is a
transient short-lived episode of low SFR. We tested this conjecture by
examining the smoothed SFR of these systems over the past
$200\units{Myr}$, calculated by tallying up the total amount of
stellar mass formed over this period. By this metric, i.e.\@ when the
star-formation rates are averaged across longer periods of time,
roughly 25 percent of these systems are actually considered
star-forming. As it stands, it appears that within the TNG model there is a quenching channel for low-mass galaxies which is likely resolution dependent. 
\label{lastpage}

\end{document}